\documentclass[amsmath,amssymb,aps,showpacs,twocolumn,superscriptaddress,letterpaper,longbibliography, floatfix]{revtex4-2}

\usepackage{graphicx}% Include figure files
\usepackage{dcolumn}% Align table columns on decimal point
\usepackage{bm}% bold math
\usepackage{hyperref}% add hypertext capabilities
\usepackage{lineno}% Enable numbering of text and display math
\linenumbers\relax % Commence numbering lines
\usepackage{natbib} % Reimplementation of the LATEX \cite command, to work with both author–year and numerical citations
\usepackage{amsmath, amssymb} % Options for displaying equations
\usepackage{color}
\usepackage[dvipsnames]{xcolor}
\usepackage{comment} % Adding comments
\usepackage{hyperref} % Hyperreferences
\usepackage{float} % floating figs
\usepackage{makecell} % moving to next line to tables
\usepackage{cleveref}
\usepackage{tikz}

\hypersetup{breaklinks = true, colorlinks = true, citecolor = blue, linkcolor = blue, urlcolor = blue}

\newcommand{\figures}{figures/}

\usepackage[textsize=tiny,%
    color=orange!25,%
    linecolor=orange,%
    bordercolor=lightgray]{todonotes}
\begin{document}

%\linenumbers
\nolinenumbers

\title{First study of neutrino angle reconstruction using quasielastic-like interactions in MicroBooNE}

% List of institutions in command form:
\newcommand{\ANL}{Argonne National Laboratory (ANL), Lemont, IL, 60439, USA}
\newcommand{\Bern}{Universit{\"a}t Bern, Bern CH-3012, Switzerland}
\newcommand{\BNL}{Brookhaven National Laboratory (BNL), Upton, NY, 11973, USA}
\newcommand{\UCSB}{University of California, Santa Barbara, CA, 93106, USA}
\newcommand{\Cambridge}{University of Cambridge, Cambridge CB3 0HE, United Kingdom}
\newcommand{\CIEMAT}{Centro de Investigaciones Energ\'{e}ticas, Medioambientales y Tecnol\'{o}gicas (CIEMAT), Madrid E-28040, Spain}
\newcommand{\Chicago}{University of Chicago, Chicago, IL, 60637, USA}
\newcommand{\Cincinnati}{University of Cincinnati, Cincinnati, OH, 45221, USA}
\newcommand{\CSU}{Colorado State University, Fort Collins, CO, 80523, USA}
\newcommand{\Columbia}{Columbia University, New York, NY, 10027, USA}
\newcommand{\Edinburgh}{University of Edinburgh, Edinburgh EH9 3FD, United Kingdom}
\newcommand{\FNAL}{Fermi National Accelerator Laboratory (FNAL), Batavia, IL 60510, USA}
\newcommand{\Granada}{Universidad de Granada, Granada E-18071, Spain}
\newcommand{\IIT}{Illinois Institute of Technology (IIT), Chicago, IL 60616, USA}
\newcommand{\ICL}{Imperial College London, London SW7 2AZ, United Kingdom}
\newcommand{\Indiana}{Indiana University, Bloomington, IN 47405, USA}
\newcommand{\Kansas}{The University of Kansas, Lawrence, KS, 66045, USA}
\newcommand{\KSU}{Kansas State University (KSU), Manhattan, KS, 66506, USA}
\newcommand{\Lancaster}{Lancaster University, Lancaster LA1 4YW, United Kingdom}
\newcommand{\LANL}{Los Alamos National Laboratory (LANL), Los Alamos, NM, 87545, USA}
\newcommand{\Louisiana}{Louisiana State University, Baton Rouge, LA, 70803, USA}
\newcommand{\Manchester}{The University of Manchester, Manchester M13 9PL, United Kingdom}
\newcommand{\MIT}{Massachusetts Institute of Technology (MIT), Cambridge, MA, 02139, USA}
\newcommand{\Michigan}{University of Michigan, Ann Arbor, MI, 48109, USA}
\newcommand{\MSU}{Michigan State University, East Lansing, MI 48824, USA}
\newcommand{\Minnesota}{University of Minnesota, Minneapolis, MN, 55455, USA}
\newcommand{\Nankai}{Nankai University, Nankai District, Tianjin 300071, China}
\newcommand{\NMSU}{New Mexico State University (NMSU), Las Cruces, NM, 88003, USA}
\newcommand{\Oxford}{University of Oxford, Oxford OX1 3RH, United Kingdom}
\newcommand{\Pitt}{University of Pittsburgh, Pittsburgh, PA, 15260, USA}
\newcommand{\QMUL}{Queen Mary University of London, London E1 4NS, United Kingdom}
\newcommand{\Rutgers}{Rutgers University, Piscataway, NJ, 08854, USA}
\newcommand{\SLAC}{SLAC National Accelerator Laboratory, Menlo Park, CA, 94025, USA}
\newcommand{\SDSMT}{South Dakota School of Mines and Technology (SDSMT), Rapid City, SD, 57701, USA}
\newcommand{\Maine}{University of Southern Maine, Portland, ME, 04104, USA}
\newcommand{\TelAviv}{Tel Aviv University, Tel Aviv, Israel, 69978}
\newcommand{\UTA}{University of Texas, Arlington, TX, 76019, USA}
\newcommand{\Tufts}{Tufts University, Medford, MA, 02155, USA}
\newcommand{\VTech}{Center for Neutrino Physics, Virginia Tech, Blacksburg, VA, 24061, USA}
\newcommand{\Warwick}{University of Warwick, Coventry CV4 7AL, United Kingdom}
%\newcommand{\Yale}{Wright Laboratory, Department of Physics, Yale University, New Haven, CT, 06520, USA}
%%\newcommand{\listerThanks}{Now at University of Wisconsin, Madison}

% So that institutions appear in alphabetical order:
\affiliation{\ANL}
\affiliation{\Bern}
\affiliation{\BNL}
\affiliation{\UCSB}
\affiliation{\Cambridge}
\affiliation{\CIEMAT}
\affiliation{\Chicago}
\affiliation{\Cincinnati}
\affiliation{\CSU}
\affiliation{\Columbia}
\affiliation{\Edinburgh}
\affiliation{\FNAL}
\affiliation{\Granada}
%\affiliation{\Harvard}
\affiliation{\IIT}
\affiliation{\ICL}
\affiliation{\Indiana}
\affiliation{\Kansas}
\affiliation{\KSU}
\affiliation{\Lancaster}
\affiliation{\LANL}
\affiliation{\Louisiana}
\affiliation{\Manchester}
\affiliation{\MIT}
\affiliation{\Michigan}
\affiliation{\MSU}
\affiliation{\Minnesota}
\affiliation{\Nankai}
\affiliation{\NMSU}
\affiliation{\Oxford}
\affiliation{\Pitt}
\affiliation{\QMUL}
\affiliation{\Rutgers}
\affiliation{\SLAC}
\affiliation{\SDSMT}
\affiliation{\Maine}
%\affiliation{\Syracuse}
\affiliation{\TelAviv}
%\affiliation{\Tennessee}
\affiliation{\UTA}
\affiliation{\Tufts}
%\affiliation{\UCL}
\affiliation{\VTech}
\affiliation{\Warwick}
%\affiliation{\Yale}

% Authors in alphabetical order
\author{P.~Abratenko} \affiliation{\Tufts}
%\author{O.~Alterkait} \affiliation{\Tufts}
\author{D.~Andrade~Aldana} \affiliation{\IIT}
\author{L.~Arellano} \affiliation{\Manchester}
\author{J.~Asaadi} \affiliation{\UTA}
\author{A.~Ashkenazi}\affiliation{\TelAviv}
\author{S.~Balasubramanian}\affiliation{\FNAL}
\author{B.~Baller} \affiliation{\FNAL}
\author{A.~Barnard} \affiliation{\Oxford}
\author{G.~Barr} \affiliation{\Oxford}
\author{D.~Barrow} \affiliation{\Oxford}
% ...

\collaboration{The MicroBooNE Collaboration}
\thanks{microboone\_info@fnal.gov}\noaffiliation
%\email[]{microboone\_info@fnal.gov}

%%%%%%%%%%%%%%%%%%%%%%%%%%%%%%%%%%%%%%%%%%%%%%%%%%%%%%%%%%%%%%%%%%%%%%%%%%%%

\begin{abstract}
\noindent
We investigate the expected precision of the reconstructed neutrino direction using a $\nu_{\mu}$-argon quasielastic-like event topology with one muon and one proton in the final state and the reconstruction capabilities of the MicroBooNE liquid argon time projection chamber.
This direction is of importance in the context of DUNE sub-GeV atmospheric oscillation studies.
MicroBooNE allows for a data-driven quantification of this resolution by investigating the deviation of the reconstructed muon-proton system orientation with respect to the well-known direction of neutrinos originating from the Booster Neutrino Beam with an exposure of $1.3 \times 10^{21}$ protons on target.
Using simulation studies, we derive the expected sub-GeV DUNE atmospheric-neutrino reconstructed simulated spectrum by developing a reweighting scheme as a function of the true neutrino energy. 
We further report flux-integrated single- and double-differential cross section measurements of charged-current $\nu_{\mu}$ quasielastic-like scattering on argon as a function of the muon-proton system angle using the full MicroBooNE data sets.
We also demonstrate the sensitivity of these results to nuclear effects and final state hadronic reinteraction modeling.
\end{abstract}

\maketitle

%%%%%%%%%%%%%%%%%%%%%%%%%%%%%%%%%%%%%%%%%%%%%%%%%%%%%%%%%%%%%%%%%%%%%%%%%%%%

\section{Introduction}\label{intro}

Atmospheric neutrinos play a crucial role in improving our understanding of neutrino oscillations and extracting mixing parameters in the lepton sector, such as the charge-parity violating phase ($\delta_{CP}$)~\cite{PhysRevD.97.072001,PhysRevLett.120.071801}. 
Of particular interest are sub-GeV atmospheric neutrinos with energies in the 100\,MeV to 1\,GeV range.
These are affected by both solar and atmospheric mass splittings, while being sensitive to nontrivial oscillation effects~\cite{PhysRevD.17.2369,Mikheyev:1985zog,Akhmedov:1988kd,KRASTEV1989341}.
A measurement of their oscillation pattern can yield important new information on $\delta_{CP}$. 
Furthermore, there are a wealth of phenomenological oscillation study efforts in this energy regime~\cite{BARGER19981,PhysRevD.70.111301,PhysRevD.71.053006,Akhmedov2008,PhysRevD.78.093003,PhysRevD.79.113002,PhysRevD.82.093011,PhysRevLett.109.091801,Akhmedov:2012ah,PhysRevD.88.013013,Kelly:2023ugn,Zhuang:2021rsg,Ternes:2019sak,Denton:2021rgt,Kelly:2021jfs,Arguelles:2022hrt,Kelly:2019itm}.

Atmospheric neutrinos' oscillatory nature will be extensively explored with precision measurements performed using data sets that will be collected, amongst others, by the forthcoming Deep Underground Neutrino Experiment (DUNE)~\cite{DUNE1:2016oaz,DUNE2:2016oaz,DUNE3:2016oaz, dunecollaboration2024supernovapointingcapabilitiesdune}.
The liquid argon time projection chamber (LArTPC) technology deployed by DUNE will be essential to that effort, since it enables excellent neutrino interaction topology and energy reconstruction by allowing the detection of the majority of the secondary charged particles with low detection thresholds for charged particles~\cite{doi:10.7566/JPSCP.12.010017}.
Yet, in addition to a good estimation of the neutrino energy, a precise reconstruction of the incoming neutrino direction and the accurate modeling of the nuclear effects are crucial to determine the event-by-event baseline necessary for studying oscillation effects and obtaining a measurement of $\delta_{CP}$~\cite{Kelly:2019itm}.

In this work, we investigate the precision of the reconstructed neutrino direction in a LArTPC detector using sub-GeV neutrinos arriving from a known direction.
The reported results use the MicroBooNE detector~\cite{Acciarri:2016smi} and data sets corresponding to an exposure of $1.3 \times 10^{21}$ protons on target.
Neutrinos from the Booster Neutrino Beam (BNB)~\cite{AguilarArevalo:2008yp} at Fermi National Accelerator Laboratory collected during 2015--2020 are used, which provide strong overlap with the sub-GeV atmospheric neutrino spectrum.
We focus on interactions where a single muon-proton pair is reconstructed with no additional detected particles, similar to previous measurements~\cite{RefPRL,RefPRD,PhysRevD.109.092007}.
We refer to these events as CC1p0$\pi$.
Such events are dominated by quasielastic (QE) interactions as it is required that there are no visible pions.
We define the direction of the muon-proton system from the sum of the muon and proton momentum, and investigate the deviation of this direction from that of the incoming neutrino.
Furthermore, we present the first flux-integrated differential cross-section measurements for muon-neutrino charged-current (CC) interactions on argon as a function of the angle between the muon-proton system and the incoming neutrino direction. 
We present both a single-differential measurement, and double-differential measurements in different ranges of variables with sensitivity to nuclear effects and undetected particles such as neutrons. 
These variables include the reconstructable energy, derived struck nucleon momentum, and derived missing momentum.

In Sec.~\ref{sec:formalism} we define the angle between the muon-proton system and the incoming neutrino direction, $\theta_{\mathrm{vis}}$.
We further define the variables with sensitivity to nuclear effects used in the double-differential cross section measurements.
In Sec.~\ref{sec:eventrate} we present $\theta_{\mathrm{vis}}$ distributions using MicroBooNE $\nu_{\mu}$-Ar CC1p0$\pi$ interactions and discuss the observed features in specific regions of phase space.
Section~\ref{sec:dune} leverages the MicroBooNE CC1p0$\pi$ event selection to make a projection for the expected DUNE atmospheric neutrino $\theta_{\mathrm{vis}}$ spectrum by deriving a reweighting factor as a function of the true neutrino energy.
In Sec.~\ref{sec:MicroBooNE} we present the first flux-integrated single- and double-differential cross section measurements in the new angular variable.
Finally, conclusions are presented in Sec.~\ref{sec:discussion}.

%%%%%%%%%%%%%%%%%%%%%%%%%%%%%%%%%%%%%%%%%%%%%%%%%%%%%%%%%%%%%%%%%%%%%%%%%%%%

\section{Observables} \label{sec:formalism}

\begin{figure}[H]
    \centering  
    \includegraphics[width=\linewidth]{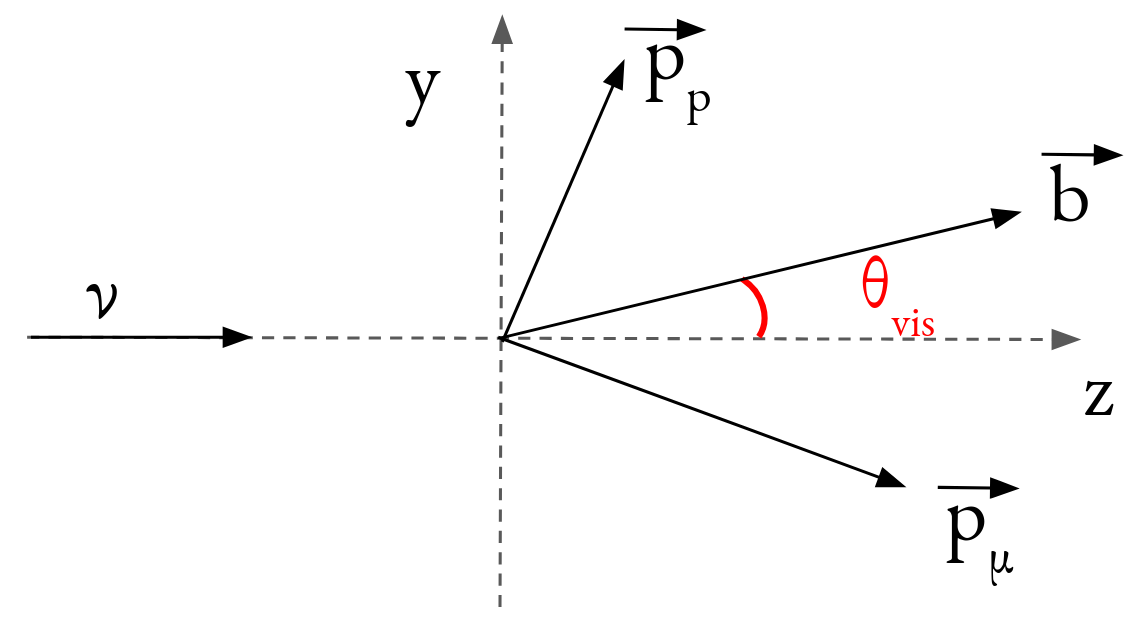}	  
    \caption{Schematic representation of $\theta_{\mathrm{vis}}$ using the muon's and proton's three-dimensional momentum vectors.}
    \label{thetavisdef}
\end{figure}

The simplest case in which to benchmark the reconstructed neutrino angular orientation precision is using charged-current quasielastic-like (CCQE-like) interactions, where the final state can be characterized by a muon and a proton.
Using the muon ($\vec{p}_{\mu}$) and proton ($\vec{p}_{p}$) momentum vectors, we define the angle between the muon-proton system and the incoming neutrino direction ($\theta_{\mathrm{vis}}$) as

\begin{align}
\theta_{\mathrm{vis}} = \mathrm{acos}(\dfrac{\vec{b} \cdot \hat{z}}{|\vec{b}|}),\\
\mathrm{with}\,\,\vec{b} = \vec{p}_{\mu} + \vec{p}_{p},
\end{align}
where $\hat{z}$ corresponds to the unit vector along the beam direction of incident neutrinos, as shown in Fig.~\ref{thetavisdef}.

\noindent This reconstructed angular orientation serves as a proxy for the neutrino direction and is studied as a function of variables with sensitivity to nuclear effects; namely the i) visible energy, ii) total struck nucleon momentum, and iii) missing momentum.

The reconstructable visible energy ($E_{\mathrm{reco}}$) is a crucial input to neutrino oscillation studies, and can be obtained following the formalism used in~\cite{PhysRevC.99.055504},

\begin{align}
E_{\mathrm{reco}} = E_{\mu} + K_{p} + B,
\end{align}
where $E_{\mu}$ is the total muon energy, $K_{p}$ is the proton kinetic energy, and $B$ the nucleon removal energy of argon set to 30.9 MeV~\cite{BodekCai2019}.
The particle energies are obtained using range-based momentum.

Assuming that the incoming neutrino travels along the $z$-direction and using conservation of momentum along that direction, the transverse and longitudinal components of the struck nucleon momentum can be obtained, as discussed in Refs.~\cite{PWIA,Furmanski2016},

\begin{align}
\delta\vec{p}_T = \vec{p}_{\mu,T} + \vec{p}_{p,T},\,\\
p_{L} = p_{\mu,L} + p_{p,L} - E_{\mathrm{reco}},
\end{align}
where $p_{\mu (p),T(L)}$ are the transverse (longitudinal) component of the muon (proton) momentum vector, respectively.
Assuming that all the particles are reconstructed, the initial struck nucleon momentum ($p_{n}$) can be obtained as the vector sum of the longitudinal and transverse components,

\begin{align}
\label{GKIpn}
p_n &= |\vec{p}_n| = \sqrt{p_{L}^{2} + \delta p_{T}^{2}}.
\end{align}

This derivation of the initial struck nucleon momentum builds on the assumption that neutrinos arrive from a known direction along the beamline.
This is not applicable for neutrinos of atmospheric origin.
To overcome this limitation, we introduce a new variable to quantify the missing momentum as

\begin{align}
p_{\mathrm{miss}} = E_{\mathrm{reco}} - b,
\end{align}
where $b$ corresponds to the magnitude of the vector $\vec{b}$.
This quantity is independent of the incoming neutrino direction since it uses no angular information.
Thus, it can be calculated for atmospheric neutrinos and leveraged to ensure that both the incoming atmospheric neutrino direction and energy will be reconstructed accurately.
In the absence of undetected particles, detection thresholds, or nuclear effects, $p_{\mathrm{miss}}$ would be equal to the proton-neutron mass difference.
Deviations from that value are indicative of the presence of these effects and introduce missing energy or momentum in the measurement.

%%%%%%%%%%%%%%%%%%%%%%%%%%%%%%%%%%%%%%%%%%%%%%%%%%%%%%%%%%%%%%%%%%%%%%%%%%%%

\section{MicroBooNE Event Selection} \label{sec:eventrate}

The CC1p0$\pi$ signal definition used in this analysis includes all $\nu_{\mu}$-Ar scattering events with a final-state muon with momentum 0.1 $< p_{\mu}<$ 1.2\,GeV/$c$, and exactly one proton with 0.3 $< p_{p} <$ 1\,GeV/$c$.
Events with final-state neutral pions of any momentum are excluded.
Signal events may contain any number of protons with momentum less than 300 MeV/$c$ or greater than 1\,GeV/$c$, neutrons of any momentum, and charged pions with momentum lower than 70 MeV/$c$.

%%%%%%%%%%%%%%%%%%%

\begin{figure}[H]
\centering 

\begin{tikzpicture} \draw (0, 0) node[inner sep=0] {
\includegraphics[width=\linewidth]{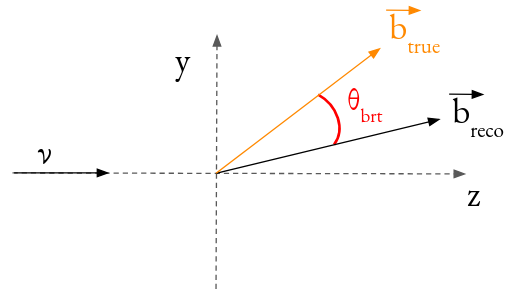}
};
\draw (0., -3.4) node {(a)};	
\end{tikzpicture}

\begin{tikzpicture} \draw (0, 0) node[inner sep=0] {
\includegraphics[width=\linewidth]{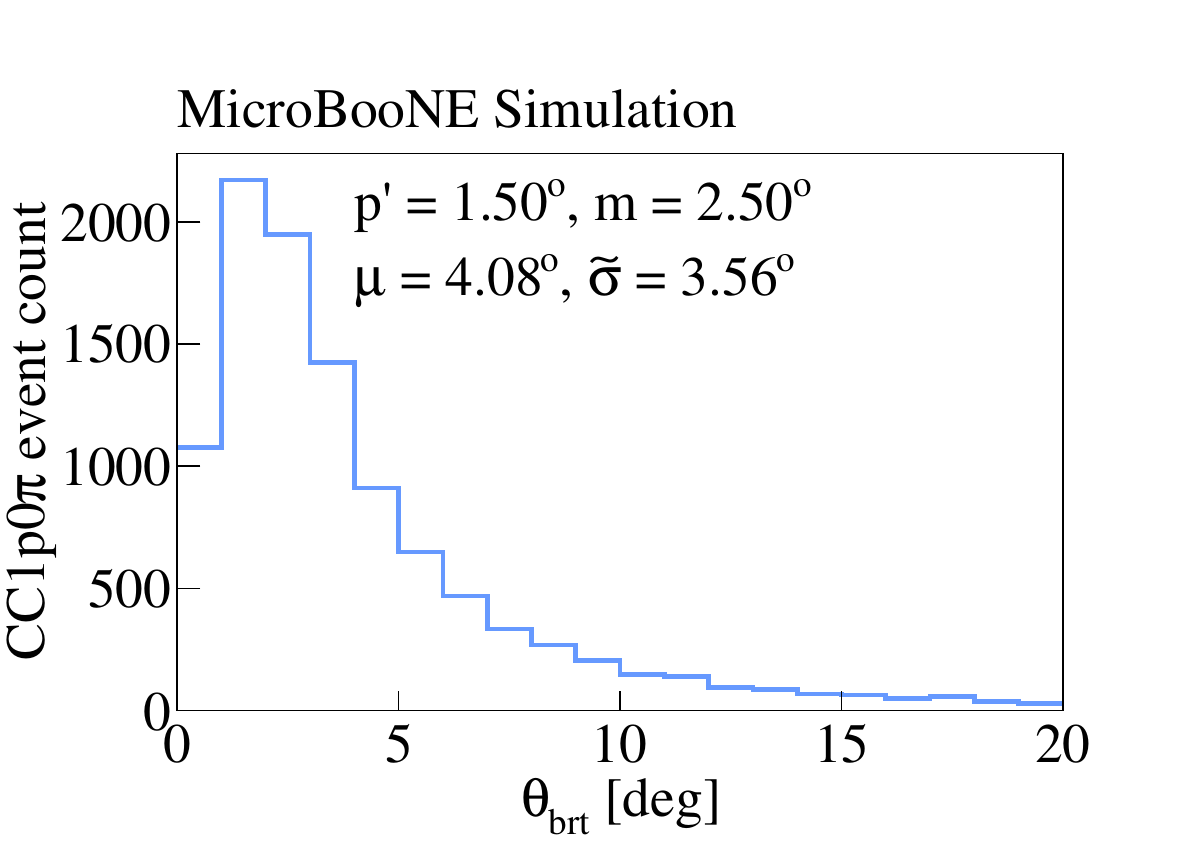}
};
\draw (0., -3.4) node {(b)};	
\end{tikzpicture}

\begin{tikzpicture} \draw (0, 0) node[inner sep=0] {
\includegraphics[width=\linewidth]{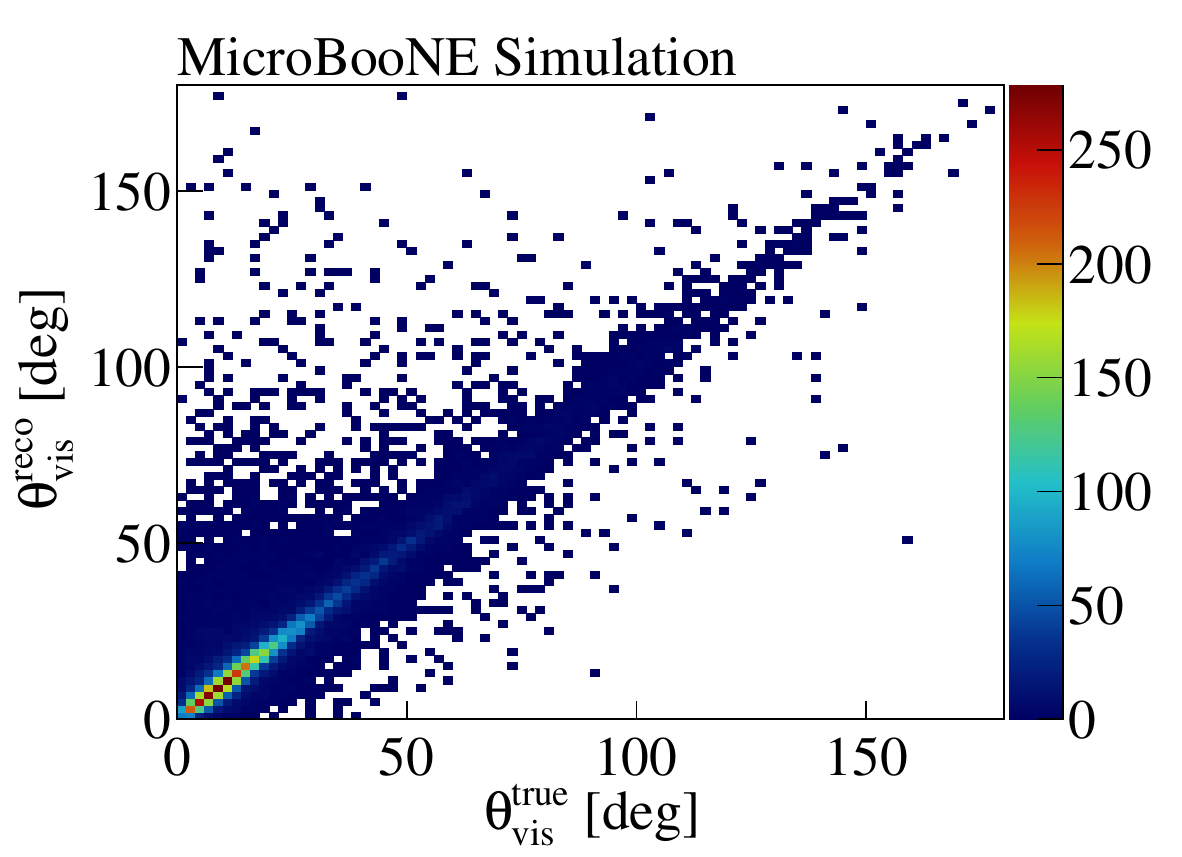}
};
\draw (0., -3.4) node {(c)};	
\end{tikzpicture}

\caption{
    (a) Schematic representation of the $\theta_{\mathrm{brt}}$ angle using the angle between the true ($\vec{b}_{\mathrm{true}}$) and reconstructed ($\vec{b}_{\mathrm{reco}}$) muon-proton system vectors with respect to the neutrino orientation for selected signal CC1p0$\pi$ simulated events.
    (b) Angular $\theta_{\mathrm{brt}}$ distribution using the selected signal CC1p0$\pi$ simulated events. The peak location ($p^{\prime}$), median ($m$), mean value ($\mu$), and standard deviation ($\tilde{\sigma}$) describing the distribution are also shown.
    (c) Two-dimensional correlation between the reconstructed and true $\theta_{\mathrm{vis}}$ using the selected signal CC1p0$\pi$ simulated events.
}
\label{brt}
\end{figure}

%%%%%%%%%%%%%%%%%%%

To report the results as a function of the $\theta_{\mathrm{vis}}$ angular orientation, we use five years of data collected by the MicroBooNE detector from its exposure to the BNB neutrino flux.
The detector is an 85 tonne active mass liquid argon time projection chamber and is described in detail in Ref.~\cite{Acciarri:2016smi}.

The selection outlined in Refs.~\cite{RefPRL,RefPRD,PhysRevD.109.092007} is used, which corresponds to the same CC1p0$\pi$ signal definition as described at the beginning of this section. 
This results in 17,130 candidate data events, a purity of CC1p0$\pi$ interactions of about 70\%, and a selection efficiency of approximately 10\%.
The final efficiency is primarily driven by the demand for exactly two fully contained track-like candidates.
Based on simulation predictions, we find that the dominant background contributions originate from events with two true protons, pion-proton pairs, and broken muon tracks.
This is attributed to reconstruction failures that either fail to reconstruct a particle track or split a particle track into multiple segments.

This data is compared against the $\texttt{GENIE v3.0.6}$ neutrino interaction generator predictions~\cite{geniev3highlights} with the $\texttt{GENIE v3.0.6 G18\_10a\_02\_11a (G18)}$ model configuration along with the MicroBooNE BNB flux prediction~\cite{AguilarArevalo:2008yp}.
This model configuration uses the local Fermi gas (LFG) model~\cite{Carrasco:1989vq}, the Nieves CCQE scattering prescription~\cite{Nieves:2012yz} which includes Coulomb corrections for the outgoing muon~\cite{Engel:1997fy}, and random phase approximation (RPA) corrections~\cite{RPA}.
Additionally, it uses the Nieves meson exchange current (MEC) model~\cite{Schwehr:2016pvn}, the Kuzmin-Lyubushkin-Naumov Berger-Sehgal resonance production (RES)~\cite{Berger:2007rq,PhysRevD.104.072009,Kuzmin:2003ji}, the Berger-Sehgal coherent production (COH)~\cite{Berger:2008xs}, and the Bodek-Yang deep inelastic scattering (DIS)~\cite{PhysRevLett.82.2467} models with the $\texttt{PYTHIA}$~\cite{Sjostrand:2006za} hadronization part, and the hA2018 final state interaction (FSI) model~\cite{Ashery:1981tq}.
The CCQE and CCMEC neutrino interaction models have been tuned to T2K $\nu_{\mu}$-$^{12}$C CC0$\pi$ data~\cite{PhysRevD.93.112012,GENIEKnobs}.
Predictions for more complex interactions, such as RES, remain unaltered and no additional Monte Carlo (MC) constraints are applied.
We refer to the corresponding tuned prediction as $\texttt{G18T}$.
In order to provide an accurate description of the dominant cosmic backgrounds pertinent to surface detectors, the full MC simulation consists of a combination of simulated neutrino interactions overlaid with background data collected when the beam is off to model cosmic ray induced interactions and detector noise.
This technique has been extensively used by previous CC1p0$\pi$ MicroBooNE analyses~\cite{Adams:2018lzd,PhysRevLett.125.201803,PhysRevLett.128.151801,PhysRevD.105.L051102}.

We first characterize the performance of the LArTPC reconstruction by comparing the true ($\vec{b}_{\mathrm{true}}$) and reconstructed ($\vec{b}_{\mathrm{reco}}$) muon-proton system vectors.
These are constructed using the true and reconstructed values for the muon and proton momentum vectors, respectively, as defined in Sec.~\ref{sec:formalism}.
The true values refer to the MC generator-level quantities, while the reconstructed values refer to the quantities as measured in the LArTPC detector.
This comparison is performed using the selected MC events that satisfy the CC1p0$\pi$ signal definition.
We refer to the relevant opening angle between the reconstructed ($r$) and true ($t$) $\vec{b}$ vectors as $\theta_{\mathrm{brt}}$, as shown in Fig.~\ref{brt}\textcolor{blue}{(a)}.
As shown in Fig.~\ref{brt}\textcolor{blue}{(b)}, for the majority of events the angle $\theta_{\mathrm{brt}}$ is better than 5$^{\mathrm{o}}$, which demonstrates the excellent reconstruction capabilities of LArTPCs.
The peak location ($p^{\prime}$), median ($m$), mean value ($\mu$), and standard deviation ($\tilde{\sigma}$) describing the distributions are also shown.
The small values that characterize this distribution indicate that the detector resolution on these quantities is not the limiting factor in determining the incoming neutrino direction.
This will also be discussed in later sections.
Finally, Fig.~\ref{brt}\textcolor{blue}{(c)} shows the two-dimensional correlation between the reconstructed and true $\theta_{\mathrm{vis}}$ and demonstrates that no major biases are observed.
The corresponding muon and proton angular correlations between the reconstructed and true quantities can be found in the Supplemental Material~\cite{suppmat}.

\begin{figure}[htb!]
\centering 
\includegraphics[width=\linewidth]{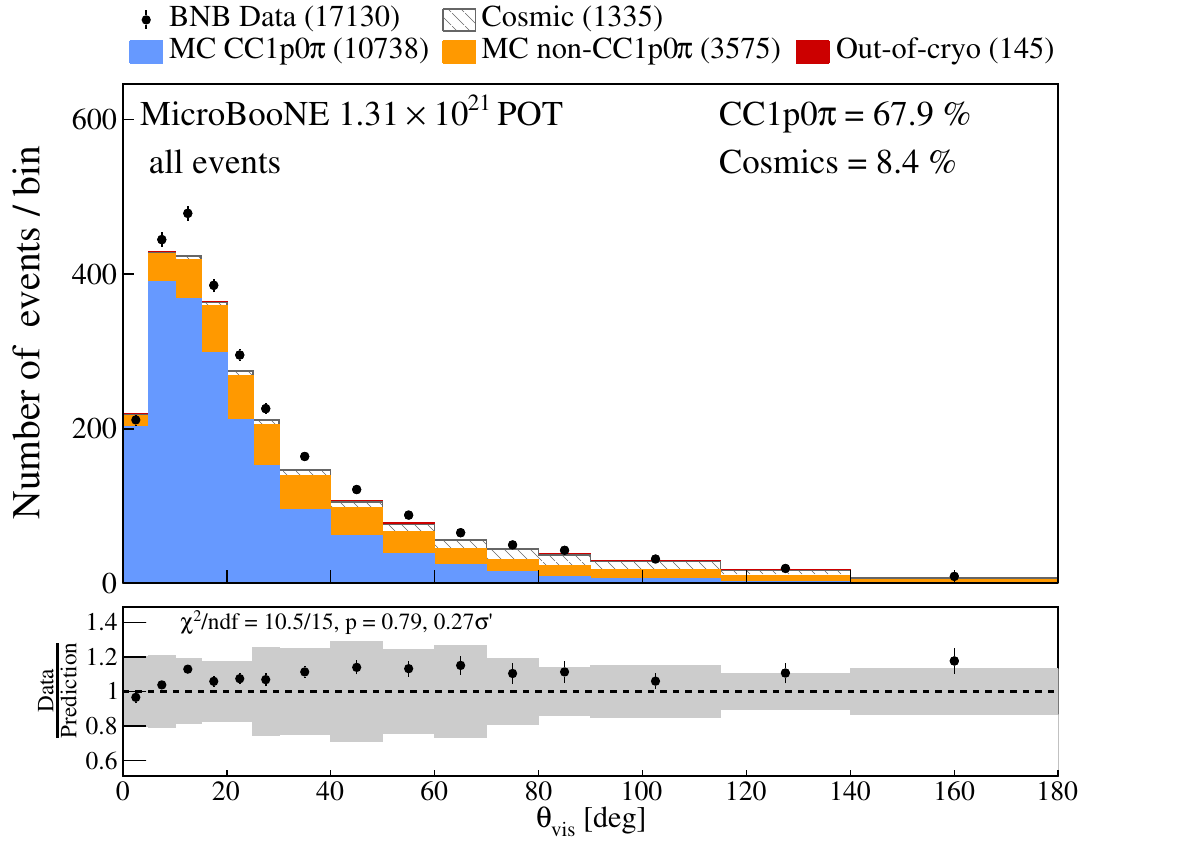}
\caption{
Distribution of the selected CC1p0$\pi$ events as a function of the angle $\theta_{\mathrm{vis}}$.
Only statistical uncertainties are shown on the data.
The bottom panel shows the ratio of data to prediction.
The prediction uncertainty is included in the bottom panel.
The data-simulation agreement is quantified across all the figures in terms of a goodness-of-fit metric ($\chi^{2}$), the corresponding p-value ($p$), and as a number of standard deviations ($\sigma'$).
}
\label{ThetaVisSlices}
\end{figure}

The reconstructed $\theta_{\mathrm{vis}}$ distribution is shown in Fig.~\ref{ThetaVisSlices}.
The uncertainty band shown in the data to MC ratio includes contributions from the neutrino flux prediction~\cite{Aguilar-Arevalo:2013dva}, neutrino interaction cross section modeling~\cite{Andreopoulos:2009rq,Andreopoulos:2015wxa,GENIEKnobs}, detector response modeling~\cite{WireMod}, beam exposure, MC statistics, number of scattering targets, reinteractions~\cite{Calcutt_2021}, and out-of-cryostat interaction modeling.
The data-simulation agreement is quantified across all the figures in terms of a goodness-of-fit metric ($\chi^{2}$), the corresponding p-value ($p$), and as a number of standard deviations ($\sigma'$).
The latter is calculated by translating the p-value to a $\chi^{2}$ value with one degree of freedom and taking the square root of that quantity.
The $\chi^{2}$ calculation includes the bin-to-bin correlations.

%%%%%%%%%%%%%%%%%%%%%%%%%%%%%%%%%%%%%%%%%%%%%%%%%%%%%%%%

%%%%%%%%%%%%%

\begin{figure}[H]
\centering 

\begin{tikzpicture} \draw (0, 0) node[inner sep=0] {
\includegraphics[width=\linewidth]{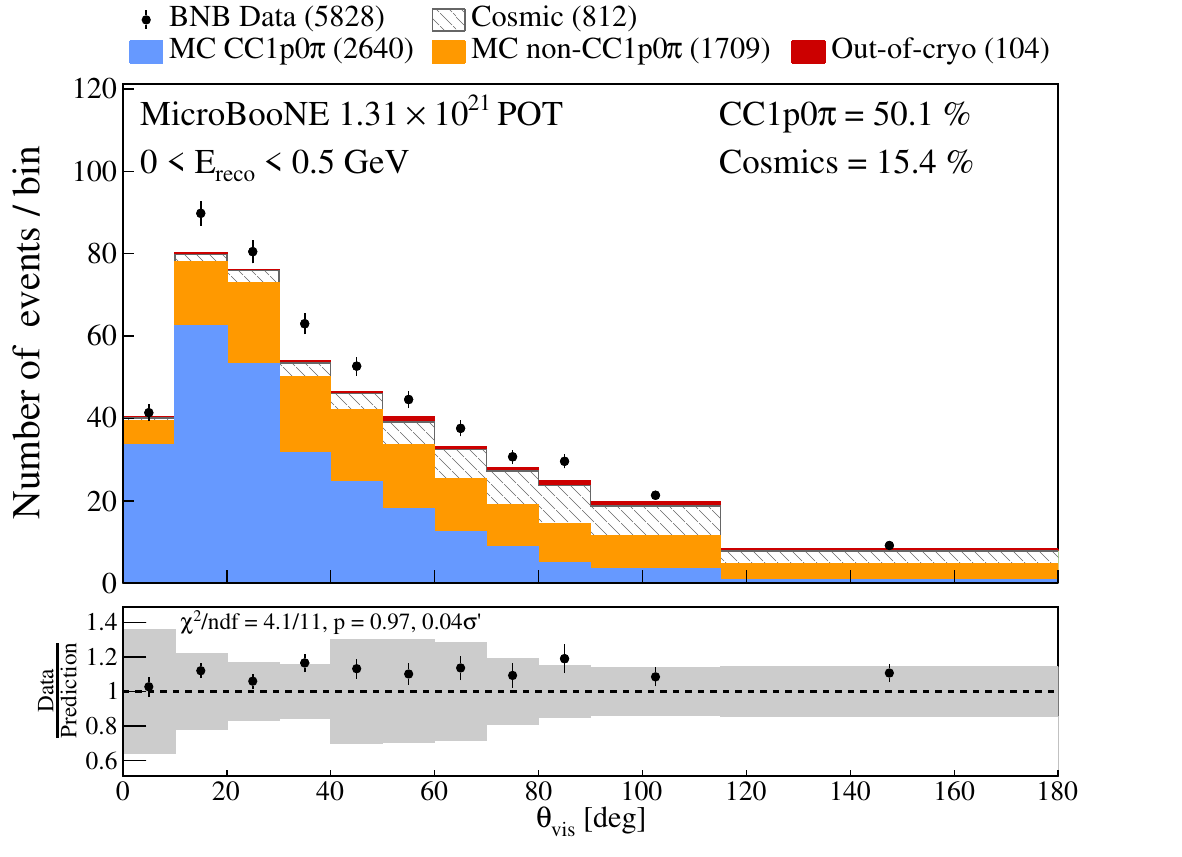}
};
\draw (0., -3.4) node {(a)};	
\end{tikzpicture}

\begin{tikzpicture} \draw (0, 0) node[inner sep=0] {
\includegraphics[width=\linewidth]{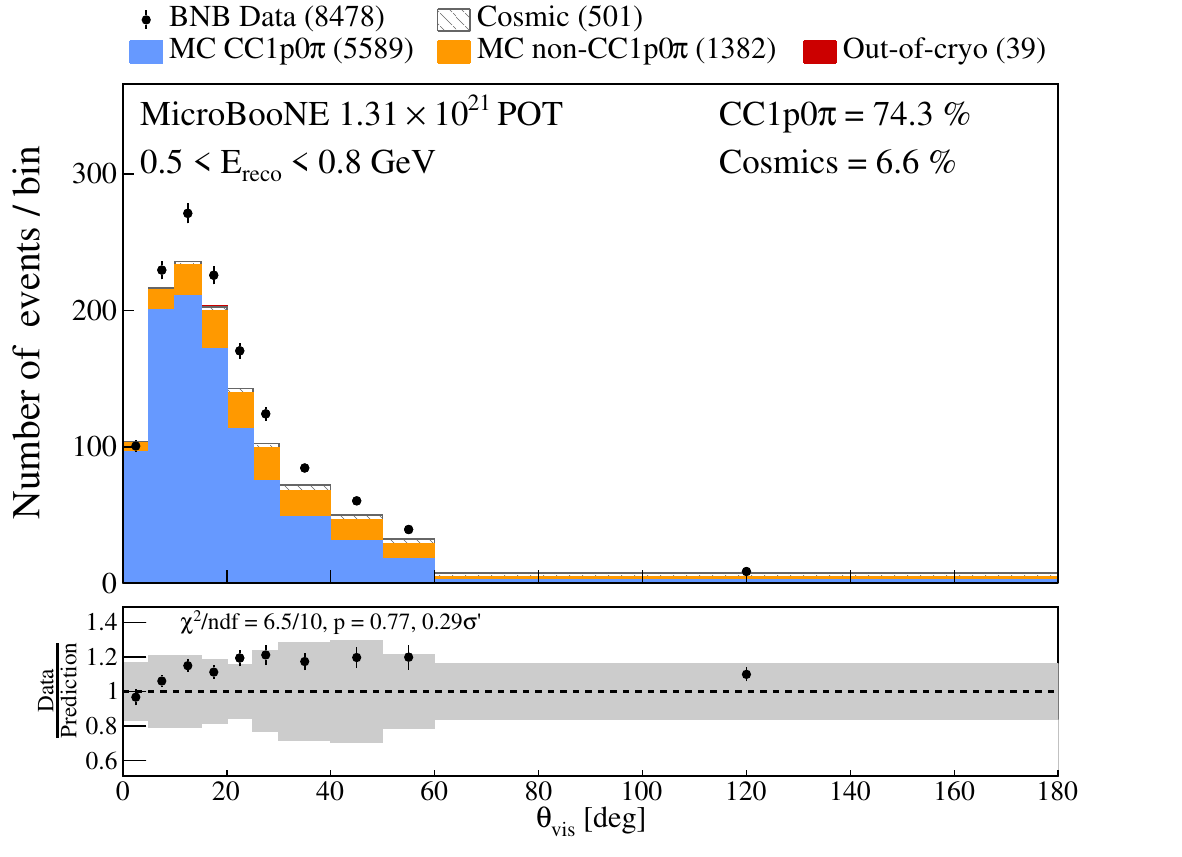}
};
\draw (0., -3.4) node {(b)};	
\end{tikzpicture}

\begin{tikzpicture} \draw (0, 0) node[inner sep=0] {
\includegraphics[width=\linewidth]{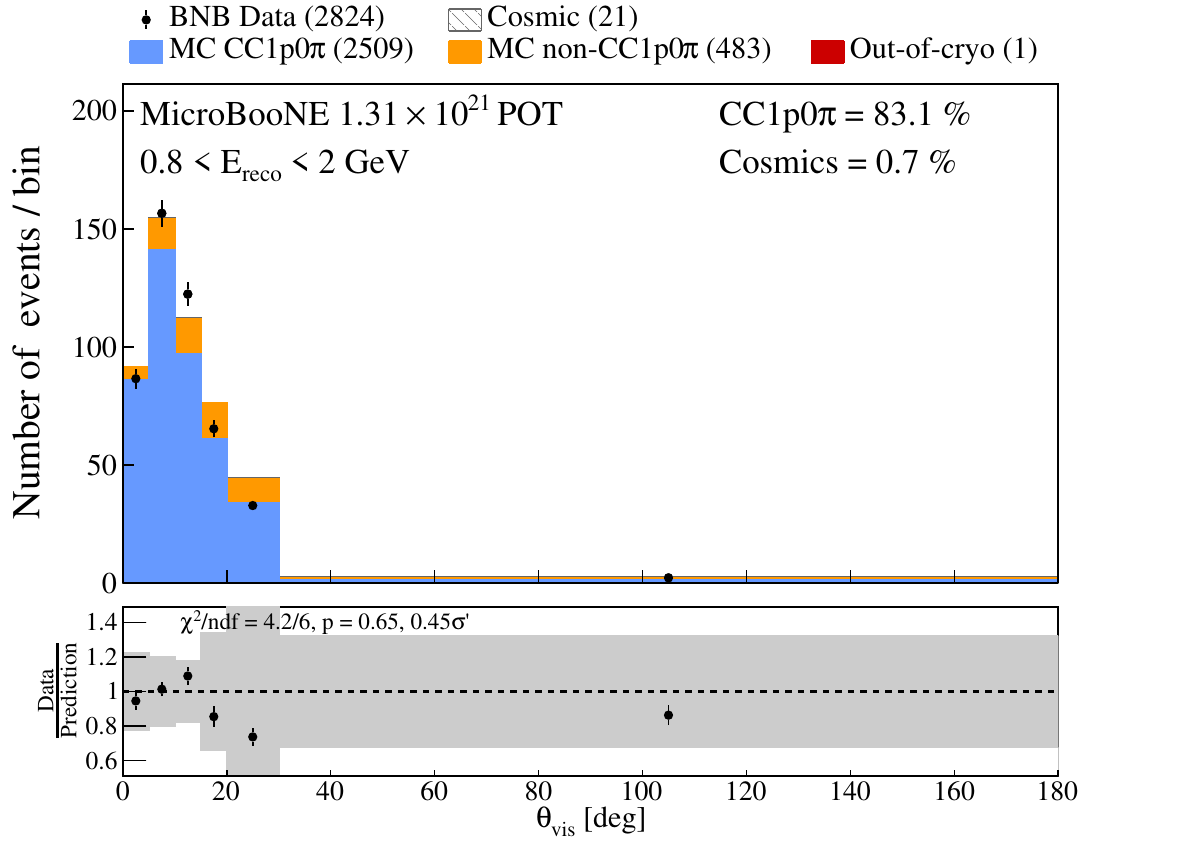}
};
\draw (0., -3.4) node {(c)};	
\end{tikzpicture}

\caption{
Distribution of the selected CC1p0$\pi$ events as a function of the muon-proton angle $\theta_{\mathrm{vis}}$ in (a) low, (b) medium, and (c) high reconstructable energy ($E_{\mathrm{reco}}$) regions.
Only statistical uncertainties are shown on the data.
The bottom panels shows the ratio of data to prediction.
The prediction uncertainty is included in the bottom panels.
The data-simulation agreement is quantified across all the figures in terms of a goodness-of-fit metric ($\chi^{2}$), the corresponding p-value ($p$), and as a number of standard deviations ($\sigma'$).
}
\label{ThetaVisInECalSlices}
\end{figure}

%%%%%%%%%%%%%%%%%%

\begin{figure}[H]
\centering 

\begin{tikzpicture} \draw (0, 0) node[inner sep=0] {
\includegraphics[width=\linewidth]{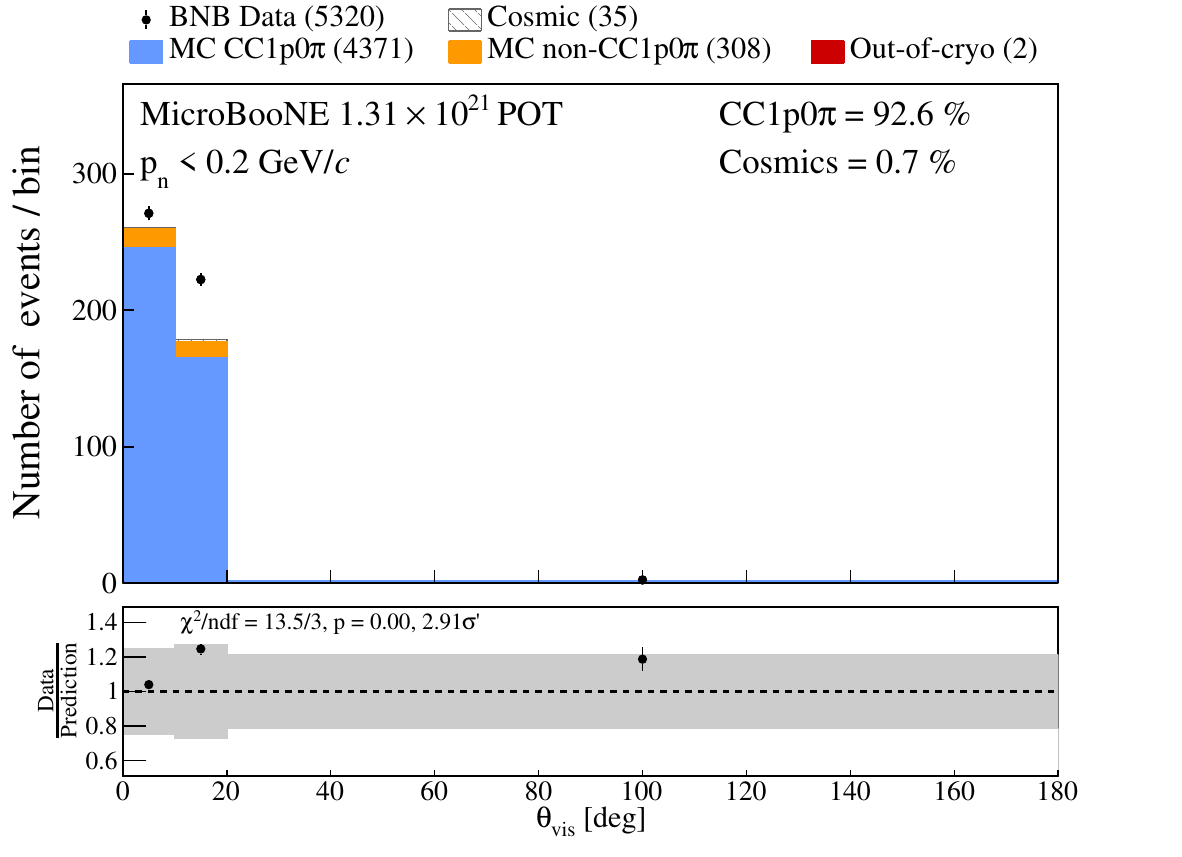}
};
\draw (0., -3.4) node {(a)};	
\end{tikzpicture}

\begin{tikzpicture} \draw (0, 0) node[inner sep=0] {
\includegraphics[width=\linewidth]{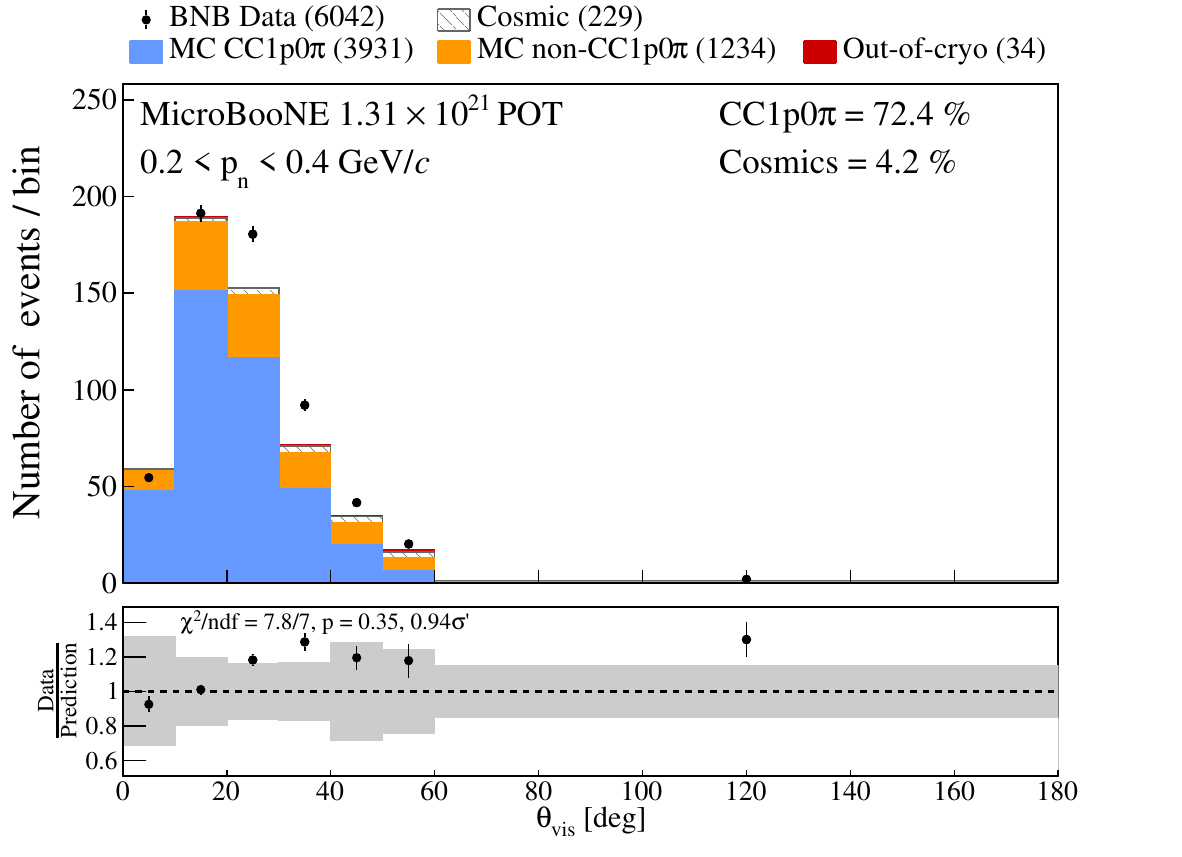}
};
\draw (0., -3.4) node {(b)};	
\end{tikzpicture}

\begin{tikzpicture} \draw (0, 0) node[inner sep=0] {
\includegraphics[width=\linewidth]{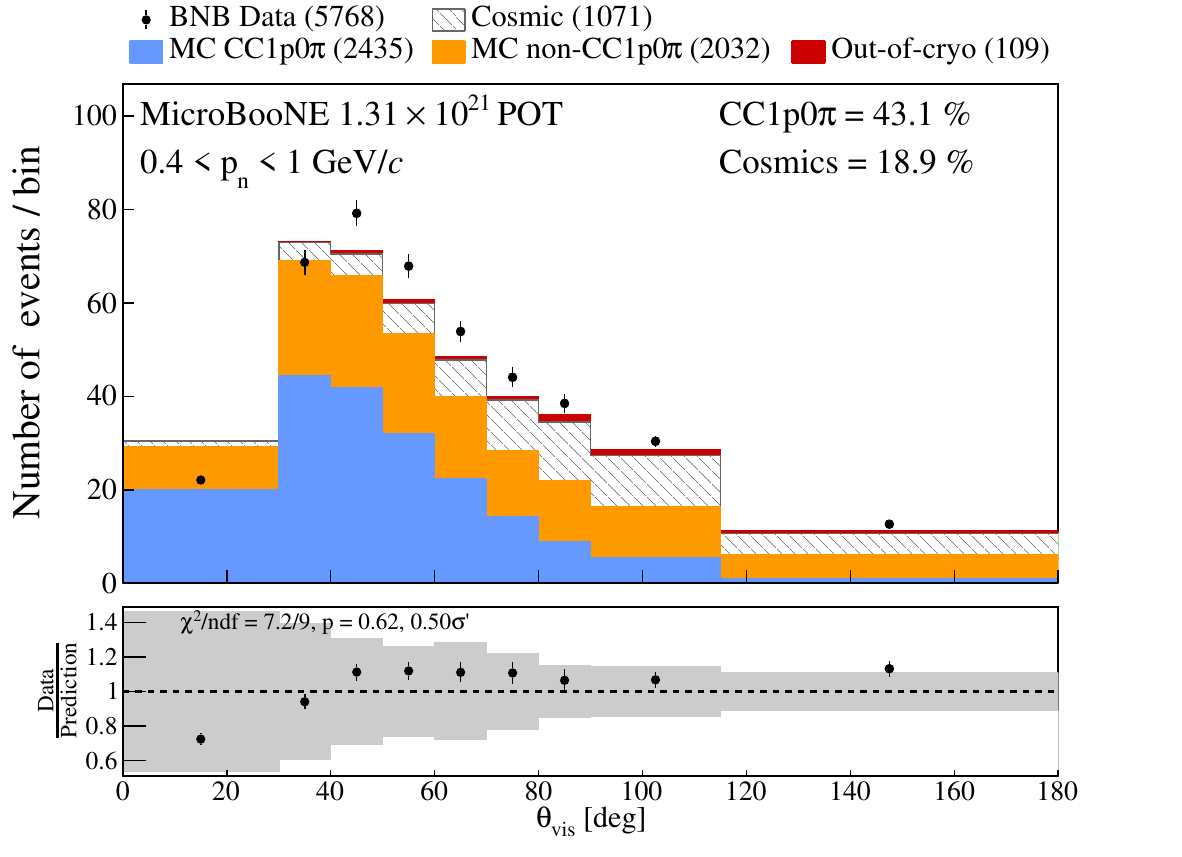}
};
\draw (0., -3.4) node {(c)};	
\end{tikzpicture}

\caption{
Distribution of the selected CC1p0$\pi$ events as a function of the muon-proton angle $\theta_{\mathrm{vis}}$ in (a) low, (b) medium, and (c) high reconstructed struck nucleon momentum ($p_{n}$) regions.
Only statistical uncertainties are shown on the data.
The bottom panels shows the ratio of data to prediction.
The prediction uncertainty is included in the bottom panels.
The data-simulation agreement is quantified across all the figures in terms of a goodness-of-fit metric ($\chi^{2}$), the corresponding p-value ($p$), and as a number of standard deviations ($\sigma'$).
}
\label{ThetaVisInDeltaPnSlices}
\end{figure}

%%%%%%%%%%%%%%%%%%

\begin{figure}[H]
\centering 

\begin{tikzpicture} \draw (0, 0) node[inner sep=0] {
\includegraphics[width=\linewidth]{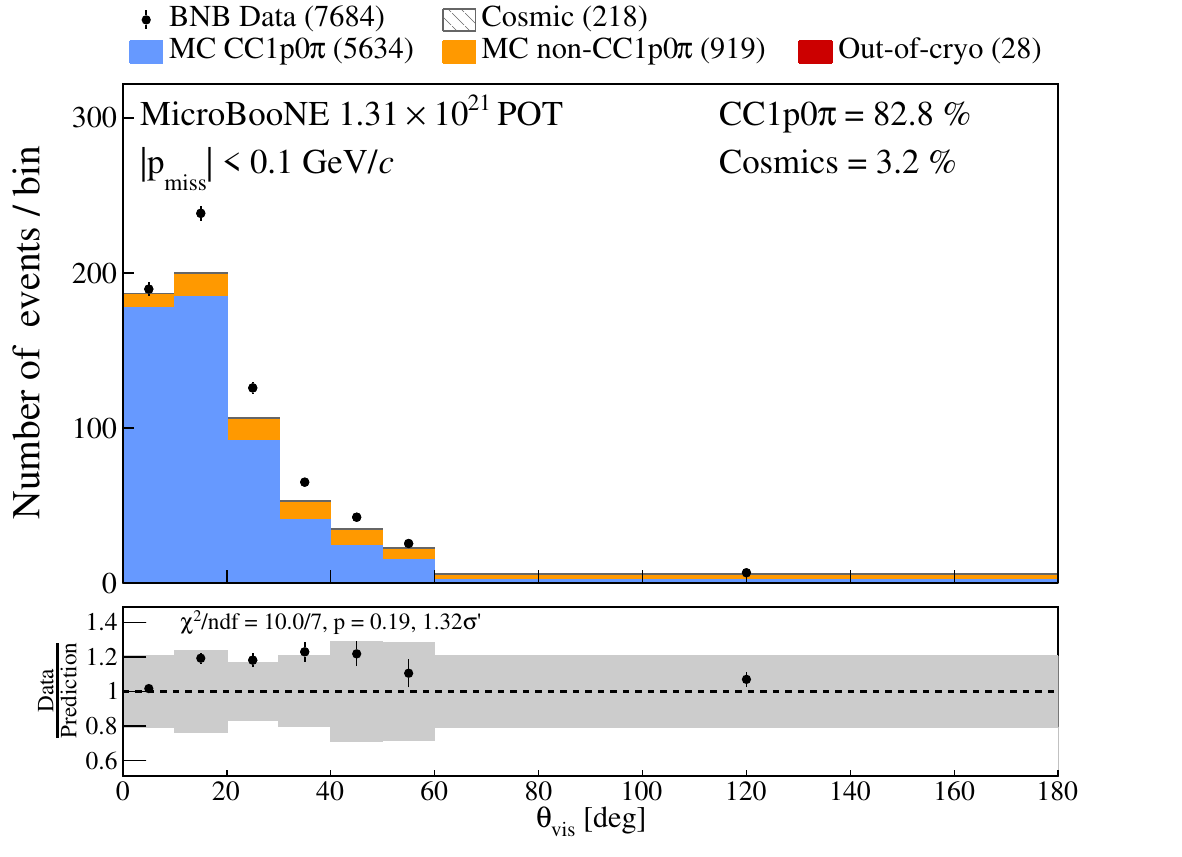}
};
\draw (0., -3.4) node {(a)};	
\end{tikzpicture}

\begin{tikzpicture} \draw (0, 0) node[inner sep=0] {
\includegraphics[width=\linewidth]{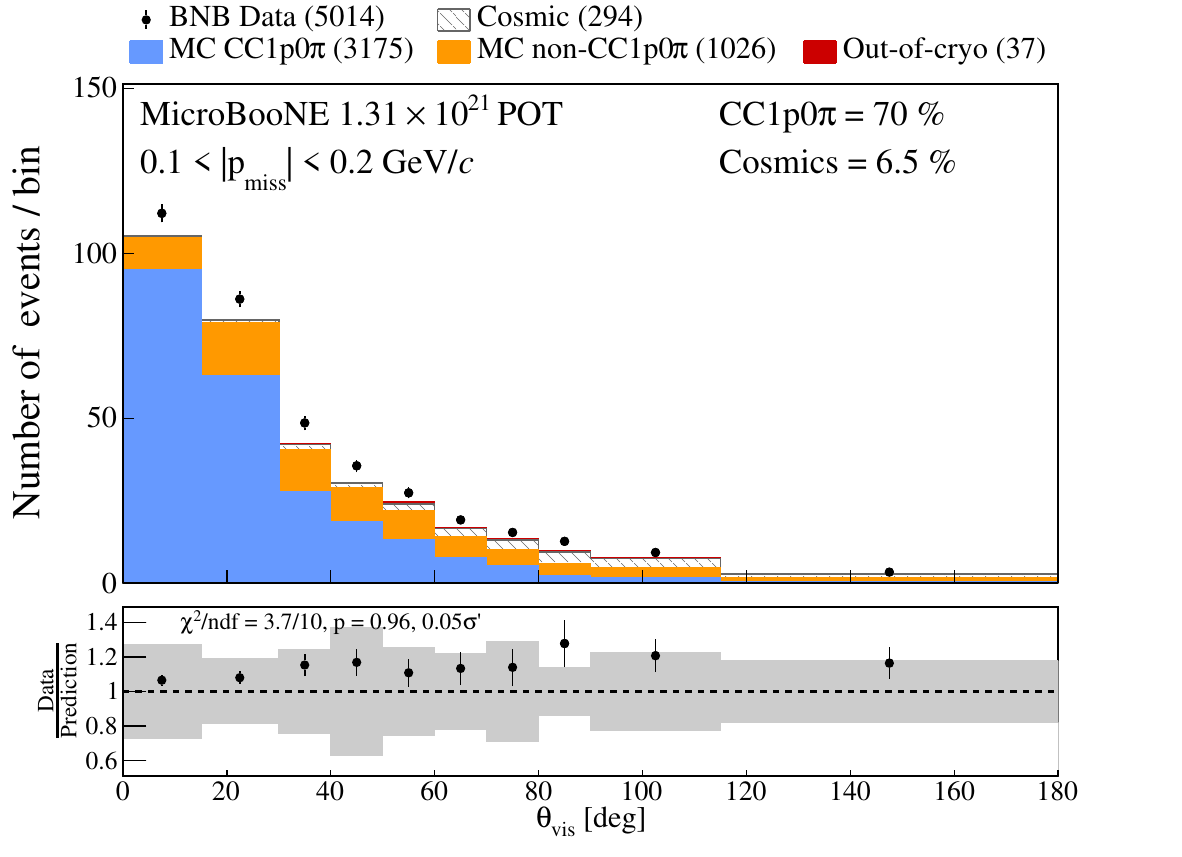}
};
\draw (0., -3.4) node {(b)};	
\end{tikzpicture}

\begin{tikzpicture} \draw (0, 0) node[inner sep=0] {
\includegraphics[width=\linewidth]{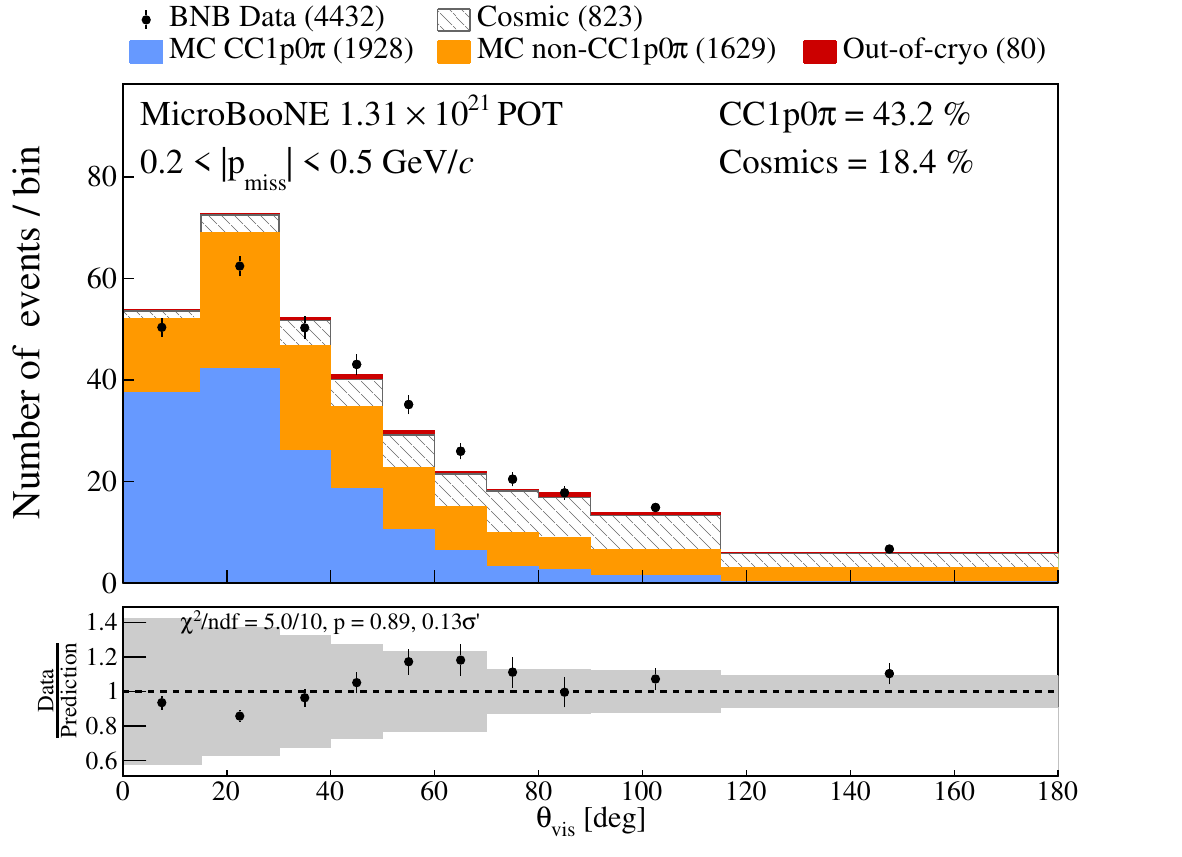}
};
\draw (0., -3.4) node {(c)};	
\end{tikzpicture}

\caption{
Distribution of the selected CC1p0$\pi$ events as a function of the muon-proton angle $\theta_{\mathrm{vis}}$ in (a) low, (b) medium, and (c) high reconstructed missing momentum ($p_{\mathrm{miss}}$) regions.
Only statistical uncertainties are shown on the data.
The bottom panels shows the ratio of data to prediction.
The prediction uncertainty is included in the bottom panels.
The data-simulation agreement is quantified across all the figures in terms of a goodness-of-fit metric ($\chi^{2}$), the corresponding p-value ($p$), and as a number of standard deviations ($\sigma'$).
}
\label{ThetaVisInPMissSlices}
\end{figure}

%%%%%%%%%%%%%%%%%%%

\begin{figure}[htb!]
\centering 

\begin{tikzpicture} \draw (0, 0) node[inner sep=0] {
\includegraphics[width=\linewidth]{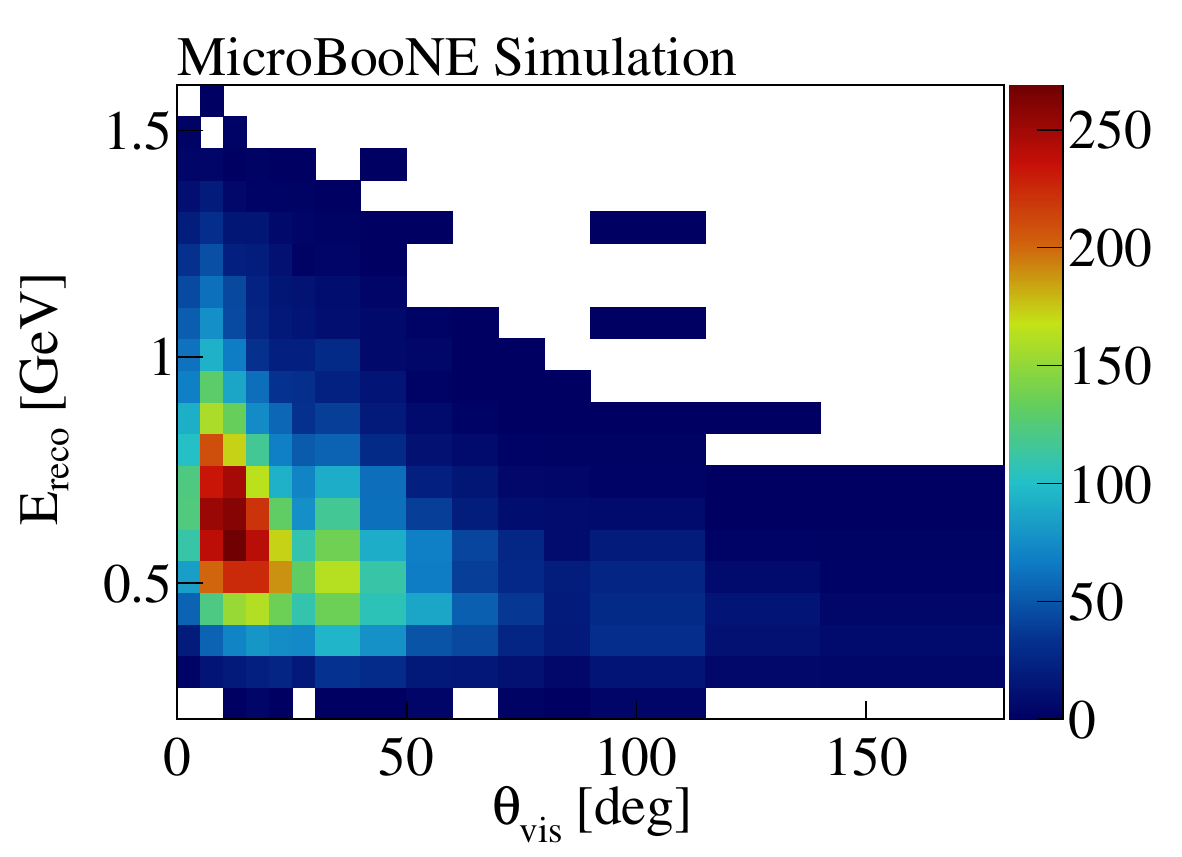}
};
\draw (0., -3.4) node {(a)};	
\end{tikzpicture}

\begin{tikzpicture} \draw (0, 0) node[inner sep=0] {
\includegraphics[width=\linewidth]{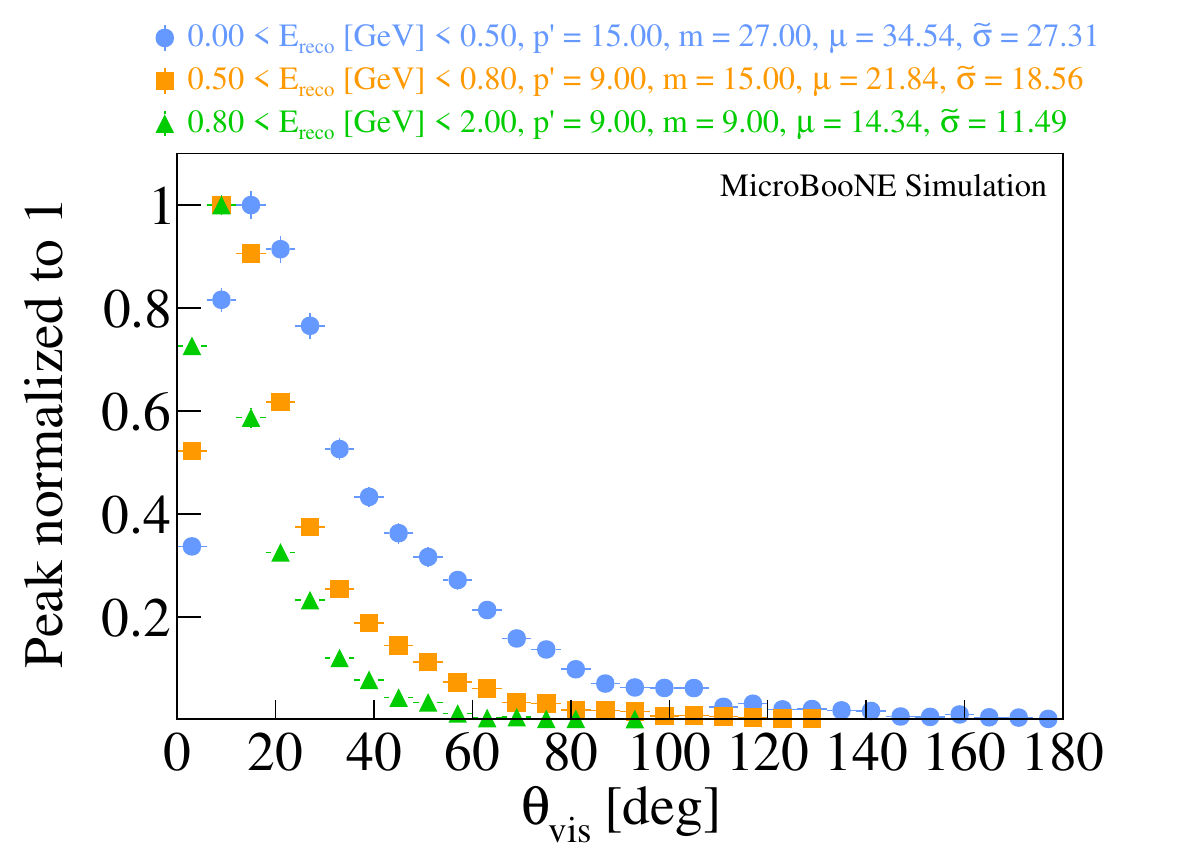}
};
\draw (0., -3.4) node {(b)};	
\end{tikzpicture}

\caption{
    (a) Angular orientation $\theta_{\mathrm{vis}}$ dependence on the reconstructable energy using simulated signal CC1p0$\pi$ events.
    (b) The $\theta_{\mathrm{vis}}$ distributions presented in reconstructable energy slices using simulated signal CC1p0$\pi$ events. The peak location ($p^{\prime}$), median ($m$), mean value ($\mu$), and standard deviation ($\tilde{\sigma}$) are also presented.
}
\label{energydependence}
\end{figure}

%%%%%%%%%%%%%%%%%%%%%%%%%%%%%%%%%%%%%%%%%%%%%%%%%%%%%%%%

In the absence of nuclear effects and final-state reinteractions, the $\theta_{\mathrm{vis}}$ distribution would be peaked at 0$^{\mathrm{o}}$ with respect to the incoming neutrino direction.
Instead it shows a smooth rise at low values until it peaks at $\approx 10^{\mathrm{o}}$, followed by a decreasing tail that extends to 180$^{\mathrm{o}}$. 
The physics contributions driving this tail are discussed in Sec.~\ref{sec:MicroBooNE}.
Candidate neutrino interactions that satisfy the CC1p0$\pi$ signal definition at a reconstruction level but not at truth level are treated as background events.
We refer to these background events as non-CC1p0$\pi$.
In addition, there are also some remaining background contributions from cosmic contamination and interactions originating outside the cryostat.

The final $\theta_{\mathrm{vis}}$ cross section results will be reported in regions of $E_{\mathrm{reco}}$, $p_{n}$, and $p_{\mathrm{miss}}$.
These regions are listed in Table~\ref{tab_mean_sigma_median} and correspond to phase-space regions with sensitivity to different nuclear effects.
The relevant event distributions in these regions are presented in Figs.~\ref{ThetaVisInECalSlices}--\ref{ThetaVisInPMissSlices}.
The $\theta_{\mathrm{vis}}$ resolutions are included in the Supplemental Material~\cite{suppmat}.
The bin-width division has already been applied to account for the irregular binning.
We further report the evolution of the $\theta_{\mathrm{vis}}$ mean value ($\mu$), standard deviation ($\tilde{\sigma}$), and median ($m$) in those regions for the simulated signal CC1p0$\pi$ events.
A representative example of the angular orientation $\theta_{\mathrm{vis}}$ for the simulated signal CC1p0$\pi$ events is presented as a function of the reconstructable energy in Fig.~\ref{energydependence}.

As can be seen both in Figs.~\ref{ThetaVisInECalSlices}--\ref{ThetaVisInPMissSlices} and in Table~\ref{tab_mean_sigma_median}, the median obtained in the $\theta_{\mathrm{vis}}$ distribution using all the events (Fig.~\ref{ThetaVisSlices}) is smaller than the median in regions with reconstructed energy less than 0.5\,GeV shown in Fig.~\ref{ThetaVisInECalSlices}\textcolor{blue}{(a)}.
The same behavior is observed for regions with high reconstructed struck nucleon momentum seen in Fig.~\ref{ThetaVisInDeltaPnSlices}\textcolor{blue}{(c)} and regions with high missing momentum shown in Fig.~\ref{ThetaVisInPMissSlices}\textcolor{blue}{(b-c)}.
Yet, the resolutions obtained with the CC1p0$\pi$ selection are, in most cases, smaller than the ones observed in the result reported in~\cite{Kopp:2024lch}.
That observation is valid even at the lower part of the neutrino energy spectrum, where the reconstruction performance is expected to worsen.

%%%%%%%%%%%%%%%%%%%%%%%

\begin{table}[H] 
\center
\caption{Evolution of the mean value ($\mu$), standard deviation ($\tilde{\sigma}$), and median ($m$) as a function of $\theta_{\mathrm{vis}}$ across various regions of $E_{\mathrm{reco}}$, $p_{n}$, and $p_{\mathrm{miss}}$ for the simulated signal CC1p0$\pi$ events.}
\vspace{5px}
\label{tab_mean_sigma_median}
\begin{tabular}{|l|c|c|c|}
\hline									
Region & $\mu$ [deg] & $\tilde{\sigma}$ [deg] & $m$ [deg]\tabularnewline
\hline	
\hline	
 All events & 23.2  &  21.1 & 16.5 \tabularnewline \hline
 \hline $E_{\mathrm{reco}} <$ 0.5\,GeV & 34.5  &  27.3 & 26.5 \tabularnewline \hline
 0.5 $ < E_{\mathrm{reco}} <$ 0.8\,GeV & 21.8  &  18.6 & 16.5 \tabularnewline \hline
 0.8 $ < E_{\mathrm{reco}} <$ 2\,GeV & 14.3  &  11.5 & 10.5 \tabularnewline \hline
 \hline $p_{n} <$ 0.2\,GeV/$c$ & 10.0  &  5.8 & 9.5 \tabularnewline \hline
 0.2 $< p_{n} <$ 0.4\,GeV/$c$ & 21.8  &  11.4 & 19.5 \tabularnewline \hline
 0.4 $ < p_{n} <$ 1\,GeV/$c$ & 49.3  &  26.3 & 44.5 \tabularnewline \hline
 \hline $|p_{\mathrm{miss}}| <$ 0.1\,GeV/$c$ & 20.7  &  19.4 & 15.5 \tabularnewline \hline
 0.1 $< |p_{\mathrm{miss}}| <$ 0.2\,GeV/$c$ & 23.6  &  21.3 & 16.5 \tabularnewline \hline
 0.2 $ < |p_{\mathrm{miss}}| <$ 0.5\,GeV/$c$ & 29.8  &  23.6 & 23.5 \tabularnewline \hline

\end{tabular}
\end{table}	

%%%%%%%%%%%%%%%%%%%%%%%%%%%%%%%%%%%%%%%%%%%%%%%%%%%%%%%%%%%%%%%%%%%%%%%%%%%%

%\vspace{-0.5cm}
\section{DUNE Atmospheric Projection} \label{sec:dune}

%%%%%%%%%%%%%%%%%%%%%%%

\begin{figure}[htb!]
\centering  

\begin{tikzpicture} \draw (0, 0) node[inner sep=0] {    
\includegraphics[width=\linewidth]{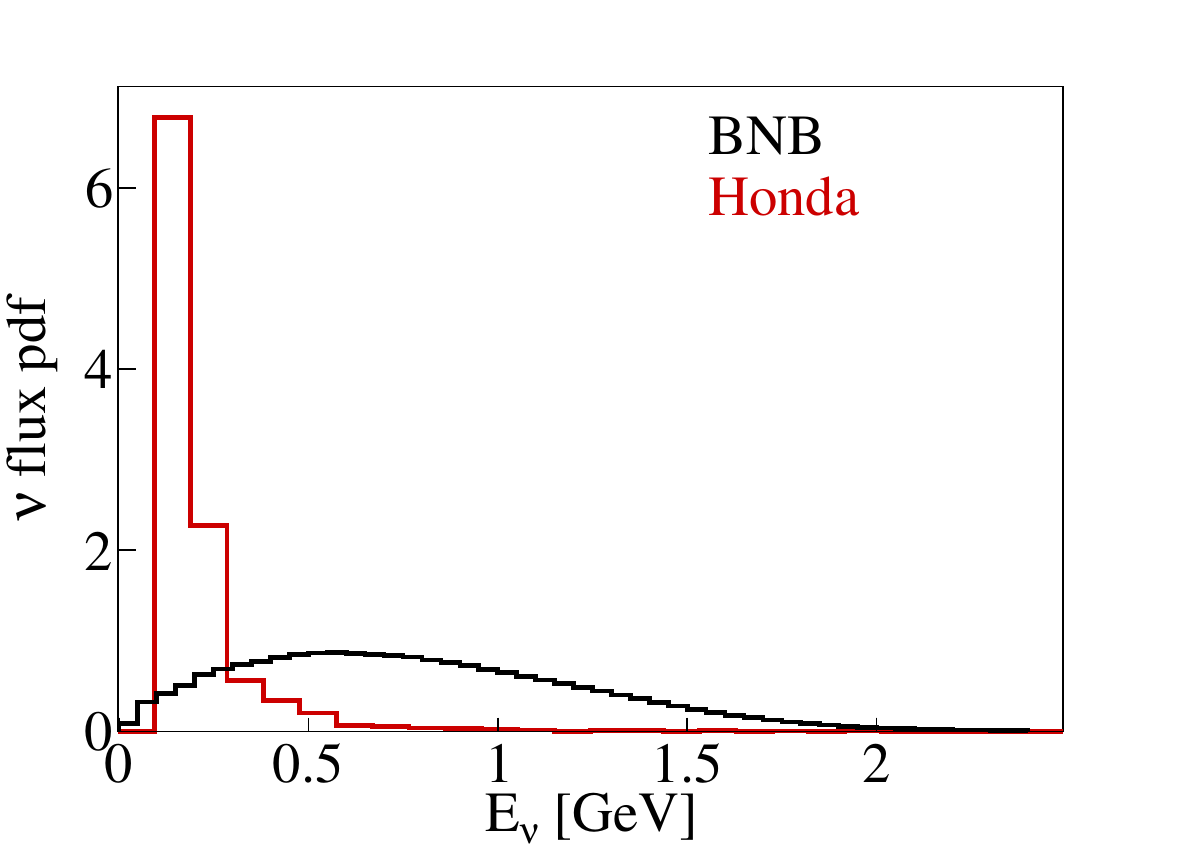}
};
\draw (0., -3.4) node {(a)};	
\end{tikzpicture}

\begin{tikzpicture} \draw (0, 0) node[inner sep=0] {
\includegraphics[width=\linewidth]{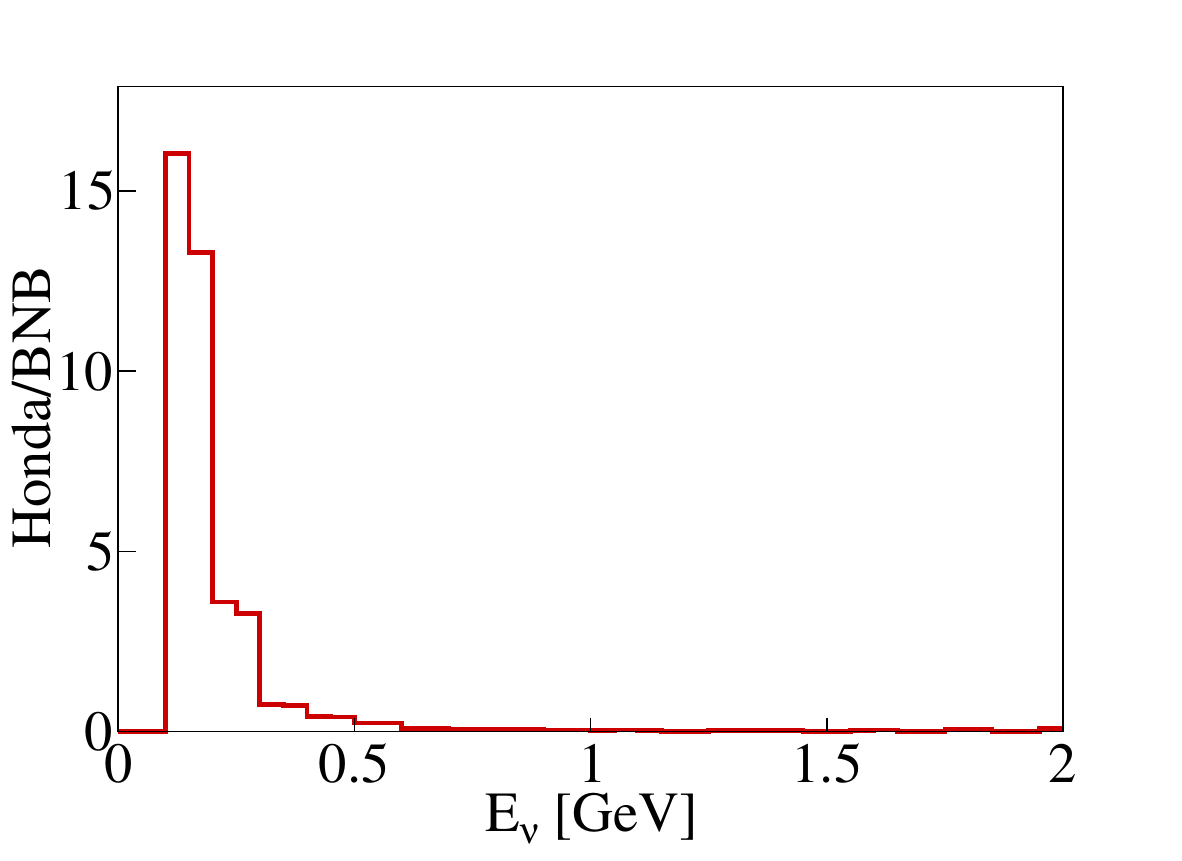}
};
\draw (0., -3.4) node {(b)};	
\end{tikzpicture}

\caption{
    (a) Comparison between the BNB and Honda $\nu_{\mu}$ flux pdfs in true neutrino energy.
    (b) Reweighting function used for the BNB-to-Honda projection using the ratio of the $\nu_{\mu}$ flux pdfs.
}
    \label{hondaflux}
\end{figure}

%%%%%%%%%%%%%%%%%%

\begin{figure}[htb!]
\centering  

\begin{tikzpicture} \draw (0, 0) node[inner sep=0] {    
\includegraphics[width=\linewidth]{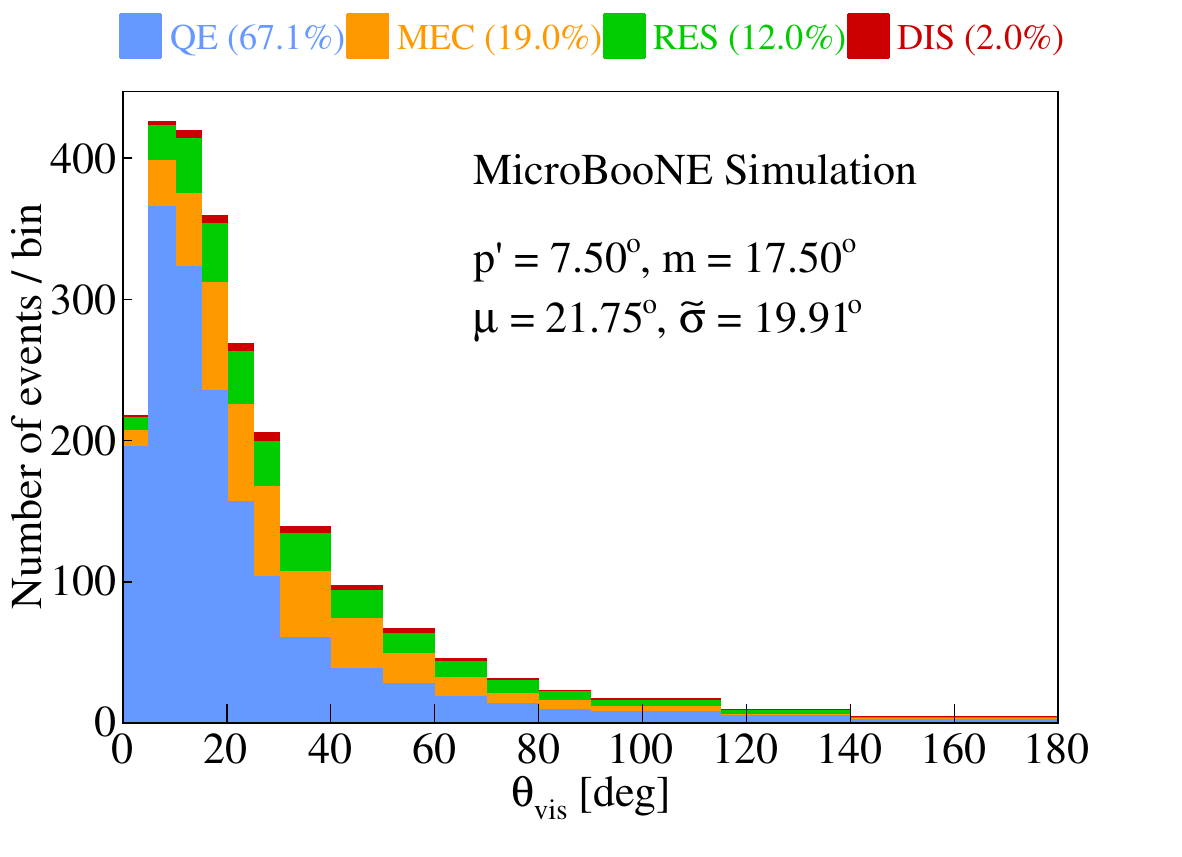}
};
\draw (0., -3.4) node {(a)};	
\end{tikzpicture}

\begin{tikzpicture} \draw (0, 0) node[inner sep=0] {
\includegraphics[width=\linewidth]{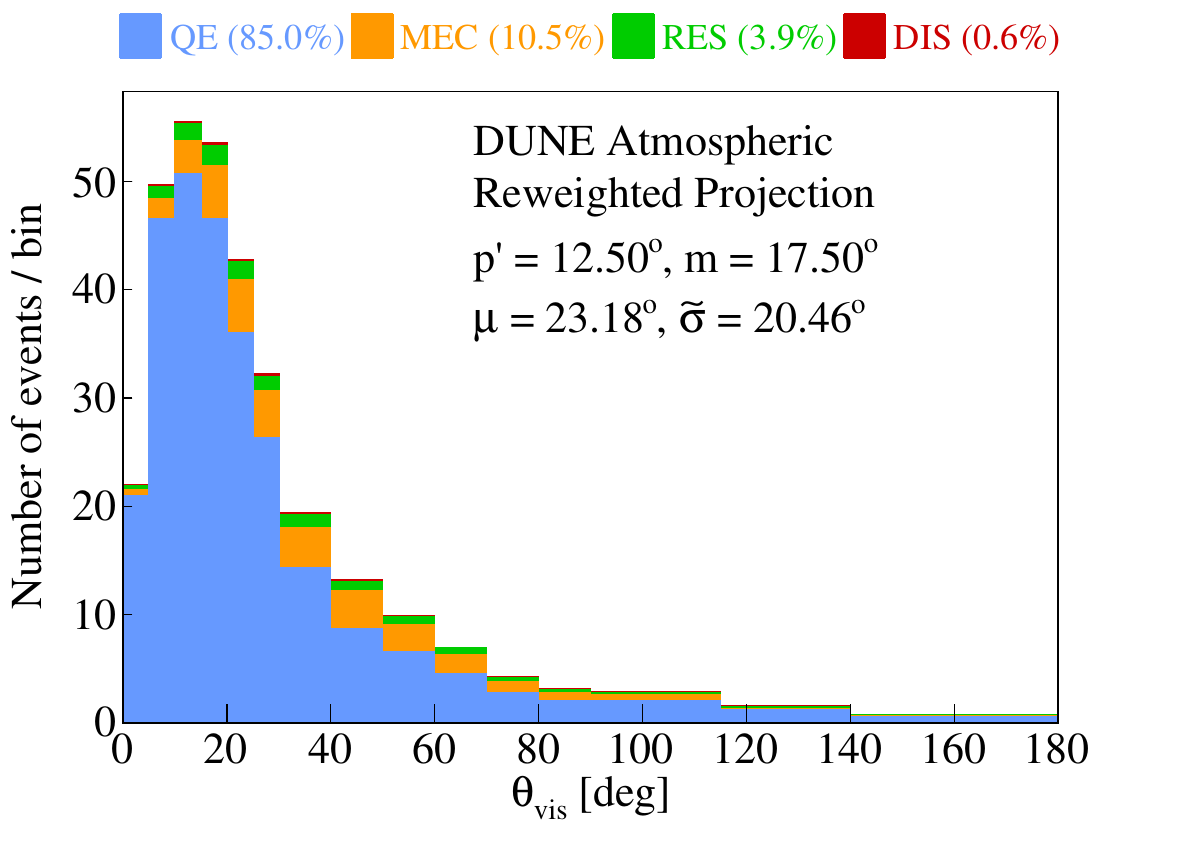}
};
\draw (0., -3.4) node {(b)};	
\end{tikzpicture}

\caption{
Distribution of $\theta_{\mathrm{vis}}$ using simulated candidate CC1p0$\pi$ events for (a) MicroBooNE and (b) a DUNE atmospheric projection using the reweighting as a function of $E_{\nu}$.
The peak location ($p^{\prime}$), median ($m$), mean value ($\mu$), and standard deviation ($\tilde{\sigma}$) describing the distributions are also shown.
}
\label{duneproj}
\end{figure}

In this section, we make a projection for the simulated $\theta_{\mathrm{vis}}$ distribution expected for DUNE atmospheric searches using the $\nu_{\mu}$ flux prediction from Honda $\textit{et al.}$~\cite{PhysRevD.92.023004} at the Homestake site.
This is motivated by the good agreement within uncertainties seen between data and prediction, which allows us to take advantage of our simulation dataset and make predictions for the expected angular resolution in DUNE.
In order to make the DUNE projection, we use the probability distribution function (pdf) for the BNB and Honda $\nu_{\mu}$ fluxes.
As shown in Fig.~\ref{hondaflux}\textcolor{blue}{(a)} showing the flux pdfs, there is a significant overlap between the Honda flux and the low energy range of the BNB flux.
Using the ratio between the two flux pdfs as a function of $E_{\nu}$, we derive a reweighting function to transition between the two fluxes, as shown in Fig.~\ref{hondaflux}\textcolor{blue}{(b)}.

This function is applied on an event-by-event basis on the simulated candidate CC1p0$\pi$ events that satisfy the MicroBooNE BNB event selection outlined in Sec.~\ref{sec:eventrate} and are shown in Fig.~\ref{duneproj}\textcolor{blue}{(a)}.
The reweighted distribution shown in Fig.~\ref{duneproj}\textcolor{blue}{(b)} corresponds to the expected $\theta_{\mathrm{vis}}$ behavior that DUNE atmospheric analyses might obtain when the Honda flux is used.
The MicroBooNE detector properties and cross section modeling are assumed to be the same for the purpose of this DUNE reweighting study.
As expected, due to the lower-energy Honda flux, the DUNE atmospheric distribution has a broader $\theta_{\mathrm{vis}}$ distribution than the MicroBooNE BNB $\theta_{\mathrm{vis}}$ one.

Figure~\ref{duneproj} also includes the interaction contributions of the simulated candidate CC1p0$\pi$ events for both the MicroBooNE result and the DUNE projection. 
Since the Honda flux peaks at lower energies compared to the BNB flux, the DUNE atmospheric projection has a higher contribution of QE events than MicroBooNE. 
%The DIS contribution is small for both fluxes.

%%%%%%%%%%%%%%%%%%%%%%%%%%%%%%%%%%%%%%%%%%%%%%%%%%%%%%%%%%%%%%%%%%%%%%%%%%%%

%\vspace{-0.5cm}

\section{MicroBooNE Cross Section Measurement} \label{sec:MicroBooNE}

%%%%%%%%%%%%%%%%%%%%%%%

\begin{figure}[htb!]
\centering  

\begin{tikzpicture} \draw (0, 0) node[inner sep=0] {    
\includegraphics[width=\linewidth]{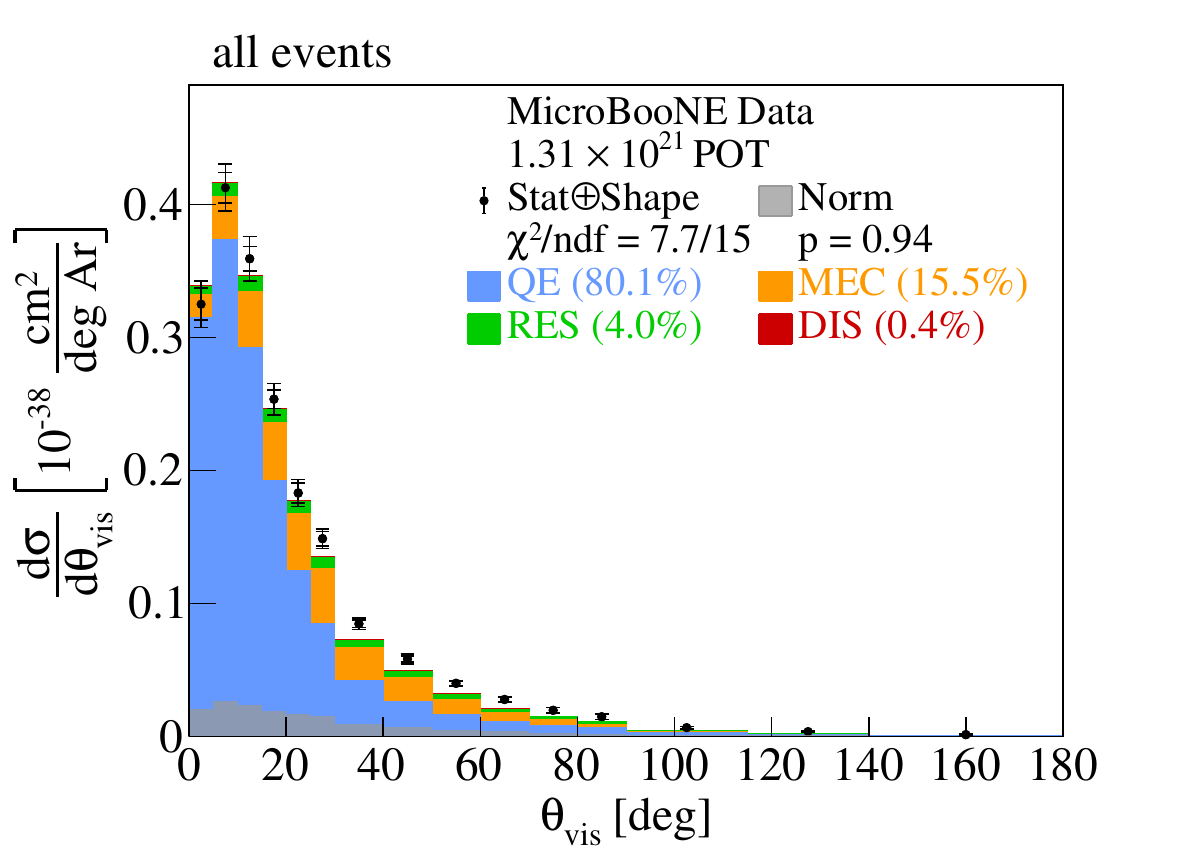}
};
\draw (0., -3.4) node {(a)};	
\end{tikzpicture}

\begin{tikzpicture} \draw (0, 0) node[inner sep=0] {
\includegraphics[width=\linewidth]{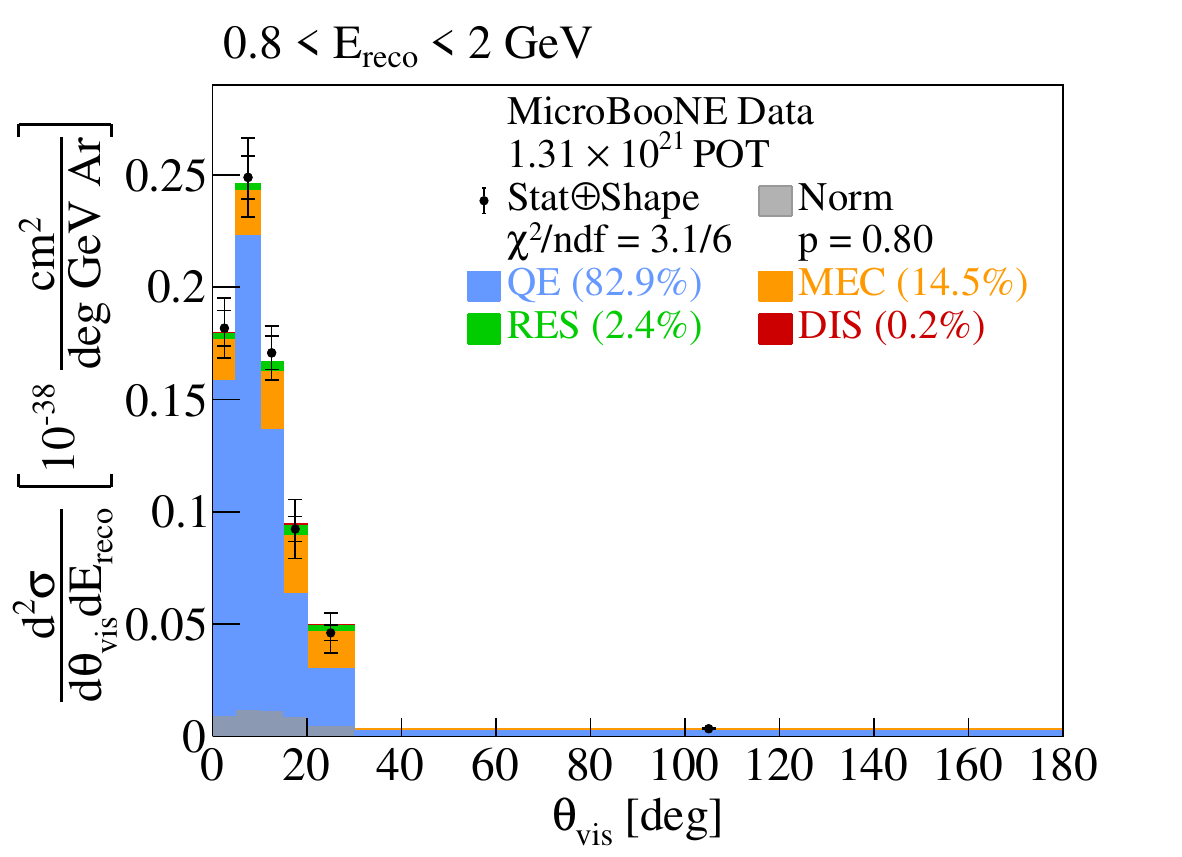}
};
\draw (0., -3.4) node {(b)};	
\end{tikzpicture}

\caption{
    (a) The unfolded cross section and the interaction contributions for the selected events for the $\texttt{G18T}$ configuration as a function of $\theta_{\mathrm{vis}}$.
    (b) The unfolded cross section and the interaction contributions for the selected events using the same configuration as a function of $\theta_{\mathrm{vis}}$ for events with 0.8 $< E_{\mathrm{reco}} <$ 2\,GeV.
    The gray band shows the normalization systematic uncertainty.
    Inner and outer error bars show the statistical and the statistical$\oplus$shape uncertainty at the 1$\sigma$, or 68\%, confidence level. 
}
\label{xsecThetaVis}
\end{figure}

%%%%%%%%%%%%%%%%%%%%%%%

\begin{figure}[htb!]
\centering  

\begin{tikzpicture} \draw (0, 0) node[inner sep=0] {    
\includegraphics[width=\linewidth]{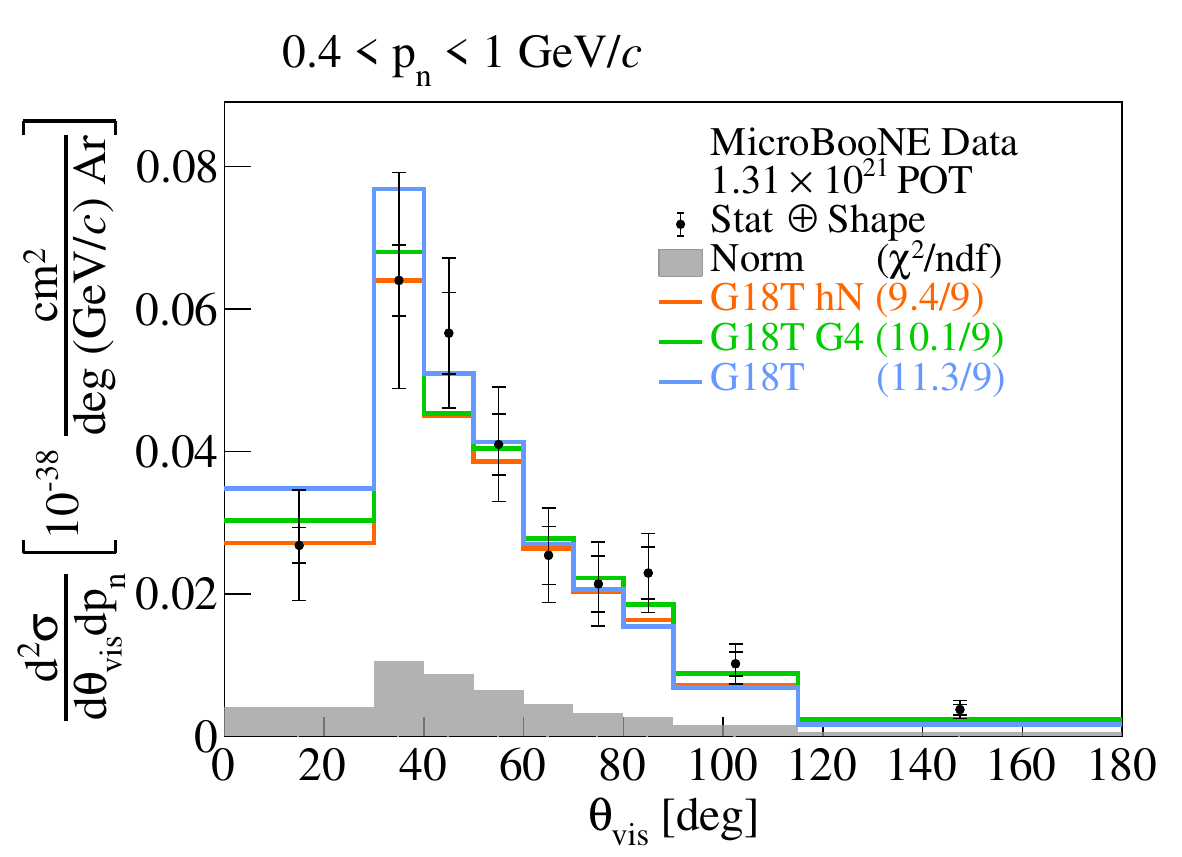}
};
\draw (0., -3.4) node {(a)};	
\end{tikzpicture}

\begin{tikzpicture} \draw (0, 0) node[inner sep=0] {
\includegraphics[width=\linewidth]{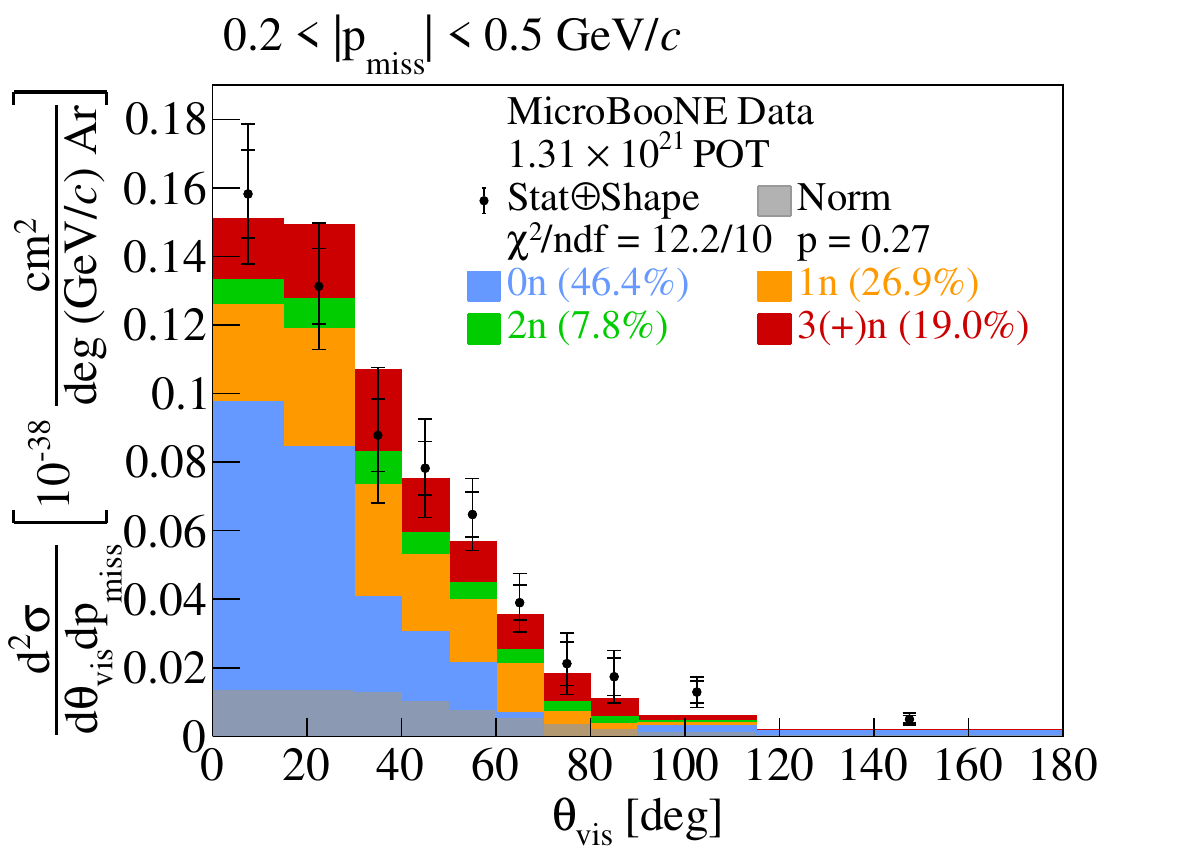}
};
\draw (0., -3.4) node {(b)};	
\end{tikzpicture}

\caption{
    (a) The flux-integrated double-differential cross sections as a function of $\theta_{\mathrm{vis}}$ for 0.4 $< p_{n} <$ 1\,GeV/$c$ reported in regularized truth space. 
Colored lines show the results of theoretical cross section calculations using the $\texttt{G18T hN}$ prediction with the hN FSI model (orange), the $\texttt{G18T G4}$ prediction with the \texttt{GEANT4} FSI model (green), and the $\texttt{G18T}$ prediction with the hA FSI model (light blue).
The gray band at the bottom shows the normalization systematic uncertainty.
The numbers in parentheses show the $\chi^{2}$/ndf calculation for each of the predictions.
    (b) The flux-integrated double-differential cross sections as a function of $\theta_{\mathrm{vis}}$ for 0.2 $< |p_{\mathrm{miss}}| <$ 0.5\,GeV/$c$ reported in regularized truth space. 
Colored stacked histograms show the results of theoretical cross section calculations for events with no neutrons, $\texttt{G18T 0n}$ (light blue), one neutron, $\texttt{G18T 1n}$ (orange), two neutrons, $\texttt{G18T 2n}$ (green), and at least three neutrons, $\texttt{G18T 3(+)n}$ (red).
The numbers in parentheses show the fractional contribution for each neutron multiplicity.
Inner and outer error bars show the statistical and the statistical$\oplus$shape uncertainty at the 1$\sigma$, or 68\%, confidence level. 
}
    \label{xsecThetaVisInSlices}
\end{figure}

%%%%%%%%%%%%%%%%%%%%%%%

We report the extracted cross sections ($\sigma$) from the MicroBooNE data as a function of true kinematic variables using the Wiener singular value decomposition (Wiener-SVD) unfolding technique~\cite{Tang_2017}.
This technique transforms both the data measurement and covariance matrix into a regularized truth space.
It requires the construction of a response matrix describing the expected detector smearing and reconstruction efficiency.
This matrix is responsible for correcting for these effects. 
The input covariance matrix is constructed using the uncertainties related to the incident neutrino flux~\cite{AguilarArevalo:2008yp}, interaction model~\cite{GENIE_tune}, particle propagation~\cite{GEANT4}, and detector response~\cite{uBooNE_det_unc}.
The binning is chosen to balance resolution and statistics.  
Each measurement is accompanied by an output regularization matrix $A_{C}$.
The $A_{C}$ matrix performs the conversion from the truth to the regularized truth space and is included in the Supplemental Material~\cite{suppmat}.
The unfolding is performed for each of the observables of interest using the $\texttt{G18T}$ model described in Sec.~\ref{sec:eventrate}.
The robustness of the unfolding method is verified using fake data studies with alternative generator predictions.
Three fake data samples were investigating by (a) using NuWro v19.02.1 events~\cite{GOLAN2012499}, (b) removing the weights corresponding to the MicroBooNE tune from the default MC, and (c) multiplying the weight for the MEC events by a factor of two in the default MC.

%%%%%%%%%%%%%%%%%

\begin{figure}[htb!]
	\centering  
	\includegraphics[width=\linewidth]{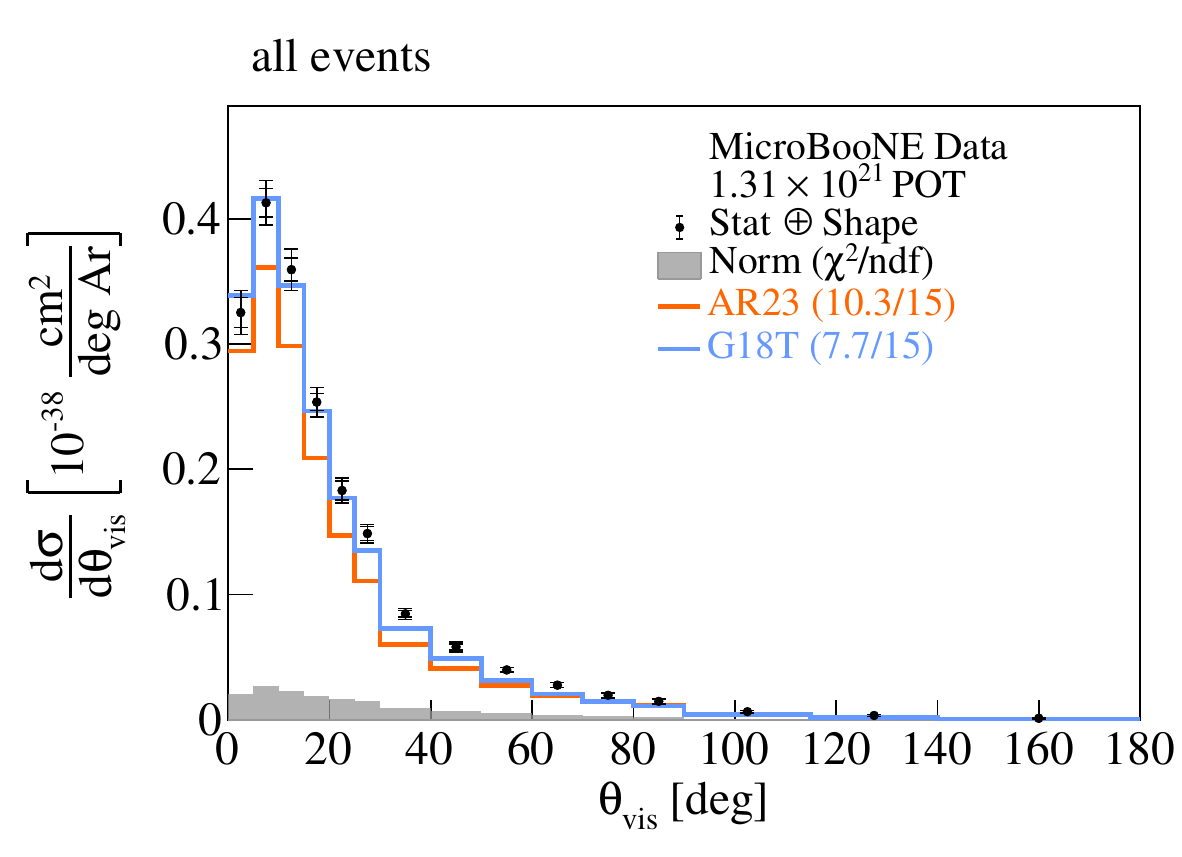}
    \caption{
The flux-integrated single-differential cross sections as a function of $\theta_\mathrm{vis}$ reported in regularized truth space. 
The $\texttt{G18T}$ (light blue) and $\texttt{AR23}$ (orange) $\texttt{GENIE}$ configuration predictions are compared to data.
Inner and outer error bars show the statistical and the statistical$\oplus$shape uncertainty at the 1$\sigma$, or 68\%, confidence level. 
The gray band shows the normalization systematic uncertainty.
The numbers in parentheses show the $\chi^{2}$/ndf calculation for each of the predictions.
    }
    \label{g18t_vs_ar23_xsec_1d}
\end{figure}

%%%%%%%%%%%%%%%%%

\begin{figure}[H]
\centering 

\begin{tikzpicture} \draw (0, 0) node[inner sep=0] {
\includegraphics[width=\linewidth]{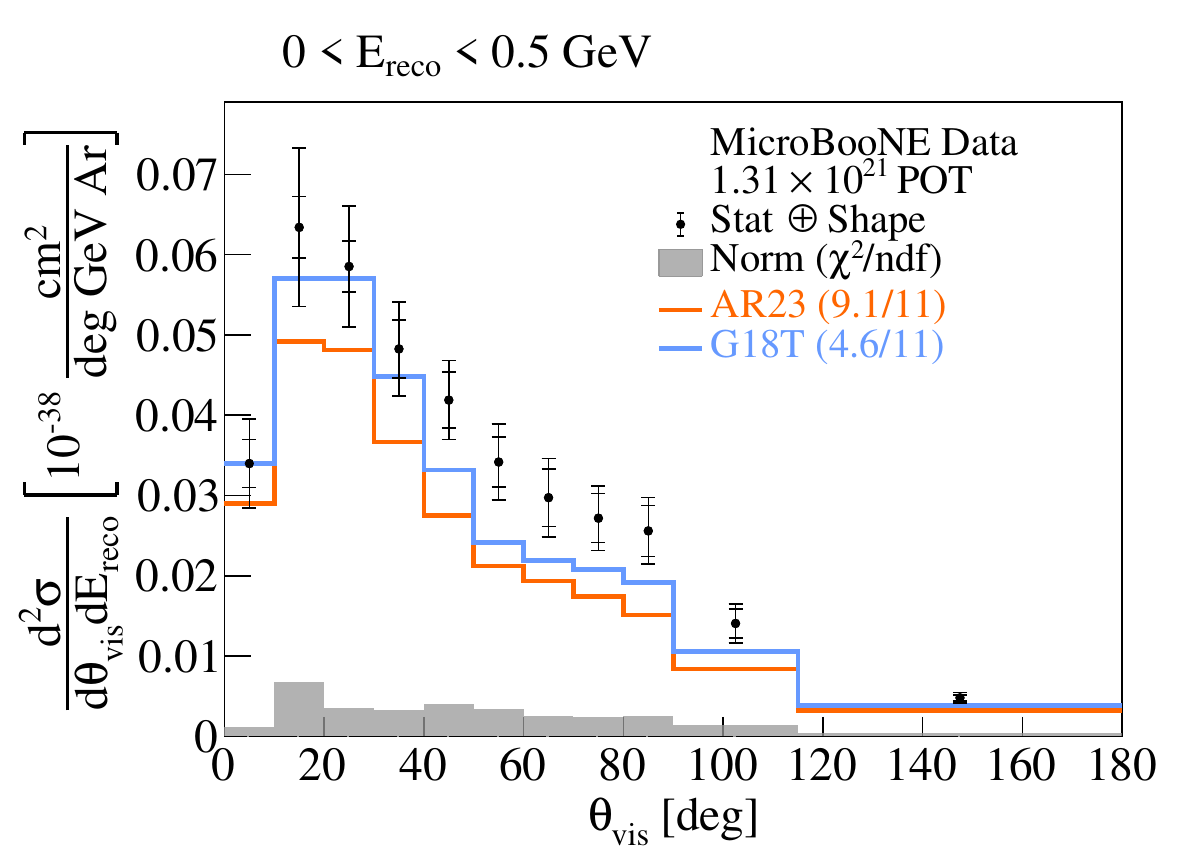}
};
\draw (0., -3.4) node {(a)};	
\end{tikzpicture}

\begin{tikzpicture} \draw (0, 0) node[inner sep=0] {
\includegraphics[width=\linewidth]{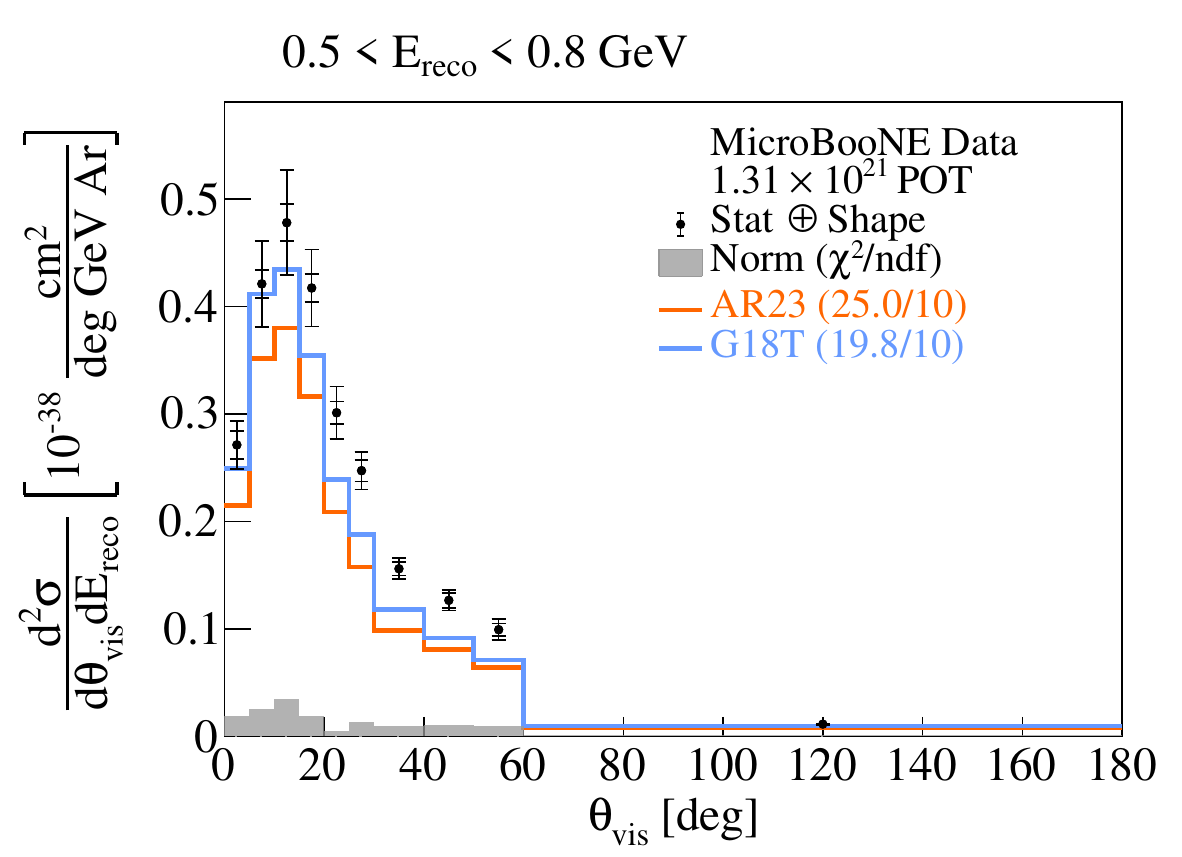}
};
\draw (0., -3.4) node {(b)};	
\end{tikzpicture}

\begin{tikzpicture} \draw (0, 0) node[inner sep=0] {
\includegraphics[width=\linewidth]{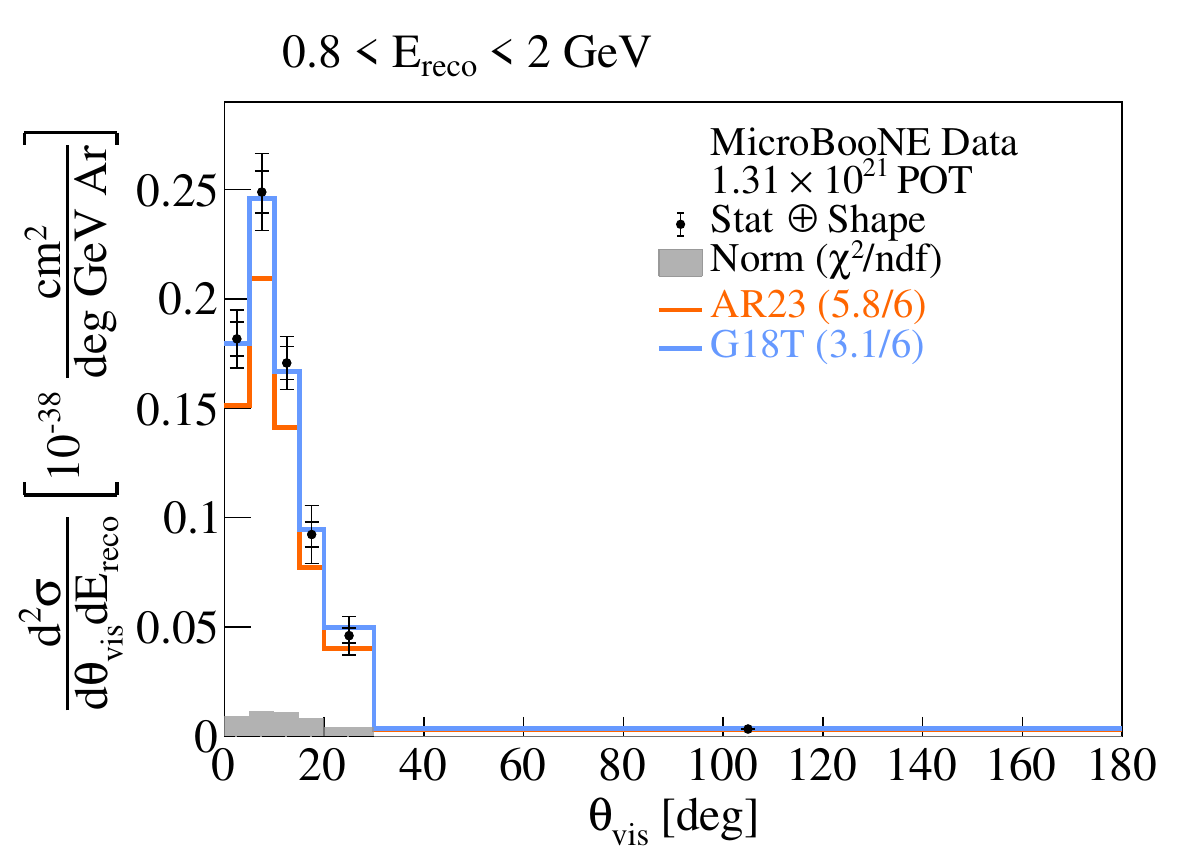}
};
\draw (0., -3.4) node {(c)};	
\end{tikzpicture}

\caption{
The flux-integrated double-differential cross sections as a function of $\theta_{\mathrm{vis}}$ for (a) $E_{\mathrm{reco}} <$ 0.5\,GeV, (b) 0.5 $ < E_{\mathrm{reco}} <$ 0.8\,GeV, and (c) 0.8 $ < E_{\mathrm{reco}} <$ 2\,GeV. 
The $\texttt{G18T}$ (light blue) and $\texttt{AR23}$ (orange) $\texttt{GENIE}$ configuration predictions are compared to data.
Inner and outer error bars show the statistical and the statistical$\oplus$shape uncertainty at the 1$\sigma$, or 68\%, confidence level. 
The gray band shows the normalization systematic uncertainty.
The numbers in parentheses show the $\chi^{2}$/ndf calculation for each  of the predictions.
}
\label{g18t_vs_ar23_xsec_2d_ecal}
\end{figure}

\begin{figure}[H]
\centering 

\begin{tikzpicture} \draw (0, 0) node[inner sep=0] {
\includegraphics[width=\linewidth]{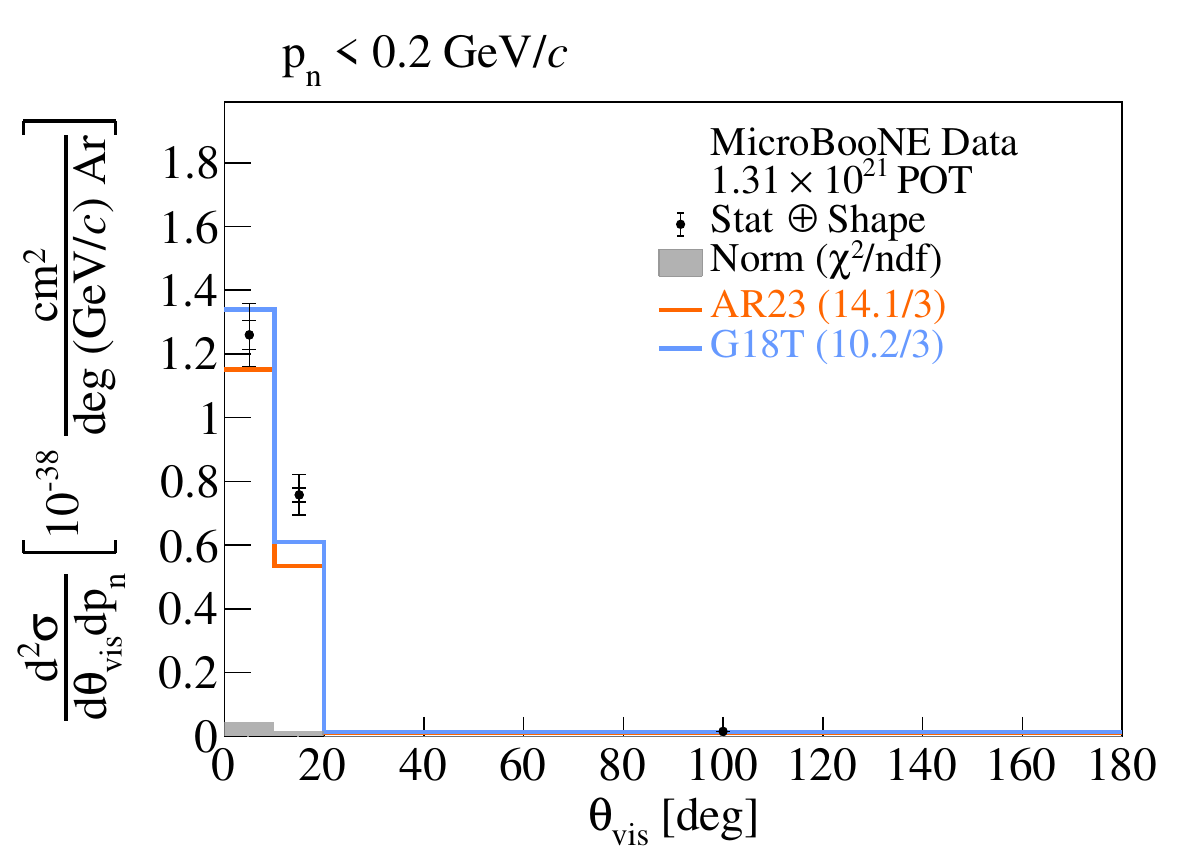}
};
\draw (0., -3.4) node {(a)};	
\end{tikzpicture}

\begin{tikzpicture} \draw (0, 0) node[inner sep=0] {
\includegraphics[width=\linewidth]{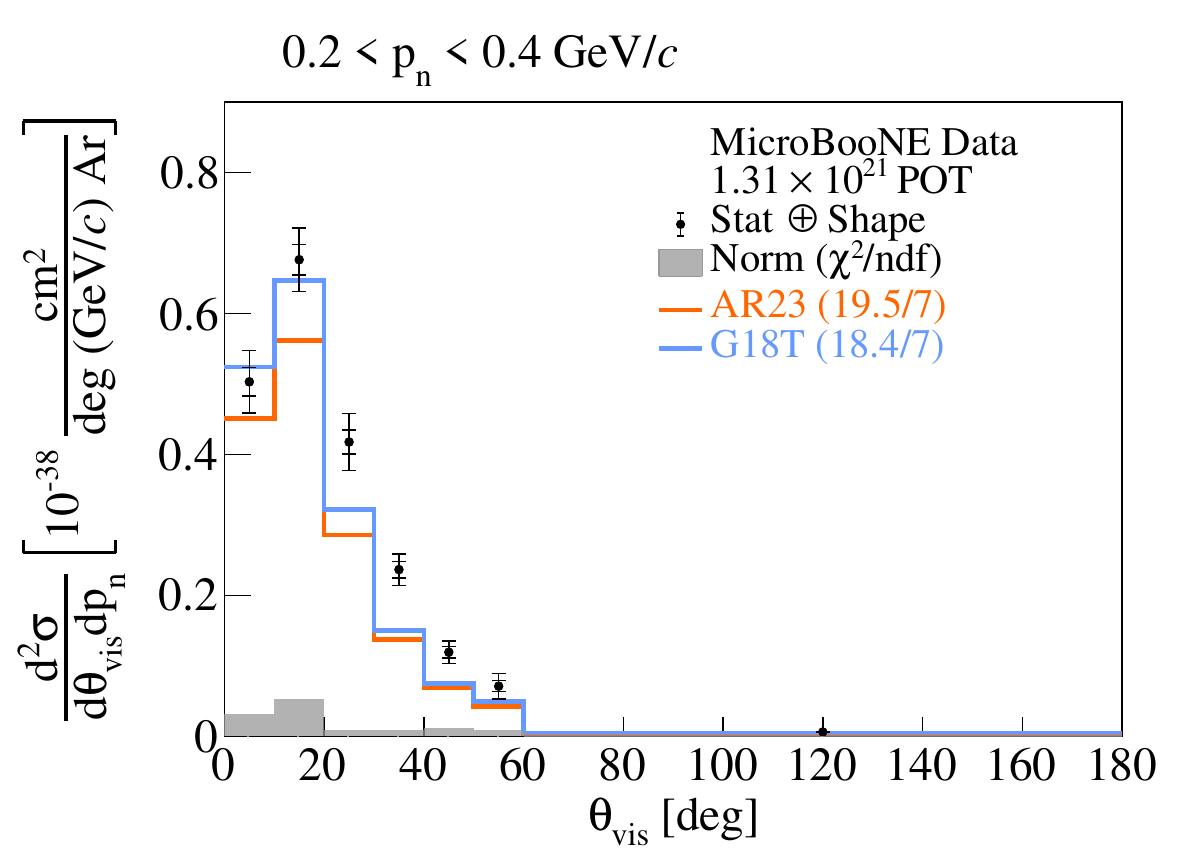}
};
\draw (0., -3.4) node {(b)};	
\end{tikzpicture}

\begin{tikzpicture} \draw (0, 0) node[inner sep=0] {
\includegraphics[width=\linewidth]{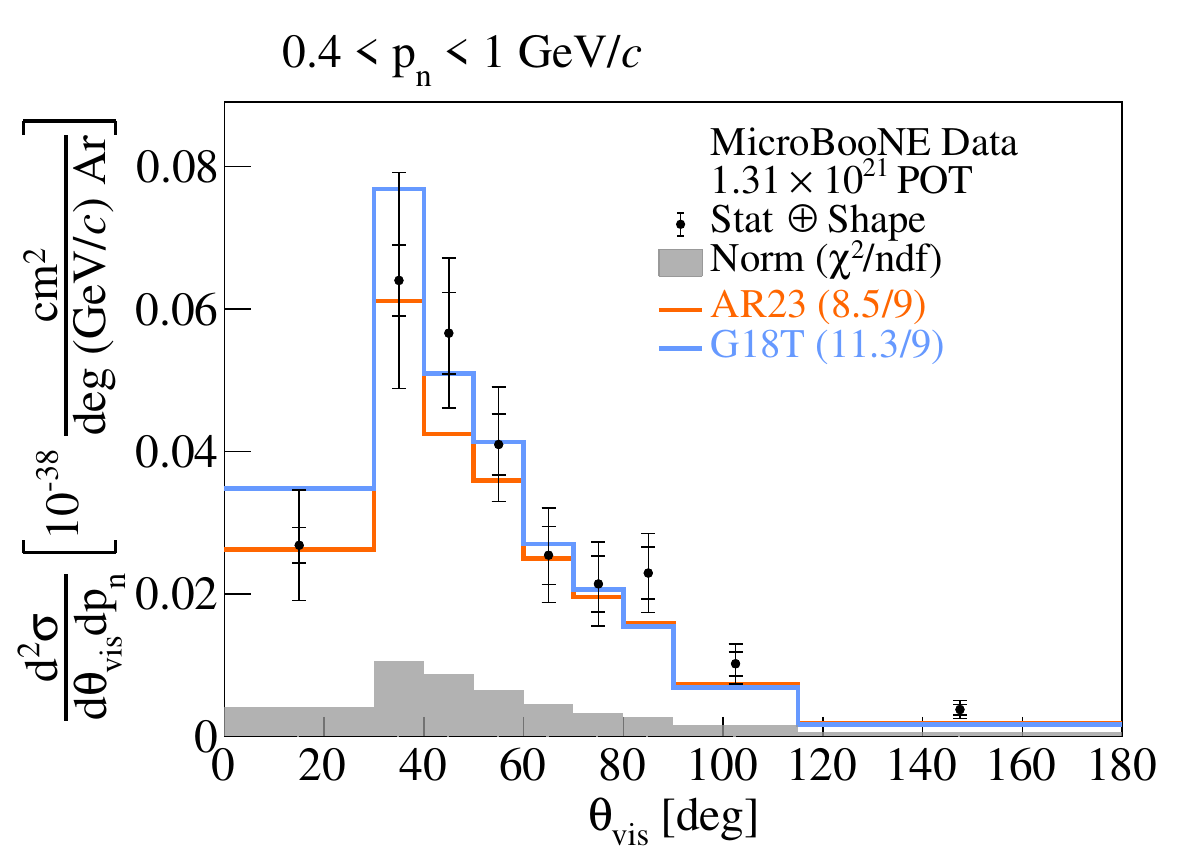}
};
\draw (0., -3.4) node {(c)};	
\end{tikzpicture}

\caption{
The flux-integrated double-differential cross sections as a function of $\theta_{\mathrm{vis}}$ for (a) $p_{n} <$ 0.2\,GeV/$c$, (b) 0.2 $ < p_{n} <$ 0.4\,GeV/$c$, and (c) 0.4 $ < p_{n} <$ 1\,GeV/$c$.
The $\texttt{G18T}$ (light blue) and $\texttt{AR23}$ (orange) $\texttt{GENIE}$ configuration predictions are compared to data.
Inner and outer error bars show the statistical and the statistical$\oplus$shape uncertainty at the 1$\sigma$, or 68\%, confidence level. 
The gray band shows the normalization systematic uncertainty.
The numbers in parentheses show the $\chi^{2}$/ndf calculation for each of the predictions.
}
\label{g18t_vs_ar23_xsec_2d_pn}
\end{figure}

\begin{figure}[H]
\centering 

\begin{tikzpicture} \draw (0, 0) node[inner sep=0] {
\includegraphics[width=\linewidth]{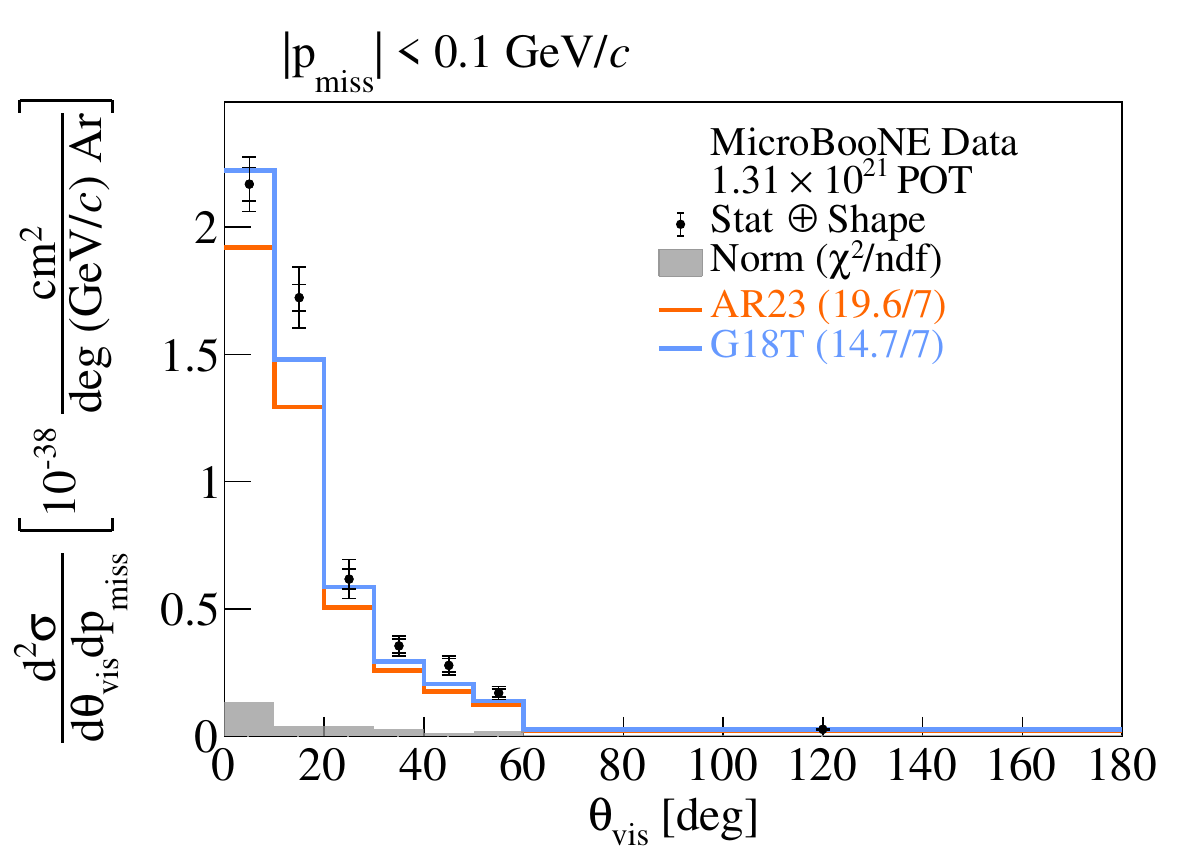}
};
\draw (0., -3.4) node {(a)};	
\end{tikzpicture}

\begin{tikzpicture} \draw (0, 0) node[inner sep=0] {
\includegraphics[width=\linewidth]{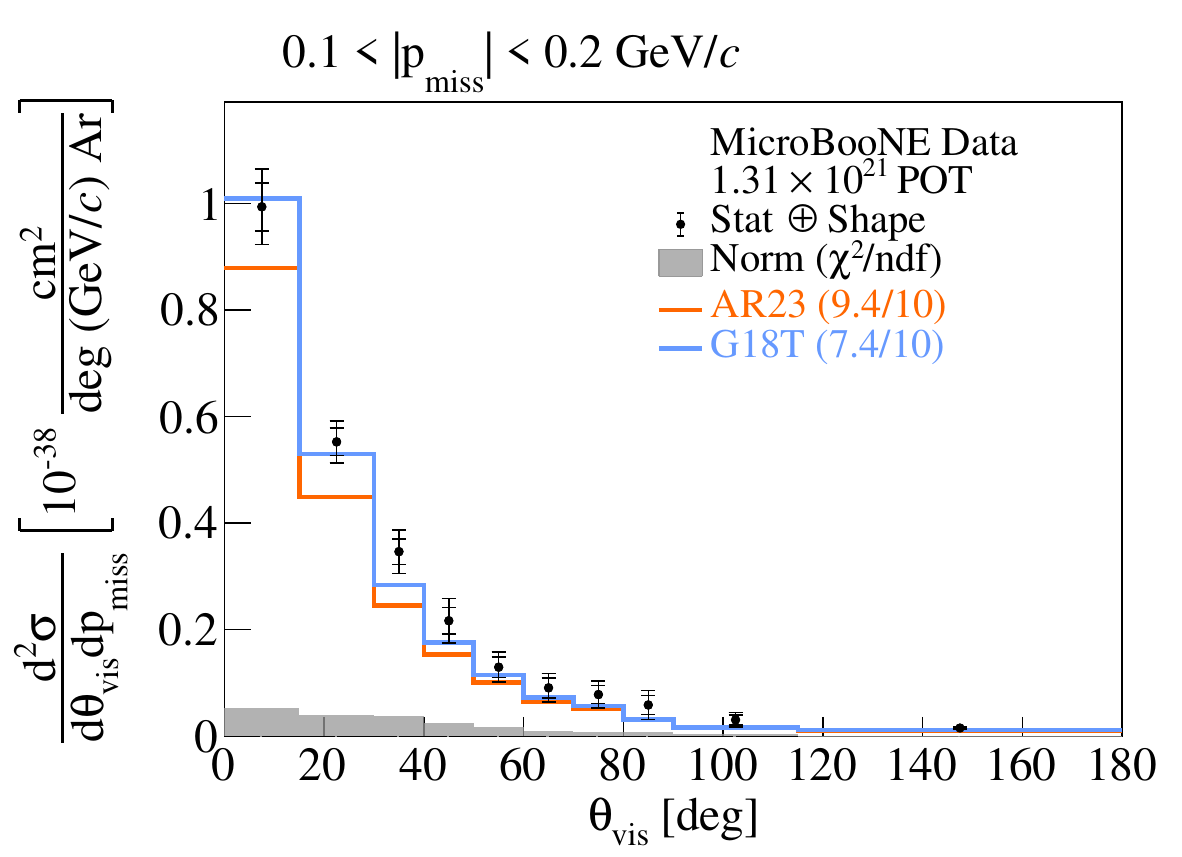}
};
\draw (0., -3.4) node {(b)};	
\end{tikzpicture}

\begin{tikzpicture} \draw (0, 0) node[inner sep=0] {
\includegraphics[width=\linewidth]{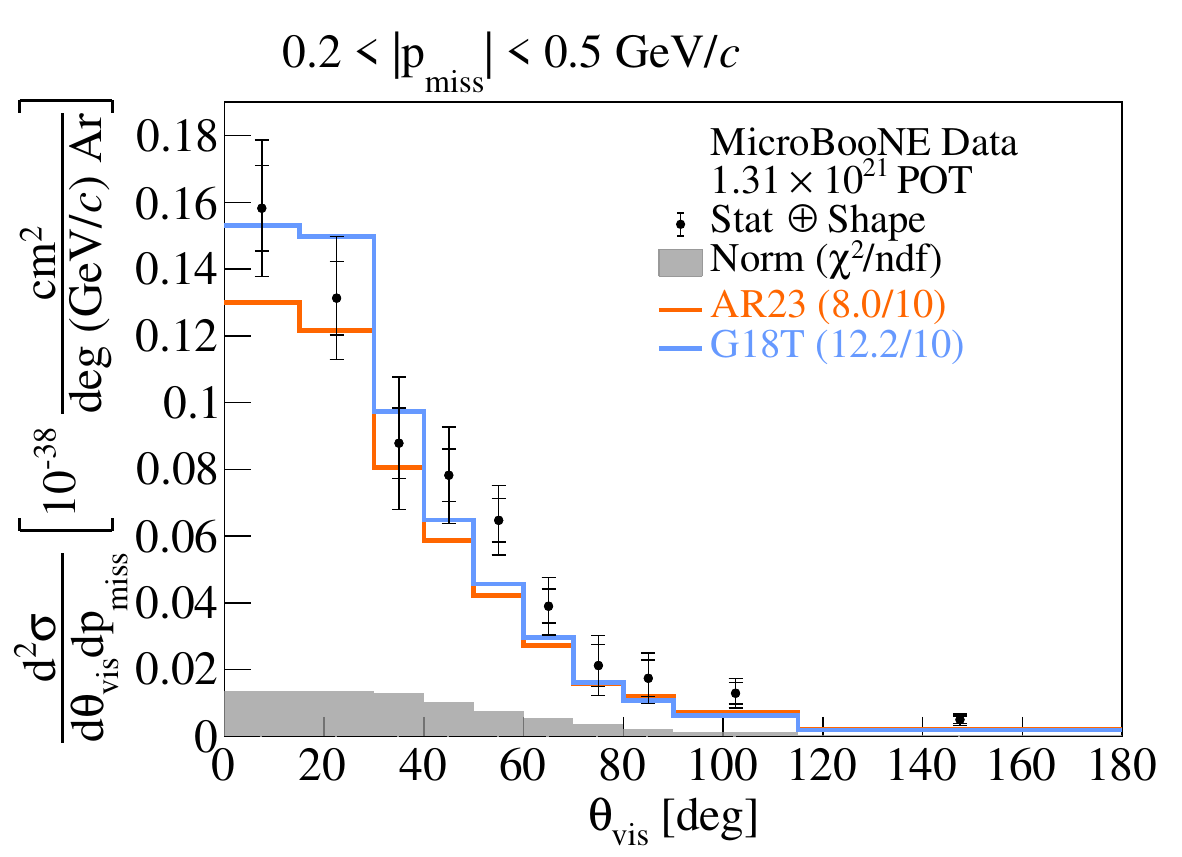}
};
\draw (0., -3.4) node {(c)};	
\end{tikzpicture}

\caption{
The flux-integrated double-differential cross sections as a function of $\theta_{\mathrm{vis}}$ for (a) $|p_{\mathrm{miss}}| <$ 0.1\,GeV/$c$, (b) 0.1 $ < |p_{\mathrm{miss}}| <$ 0.2\,GeV/$c$, and (c) 0.2 $ < |p_{\mathrm{miss}}| <$ 0.5\,GeV/$c$.
The $\texttt{G18T}$ (light blue) and $\texttt{AR23}$ (orange) $\texttt{GENIE}$ configuration predictions are compared to data.
Inner and outer error bars show the statistical and the statistical$\oplus$shape uncertainty at the 1$\sigma$, or 68\%, confidence level. 
The gray band shows the normalization systematic uncertainty.
The numbers in parentheses show the $\chi^{2}$/ndf calculation for each of the predictions.
}
\label{g18t_vs_ar23_xsec_2d_pmiss}
\end{figure}

The event rate has the predicted background subtracted before the unfolding.
It is further divided by the integrated neutrino flux and number of argon nuclei in the fiducial volume to report a differential cross section.
In the results presented in Figs.~\ref{xsecThetaVis}--\ref{g18t_vs_ar23_xsec_2d_pmiss}, the inner error bars on the cross sections correspond to the data statistical uncertainties.
The systematic uncertainties are decomposed into data shape and normalization sources following the procedure outlined in Ref.~\cite{MatrixDecomv}.
The cross-term uncertainties are incorporated in the normalization.
The outer error bars on the reported cross sections correspond to data statistical and shape uncertainties added in quadrature.
The data normalization uncertainties are presented as a band at the bottom of each plot.
The degrees of freedom (ndf) correspond to the number of bins.
The $\chi^{2}$/ndf and p-value ($p$) data comparison for each generator prediction shown on all the figures takes into account the total covariance matrix.
More details on the systematic uncertainties and the cross section extraction technique can be found in Ref.~\cite{RefPRD}.
All the extracted cross sections are reported in the Supplemental Material~\cite{suppmat}.
They are compared to the $\texttt{G18T}$ model set used by MicroBooNE, as well as the model set used by DUNE. 

The single-differential cross section as a function of $\theta_{\mathrm{vis}}$ is shown in Fig.~\ref{xsecThetaVis}\textcolor{blue}{(a)}.
As expected from the reconstructed event spectrum shown in Fig.~\ref{ThetaVisSlices}, it is a distribution that peaks at a non-zero value of $\sim$10$^{\mathrm{o}}$ and extends to 180$^{\mathrm{o}}$.
The low-$\theta_{\mathrm{vis}}$ part of the distribution is dominated by QE interactions.
The higher-$\theta_{\mathrm{vis}}$ part of the spectrum has strong contributions from interactions with multi-nucleon effects, namely MEC and RES along with small DIS contributions.
In this case where all events are considered, the $\texttt{G18T}$ prediction yields good data-simulation agreement with a $\chi^{2}$/ndf less than one and a p-value close to unity.

%%%%%%%%%%%%%%%%%

%%%%%%%%%%%%%%%%%

%%%%%%%%%%%%%%%%%

%%%%%%%%%%%%%%%%%

As discussed in Sec.~\ref{intro}, the neutrino angular orientation is of great importance for atmospheric neutrino oscillation studies.  
Thus, an extended $\theta_{\mathrm{vis}}$ tail can be a limiting factor that introduces significant uncertainties and limits the experimental sensitivity.
To that end, experimental efforts might want to primarily study events of interest in regions of the phase-space that result in a smaller spread of the $\theta_{\mathrm{vis}}$ distribution.
An example of such a phase-space region is shown in Fig.~\ref{xsecThetaVis}\textcolor{blue}{(b)} and corresponds to higher reconstructable energies (0.8 $< E_{\mathrm{reco}} <$ 2\,GeV) averaged over the relevant energy range.
In this limited phase space, $\theta_{\mathrm{vis}}$ rarely exceeds 30$^{\mathrm{o}}$ and the $\texttt{G18T}$ prediction results in good agreement when compared to the data.
Other phase space regions with small $\theta_{\mathrm{vis}}$ spread are those with $p_{n} <$ 0.2\,GeV/$c$ and $|p_{\mathrm{miss}}| <$ 0.1\,GeV/$c$, as can be seen in Figs.~\ref{g18t_vs_ar23_xsec_2d_pn} and~\ref{g18t_vs_ar23_xsec_2d_pmiss}, which will be shown later.

In contrast to these phase space restrictions that yield noticeably narrower $\theta_{\mathrm{vis}}$ distributions, regions with a wide angular spread have also been identified.
Figure~\ref{xsecThetaVisInSlices}\textcolor{blue}{(a)} shows an example corresponding to high total reconstructed struck nucleon momentum values (0.4 $< p_{n} <$ 1\,GeV/$c$).
This phase space region has been shown to be dominated by events that undergo FSI~\cite{PhysRevD.109.092007}.
To test the $\theta_{\mathrm{vis}}$ FSI sensitivity in that region, we study the performance of different FSI modeling options using T2K-tuned $\texttt{GENIE}$ configurations and compare them to the data results.
These tuned configurations include the $\texttt{GENIE v3.0.6 G18\_10a\_02\_11}$ configuration with the empirical \texttt{hA2018} model ($\texttt{G18T}$)~\cite{Ashery:1981tq},
the $\texttt{GENIE v3.0.6 G18\_10b\_02\_11}$ configuration with the \texttt{hN2018} cascade model ($\texttt{G18b}$)~\cite{hN2018},
and the $\texttt{GENIE v3.0.6 G18\_10d\_02\_11}$ configuration with the \texttt{GEANT4}-Bertini model ($\texttt{G18d}$)~\cite{Wright:2015xia}.
A comparison across the relevant predictions is shown in Fig.~\ref{xsecThetaVisInSlices}\textcolor{blue}{(a)} and reveals that the three FSI models yield comparable predictions that describe the shape of the $\theta_{\mathrm{vis}}$ distribution well.
Discrimination power across the three FSI models could be established in future iterations of the analysis with reduced uncertainties in the first bins.
A similar broadband $\theta_{\mathrm{vis}}$ distribution is observed for low $E_{\mathrm{reco}}$ values and is presented in Fig.~\ref{g18t_vs_ar23_xsec_2d_ecal}, which will be discussed later.

%%%%%%%%%%%%%%%%%

Other broad-spectrum features are observed for high missing momentum values (0.2 $< |p_\mathrm{miss}|<$ 0.5\,GeV/$c$) in Fig.~\ref{xsecThetaVisInSlices}\textcolor{blue}{(b)}. 
The figure shows that these high missing momenta are obtained due to the presence of neutrons that deposit little to no energy in the detector, resulting in a tail that extends beyond $30^\circ$ in $\theta_{\mathrm{vis}}$. 
Multi-neutron contributions account for more than half of the events in the high $|p_\mathrm{miss}|$ sample. 
On the other hand, this reduces to 30\% for events with low $|p_\mathrm{miss}|$ values.
The neutron contributions for the $\theta_{\mathrm{vis}}$ distribution using all the selected events and those events with $|p_{\mathrm{miss}}| <$ 0.2\,GeV/$c$ can be found in the Supplemental Material~\cite{suppmat}.
The importance of neutron detection in liquid argon has motivated a number of efforts and identification techniques~\cite{refId0,PhysRevD.107.072009,Rivera:2021dcf}.

%%%%%%%%%%%%%%%%%

For completeness, the data results are also compared against the $\texttt{v3.4.2 GENIE AR23\_20i\_00\_000 (AR23)}$ model prediction that is currently used by DUNE and other Short Baseline Neutrino (SBN) experiments.
The $\texttt{AR23}$ model set shares many common features with the $\texttt{G18T}$ one, but it includes some notable differences.
Namely, $\texttt{AR23}$ uses the local Fermi gas ground state modeling along with a correlated high-momentum nucleon tail~\cite{geniev3highlights};
the $z$-expansion form factors for CCQE interactions~\cite{Meyer:2016oeg};
the SuSAv2 modeling  for MEC interactions~\cite{PhysRevD.101.033003};
emission of de-excitation photons for argon nuclei~\cite{geniev3highlights};
and the free nucleon tune~\cite{PhysRevD.104.072009}. 

Figure~\ref{g18t_vs_ar23_xsec_1d} shows the single-differential cross section measurement, while Figs.~\ref{g18t_vs_ar23_xsec_2d_ecal}--\ref{g18t_vs_ar23_xsec_2d_pmiss} show the double-differential cross sections as functions of $\theta_\mathrm{vis}$ and $E_{\mathrm{reco}}$, $p_{n}$ and $|p_{\mathrm{miss}}|$.
$\texttt{AR23}$ yields a systematically lower prediction than $\texttt{G18T}$ due to the fact that, unlike $\texttt{G18T}$, no tuning is applied on the $\texttt{AR23}$ QE contribution.
Yet, there are also parts of the phase space where $\texttt{G18T}$ demonstrates a poor performance with $\chi^{2}$/ndf greater than unity.

%%%%%%%%%%%%%%%%%

\section{Conclusions} \label{sec:discussion}

We report on the precision of the reconstructed neutrino orientation $\theta_{\mathrm{vis}}$ for event topologies with a single muon and a single proton in the final state.
The data are recorded with the MicroBooNE LArTPC detector using the Booster Neutrino Beam at Fermi National Accelerator Laboratory with an exposure of $1.3 \times 10^{21}$ protons on target.
We find that the neutrino direction reconstruction performance using the single-proton selection is, in most cases, better than assumed in already published literature using an inclusive selection.
Using a reweighting function in $E_{\nu}$, we use the reconstructed simulated events in MicroBooNE to make a projection for the spectrum that the DUNE atmospheric studies might observe.
The $\theta_{\mathrm{vis}}$ cross sections are studied in phase-space regions of reconstructable neutrino energy, total struck nucleon momentum, and total missing momentum.
The latter is agnostic to the angular orientation of the incoming neutrino and, therefore, can be used in atmospheric neutrino studies to separate events with better and worse directional reconstruction.
The $\texttt{G18T}$ modeling performance is found to be satisfactory within the extracted uncertainties and able to describe the majority of the nuclear effects driving the $\theta_{\mathrm{vis}}$ distribution shape in different parts of the phase-space.
We also report single-differential cross section measurements as a function of $\theta_{\mathrm{vis}}$, and double-differential cross section measurements as functions of $\theta_{\mathrm{vis}}$ and the reconstructable visible energy, the total struck nucleon momentum, and the missing momentum.
These results can be used to inform the sub-GeV atmospheric oscillation studies that will be reported by forthcoming experiments like DUNE.

%%%%%%%%%%%%%%%%%%%%%%%%%%%%%%%%%%%%%%%%%%%%%%%%%%%%%%%%%%%%%%%%%%%%%%%%%%%%

\section{Acknowledgements} \label{sec:ack}

This document was prepared by the MicroBooNE collaboration using the
resources of the Fermi National Accelerator Laboratory (Fermilab), a
U.S. Department of Energy, Office of Science, Office of High Energy Physics HEP User Facility.
Fermilab is managed by Fermi Forward Discovery Group, LLC, acting
under Contract No. 89243024CSC000002.  MicroBooNE is supported by the
following: 
the U.S. Department of Energy, Office of Science, Offices of High Energy Physics and Nuclear Physics; 
the U.S. National Science Foundation; 
the Swiss National Science Foundation; 
the Science and Technology Facilities Council (STFC), part of the United Kingdom Research and Innovation; 
the Royal Society (United Kingdom); 
the UK Research and Innovation (UKRI) Future Leaders Fellowship; 
and the NSF AI Institute for Artificial Intelligence and Fundamental Interactions. 
Additional support for 
the laser calibration system and cosmic ray tagger was provided by the 
Albert Einstein Center for Fundamental Physics, Bern, Switzerland. We 
also acknowledge the contributions of technical and scientific staff 
to the design, construction, and operation of the MicroBooNE detector 
as well as the contributions of past collaborators to the development 
of MicroBooNE analyses, without whom this work would not have been 
possible. 
For the purpose of open access, the authors have applied 
a Creative Commons Attribution (CC BY) public copyright license to 
any Author Accepted Manuscript version arising from this submission.

%%%%%%%%%%%%%%%%%%%%%%%%%%%%%%%%%%%%%%%%%%%%%%%%%%%%%%%%%%%%%%%%%%%%%%%%%%%%

%\clearpage
\bibliography{main}
%\bibliography{apssamp}

%%%%%%%%%%%%%%%%%%%%%%%%%%%%%%%%%%%%%%%%%%%%%%%%%%%%%%%%%%%%%%%%%%%%%%%%%%%%

\end{document}

\centering
\large{First study of neutrino angle reconstruction using quasielastic-like interactions in MicroBooNE}

(Dated: \today)

%%%%%%%%%%%%%%%%%%%%%%%%%%%%%%%%%%%%%%%%%%%%%%%%%%%%

%\linenumbers

\justify
\section{Data Release}\label{datarel}

Overflow values are included in the last bin.
To compare an independent theoretical prediction to the data, one must first left-multiply the predicted cross section by the additional smearing matrix $A_{C}$ and then divide the content of each bin by the corresponding bin width or area to obtain a differential cross section.
The $A_{C}$ matrices are dimensionless.
The double-differential cross sections include correlations between the phase-space slices.
The data release with the data results, the covariance matrices, and the additional smearing matrices are included in the DataRelease.root file.
These are also included in the Supplemental Material in Sec.~\ref{datarel}, Sec.~\ref{cov}, and Sec.~\ref{smear}, respectively.
Instructions on how to use the data release and the description of the binning scheme are included in the README file.
\raggedbottom

\begin{table}[H]
\raggedright
\begin{adjustbox}{width=\textwidth}
\small
\begin{tabular}{ |c|c|c|c|c| }
\hline
\multicolumn{5}{|c|}{Cross section as a function of $\theta_{\mathrm{vis}}$, $\mathrm{\,all\,events}$} \\
\hline
\hline
Bin \# & $\theta_{\mathrm{vis}}$ low edge [deg] & $\theta_{\mathrm{vis}}$ high edge [deg] & Cross section [$10^{-38}\frac{\mathrm{cm}^{2}}{\mathrm{deg}\,\mathrm{Ar}}$] & Uncertainty [$10^{-38}\frac{\mathrm{cm}^{2}}{\mathrm{deg}\,\mathrm{Ar}}$] \\
\hline
\hline
1 & 0 & 5 & 0.3251 & 0.0175\\
2 & 5.0000 & 10.0000 & 0.4128 & 0.0177\\
3 & 10.0000 & 15.0000 & 0.3593 & 0.0166\\
4 & 15.0000 & 20.0000 & 0.2536 & 0.0117\\
5 & 20.0000 & 25.0000 & 0.1831 & 0.0101\\
6 & 25.0000 & 30.0000 & 0.1487 & 0.0073\\
7 & 30.0000 & 40.0000 & 0.0846 & 0.0044\\
8 & 40.0000 & 50.0000 & 0.0582 & 0.0039\\
9 & 50.0000 & 60.0000 & 0.0398 & 0.0033\\
10 & 60.0000 & 70.0000 & 0.0277 & 0.0034\\
11 & 70.0000 & 80.0000 & 0.0195 & 0.0032\\
12 & 80.0000 & 90.0000 & 0.0147 & 0.0032\\
13 & 90.0000 & 115.0000 & 0.0064 & 0.0015\\
14 & 115.0000 & 140.0000 & 0.0034 & 0.0013\\
15 & 140.0000 & 180.0000 & 0.0012 & 0.0007\\
\hline
\end{tabular}
\end{adjustbox}
\end{table}

\begin{table}[H]
\raggedright
\begin{adjustbox}{width=\textwidth}
\small
\begin{tabular}{ |c|c|c|c|c| }
\hline
\multicolumn{5}{|c|}{Cross section as a function of $\theta_{\mathrm{vis}}$, $\mathrm{0\,<\,E_{re\textit{c}o}\,<\,0.5\,GeV}$} \\
\hline
\hline
Bin \# & $\theta_{\mathrm{vis}}$ low edge [deg] & $\theta_{\mathrm{vis}}$ high edge [deg] & Cross section [$10^{-38}\mathrm{\frac{cm^{2}}{deg\,GeV\,Ar}}$] & Uncertainty [$10^{-38}\mathrm{\frac{cm^{2}}{deg\,GeV\,Ar}}$] \\
\hline
\hline
1 & 0 & 10 & 0.0340 & 0.0054\\
2 & 10.0000 & 20.0000 & 0.0634 & 0.0072\\
3 & 20.0000 & 30.0000 & 0.0585 & 0.0067\\
4 & 30.0000 & 40.0000 & 0.0483 & 0.0067\\
5 & 40.0000 & 50.0000 & 0.0419 & 0.0064\\
6 & 50.0000 & 60.0000 & 0.0342 & 0.0058\\
7 & 60.0000 & 70.0000 & 0.0297 & 0.0055\\
8 & 70.0000 & 80.0000 & 0.0272 & 0.0047\\
9 & 80.0000 & 90.0000 & 0.0256 & 0.0048\\
10 & 90.0000 & 115.0000 & 0.0141 & 0.0028\\
11 & 115.0000 & 180.0000 & 0.0048 & 0.0008\\
\hline
\end{tabular}
\end{adjustbox}
\end{table}

\begin{table}[H]
\raggedright
\begin{adjustbox}{width=\textwidth}
\small
\begin{tabular}{ |c|c|c|c|c| }
\hline
\multicolumn{5}{|c|}{Cross section as a function of $\theta_{\mathrm{vis}}$, $\mathrm{0.5\,<\,E_{re\textit{c}o}\,<\,0.8\,GeV}$} \\
\hline
\hline
Bin \# & $\theta_{\mathrm{vis}}$ low edge [deg] & $\theta_{\mathrm{vis}}$ high edge [deg] & Cross section [$10^{-38}\mathrm{\frac{cm^{2}}{deg\,GeV\,Ar}}$] & Uncertainty [$10^{-38}\mathrm{\frac{cm^{2}}{deg\,GeV\,Ar}}$] \\
\hline
\hline
1 & 0.0000 & 5.0000 & 0.2711 & 0.0289\\
2 & 5.0000 & 10.0000 & 0.4210 & 0.0311\\
3 & 10.0000 & 15.0000 & 0.4779 & 0.0346\\
4 & 15.0000 & 20.0000 & 0.4171 & 0.0307\\
5 & 20.0000 & 25.0000 & 0.3010 & 0.0252\\
6 & 25.0000 & 30.0000 & 0.2471 & 0.0222\\
7 & 30.0000 & 40.0000 & 0.1559 & 0.0135\\
8 & 40.0000 & 50.0000 & 0.1265 & 0.0141\\
9 & 50.0000 & 60.0000 & 0.0992 & 0.0137\\
10 & 60.0000 & 180.0000 & 0.0112 & 0.0015\\
\hline
\end{tabular}
\end{adjustbox}
\end{table}

\begin{table}[H]
\raggedright
\begin{adjustbox}{width=\textwidth}
\small
\begin{tabular}{ |c|c|c|c|c| }
\hline
\multicolumn{5}{|c|}{Cross section as a function of $\theta_{\mathrm{vis}}$, $\mathrm{0.8\,<\,E_{re\textit{c}o}\,<\,2\,GeV}$} \\
\hline
\hline
Bin \# & $\theta_{\mathrm{vis}}$ low edge [deg] & $\theta_{\mathrm{vis}}$ high edge [deg] & Cross section [$10^{-38}\mathrm{\frac{cm^{2}}{deg\,GeV\,Ar}}$] & Uncertainty [$10^{-38}\mathrm{\frac{cm^{2}}{deg\,GeV\,Ar}}$] \\
\hline
\hline
1 & 0.0000 & 5.0000 & 0.1817 & 0.0161\\
2 & 5.0000 & 10.0000 & 0.2489 & 0.0211\\
3 & 10.0000 & 15.0000 & 0.1707 & 0.0165\\
4 & 15.0000 & 20.0000 & 0.0923 & 0.0157\\
5 & 20.0000 & 30.0000 & 0.0460 & 0.0099\\
6 & 30.0000 & 180.0000 & 0.0033 & 0.0006\\
\hline
\end{tabular}
\end{adjustbox}
\end{table}

\begin{table}[H]
\raggedright
\begin{adjustbox}{width=\textwidth}
\small
\begin{tabular}{ |c|c|c|c|c| }
\hline
\multicolumn{5}{|c|}{Cross section as a function of $\theta_{\mathrm{vis}}$, $\mathrm{p_{n}\,<\,0.2\,GeV/\textit{c}}$} \\
\hline
\hline
Bin \# & $\theta_{\mathrm{vis}}$ low edge [deg] & $\theta_{\mathrm{vis}}$ high edge [deg] & Cross section [$10^{-38}\mathrm{\frac{cm^{2}}{deg\,(GeV/\textit{c})\,Ar}}$] & Uncertainty [$10^{-38}\mathrm{\frac{cm^{2}}{deg\,(GeV/\textit{c})\,Ar}}$] \\
\hline
\hline
1 & 0.0000 & 10.0000 & 1.2597 & 0.1080\\
2 & 10.0000 & 20.0000 & 0.7576 & 0.0612\\
3 & 20.0000 & 180.0000 & 0.0151 & 0.0020\\
\hline
\end{tabular}
\end{adjustbox}
\end{table}

\begin{table}[H]
\raggedright
\begin{adjustbox}{width=\textwidth}
\small
\begin{tabular}{ |c|c|c|c|c| }
\hline
\multicolumn{5}{|c|}{Cross section as a function of $\theta_{\mathrm{vis}}$, $\mathrm{0.2\,<\,p_{n}\,<\,0.4\,GeV/\textit{c}}$} \\
\hline
\hline
Bin \# & $\theta_{\mathrm{vis}}$ low edge [deg] & $\theta_{\mathrm{vis}}$ high edge [deg] & Cross section [$10^{-38}\mathrm{\frac{cm^{2}}{deg\,(GeV/\textit{c})\,Ar}}$] & Uncertainty [$10^{-38}\mathrm{\frac{cm^{2}}{deg\,(GeV/\textit{c})\,Ar}}$] \\
\hline
\hline
1 & 0.0000 & 10.0000 & 0.5031 & 0.0549\\
2 & 10.0000 & 20.0000 & 0.6764 & 0.0698\\
3 & 20.0000 & 30.0000 & 0.4177 & 0.0396\\
4 & 30.0000 & 40.0000 & 0.2366 & 0.0244\\
5 & 40.0000 & 50.0000 & 0.1193 & 0.0194\\
6 & 50.0000 & 60.0000 & 0.0711 & 0.0198\\
7 & 60.0000 & 180.0000 & 0.0060 & 0.0018\\
\hline
\end{tabular}
\end{adjustbox}
\end{table}

\begin{table}[H]
\raggedright
\begin{adjustbox}{width=\textwidth}
\small
\begin{tabular}{ |c|c|c|c|c| }
\hline
\multicolumn{5}{|c|}{Cross section as a function of $\theta_{\mathrm{vis}}$, $\mathrm{0.4\,<\,p_{n}\,<\,1\,GeV/\textit{c}}$} \\
\hline
\hline
Bin \# & $\theta_{\mathrm{vis}}$ low edge [deg] & $\theta_{\mathrm{vis}}$ high edge [deg] & Cross section [$10^{-38}\mathrm{\frac{cm^{2}}{deg\,(GeV/\textit{c})\,Ar}}$] & Uncertainty [$10^{-38}\mathrm{\frac{cm^{2}}{deg\,(GeV/\textit{c})\,Ar}}$] \\
\hline
\hline
1 & 0.0000 & 30.0000 & 0.0268 & 0.0088\\
2 & 30.0000 & 40.0000 & 0.0640 & 0.0185\\
3 & 40.0000 & 50.0000 & 0.0566 & 0.0137\\
4 & 50.0000 & 60.0000 & 0.0410 & 0.0104\\
5 & 60.0000 & 70.0000 & 0.0254 & 0.0080\\
6 & 70.0000 & 80.0000 & 0.0214 & 0.0067\\
7 & 80.0000 & 90.0000 & 0.0229 & 0.0062\\
8 & 90.0000 & 115.0000 & 0.0102 & 0.0033\\
9 & 115.0000 & 180.0000 & 0.0038 & 0.0014\\
\hline
\end{tabular}
\end{adjustbox}
\end{table}

\begin{table}[H]
\raggedright
\begin{adjustbox}{width=\textwidth}
\small
\begin{tabular}{ |c|c|c|c|c| }
\hline
\multicolumn{5}{|c|}{Cross section as a function of $\theta_{\mathrm{vis}}$, $\mathrm{|p_{miss}|\,<\,0.10\,GeV/\textit{c}}$} \\
\hline
\hline
Bin \# & $\theta_{\mathrm{vis}}$ low edge [deg] & $\theta_{\mathrm{vis}}$ high edge [deg] & Cross section [$10^{-38}\mathrm{\frac{cm^{2}}{deg\,(GeV/\textit{c})\,Ar}}$] & Uncertainty [$10^{-38}\mathrm{\frac{cm^{2}}{deg\,(GeV/\textit{c})\,Ar}}$] \\
\hline
\hline
1 & 0.0000 & 10.0000 & 2.1682 & 0.1707\\
2 & 10.0000 & 20.0000 & 1.7225 & 0.1269\\
3 & 20.0000 & 30.0000 & 0.6172 & 0.0860\\
4 & 30.0000 & 40.0000 & 0.3550 & 0.0492\\
5 & 40.0000 & 50.0000 & 0.2783 & 0.0394\\
6 & 50.0000 & 60.0000 & 0.1702 & 0.0328\\
7 & 60.0000 & 180.0000 & 0.0273 & 0.0052\\
\hline
\end{tabular}
\end{adjustbox}
\end{table}

\begin{table}[H]
\raggedright
\begin{adjustbox}{width=\textwidth}
\small
\begin{tabular}{ |c|c|c|c|c| }
\hline
\multicolumn{5}{|c|}{Cross section as a function of $\theta_{\mathrm{vis}}$, $\mathrm{0.10\,<\,|p_{miss}|\,<\,0.20\,GeV/\textit{c}}$} \\
\hline
\hline
Bin \# & $\theta_{\mathrm{vis}}$ low edge [deg] & $\theta_{\mathrm{vis}}$ high edge [deg] & Cross section [$10^{-38}\mathrm{\frac{cm^{2}}{deg\,(GeV/\textit{c})\,Ar}}$] & Uncertainty [$10^{-38}\mathrm{\frac{cm^{2}}{deg\,(GeV/\textit{c})\,Ar}}$] \\
\hline
\hline
1 & 0.0000 & 15.0000 & 0.9937 & 0.0888\\
2 & 15.0000 & 30.0000 & 0.5525 & 0.0557\\
3 & 30.0000 & 40.0000 & 0.3465 & 0.0553\\
4 & 40.0000 & 50.0000 & 0.2168 & 0.0485\\
5 & 50.0000 & 60.0000 & 0.1297 & 0.0328\\
6 & 60.0000 & 70.0000 & 0.0909 & 0.0289\\
7 & 70.0000 & 80.0000 & 0.0783 & 0.0264\\
8 & 80.0000 & 90.0000 & 0.0588 & 0.0284\\
9 & 90.0000 & 115.0000 & 0.0308 & 0.0143\\
10 & 115.0000 & 180.0000 & 0.0154 & 0.0039\\
\hline
\end{tabular}
\end{adjustbox}
\end{table}

\begin{table}[H]
\raggedright
\begin{adjustbox}{width=\textwidth}
\small
\begin{tabular}{ |c|c|c|c|c| }
\hline
\multicolumn{5}{|c|}{Cross section as a function of $\theta_{\mathrm{vis}}$, $\mathrm{0.20\,<\,|p_{miss}|\,<\,0.50\,GeV/\textit{c}}$} \\
\hline
\hline
Bin \# & $\theta_{\mathrm{vis}}$ low edge [deg] & $\theta_{\mathrm{vis}}$ high edge [deg] & Cross section [$10^{-38}\mathrm{\frac{cm^{2}}{deg\,(GeV/\textit{c})\,Ar}}$] & Uncertainty [$10^{-38}\mathrm{\frac{cm^{2}}{deg\,(GeV/\textit{c})\,Ar}}$] \\
\hline
\hline
1 & 0.0000 & 15.0000 & 0.1582 & 0.0245\\
2 & 15.0000 & 30.0000 & 0.1313 & 0.0229\\
3 & 30.0000 & 40.0000 & 0.0878 & 0.0237\\
4 & 40.0000 & 50.0000 & 0.0782 & 0.0177\\
5 & 50.0000 & 60.0000 & 0.0647 & 0.0129\\
6 & 60.0000 & 70.0000 & 0.0390 & 0.0101\\
7 & 70.0000 & 80.0000 & 0.0212 & 0.0097\\
8 & 80.0000 & 90.0000 & 0.0174 & 0.0079\\
9 & 90.0000 & 115.0000 & 0.0129 & 0.0046\\
10 & 115.0000 & 180.0000 & 0.0050 & 0.0018\\
\hline
\end{tabular}
\end{adjustbox}
\end{table}

\section{Resolution study}\label{reso}

\begin{figure}[H]
	\centering  
	\includegraphics[width=0.49\linewidth]{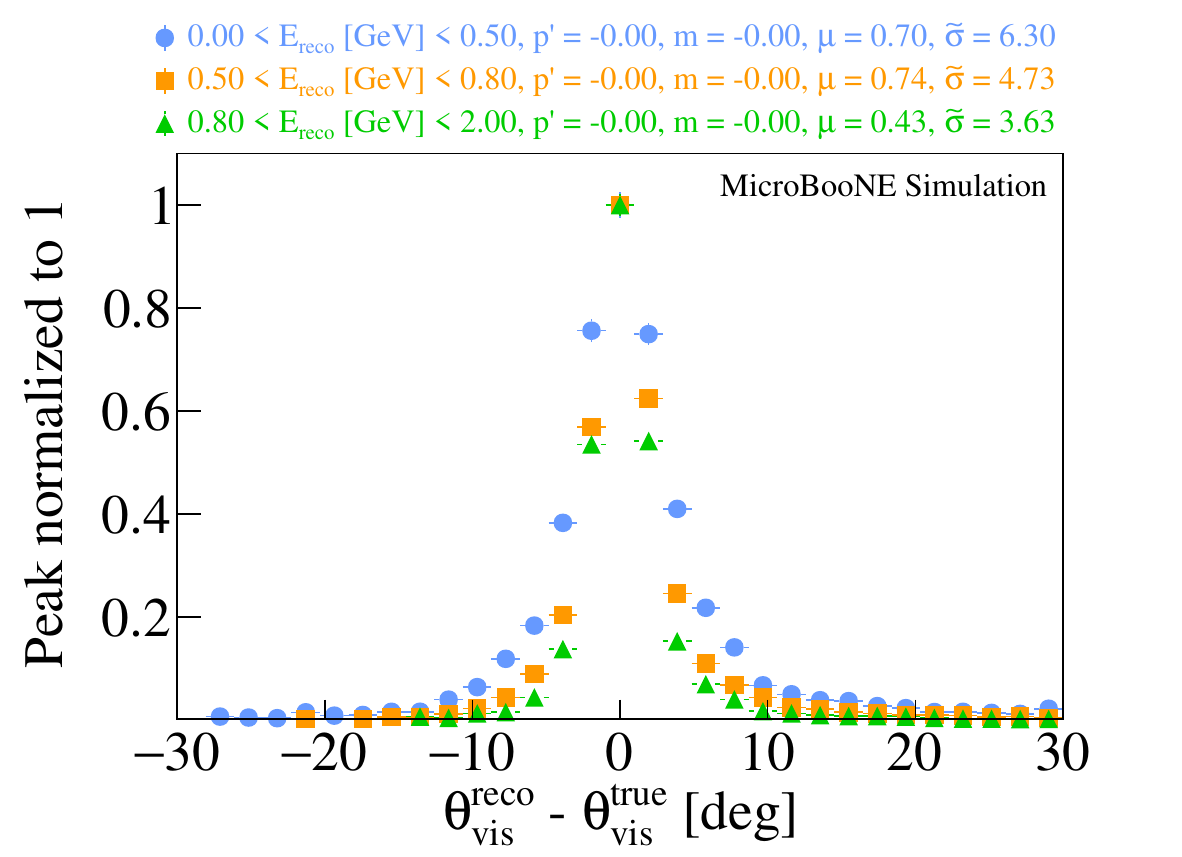}
	\includegraphics[width=0.49\linewidth]{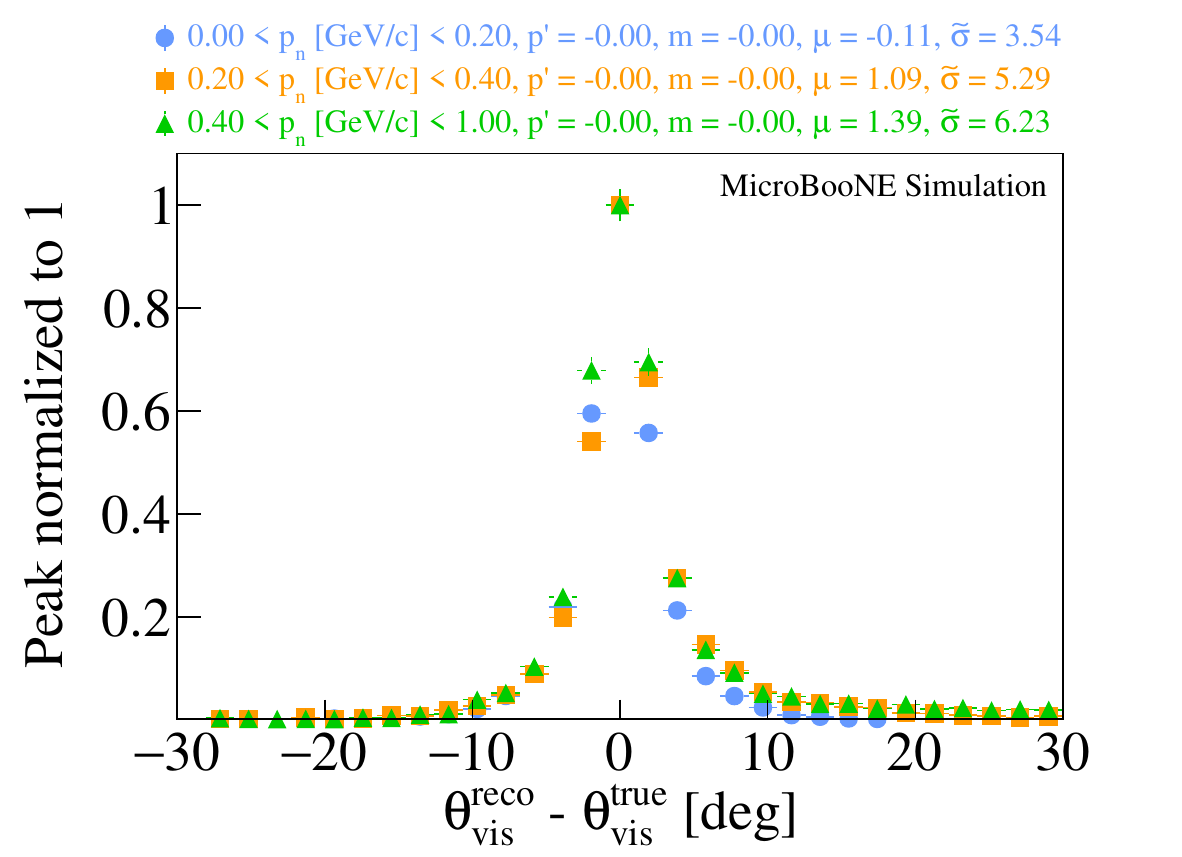}
    \includegraphics[width=0.49\linewidth]{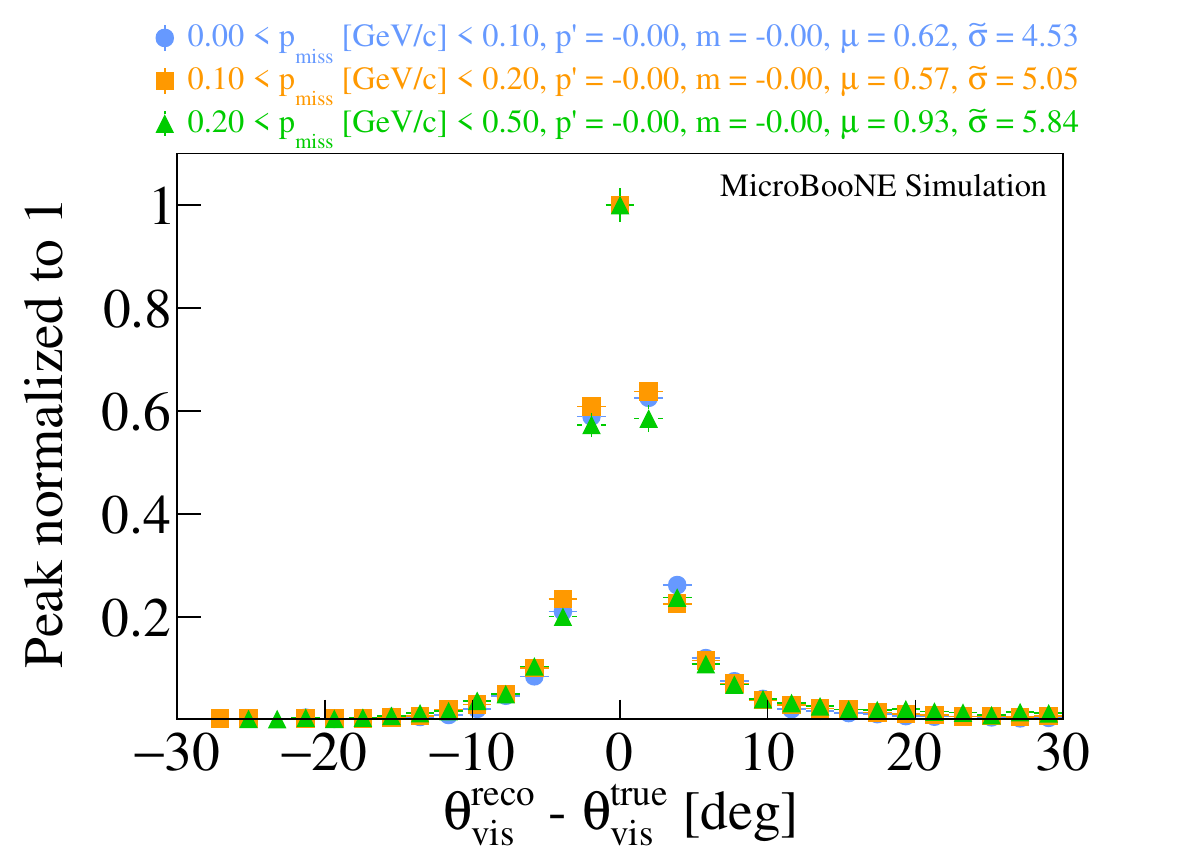}
    \caption{
The $\theta_{\mathrm{vis}}$ resolutions and biases presented in (top left) reconstructable energy slices, (top right) total struck nucleon momentum, and (bottom) total missing momentum. 
The peak location ($p^{\prime}$), median ($m$), mean value ($\mu$), and standard deviation ($\tilde{\sigma}$) are also presented.
    }
    \label{resofig}
\end{figure}

The dependence of the angular orientation $\theta_{\mathrm{vis}}$ is studied as a function of the reconstructable energy, total struck nucleon momentum, and total missing momentum in Fig.~\ref{resofig}.
The peak location ($p^{\prime}$), median ($m$), mean value ($\mu$), and standard deviation ($\tilde{\sigma}$) describing the $\theta_{\mathrm{vis}}$ distributions are also shown for each of the reconstructable energy regions of interest.

%%%%%%%%%%%%%%%%%%%%%%%%%%%%%%%%%%%%%%%%%%%%%%%%%%%%

%\clearpage
\section{Muon and proton angular correlations}\label{angle}

\begin{figure}[H]
	\centering  
	\includegraphics[width=0.49\linewidth]{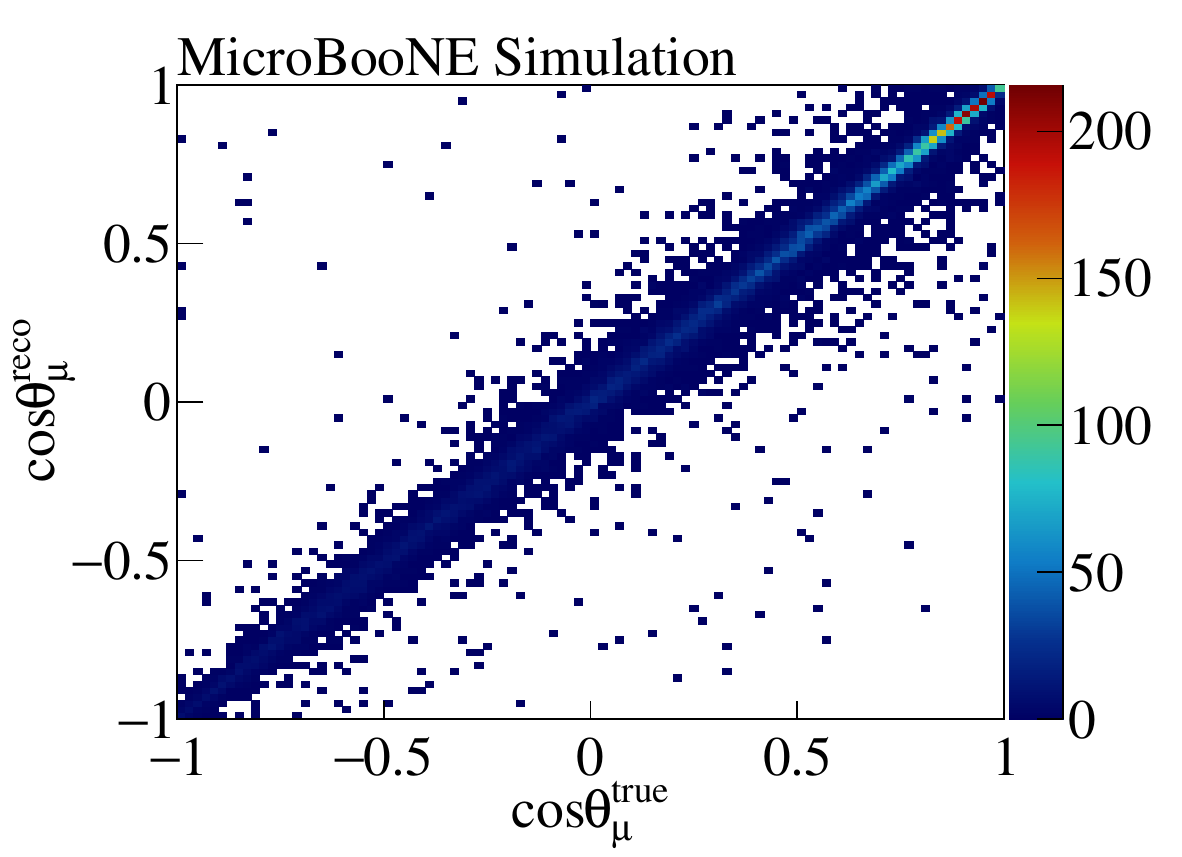}
	\includegraphics[width=0.49\linewidth]{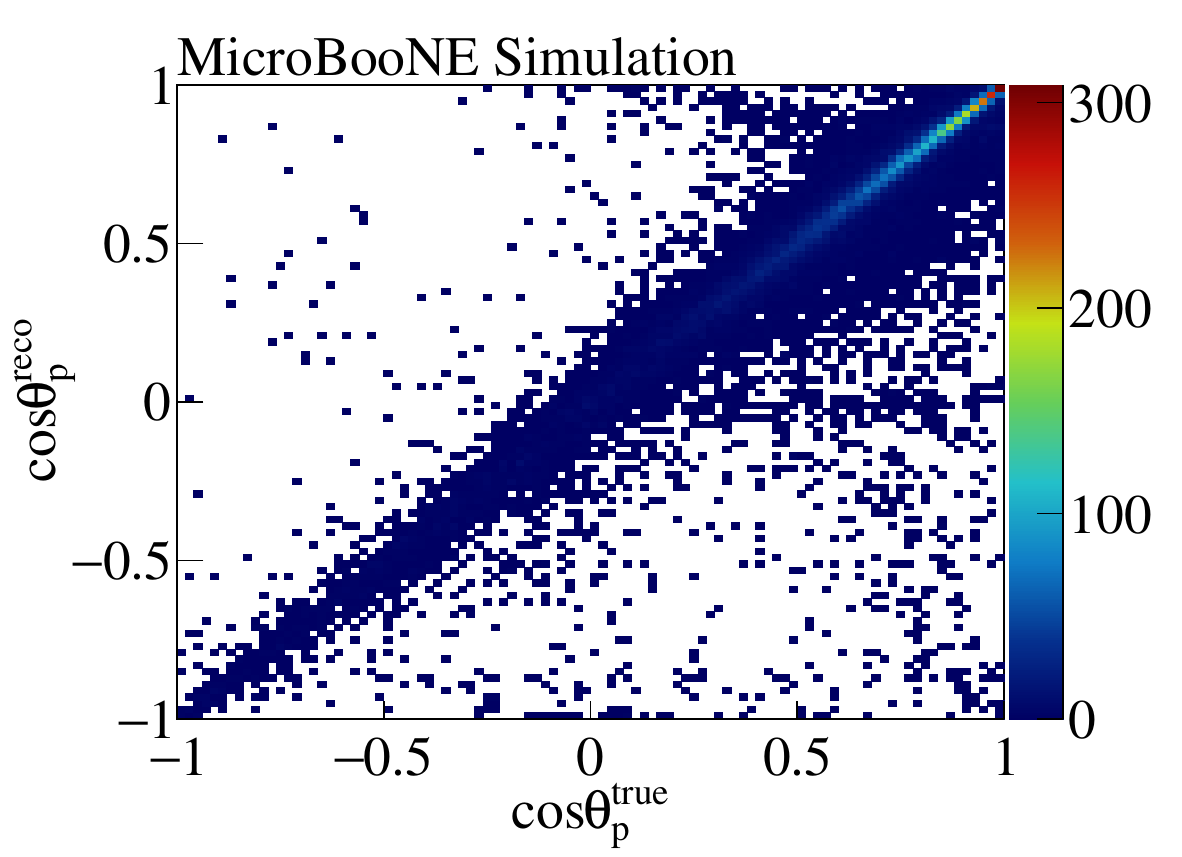}
	\includegraphics[width=0.49\linewidth]{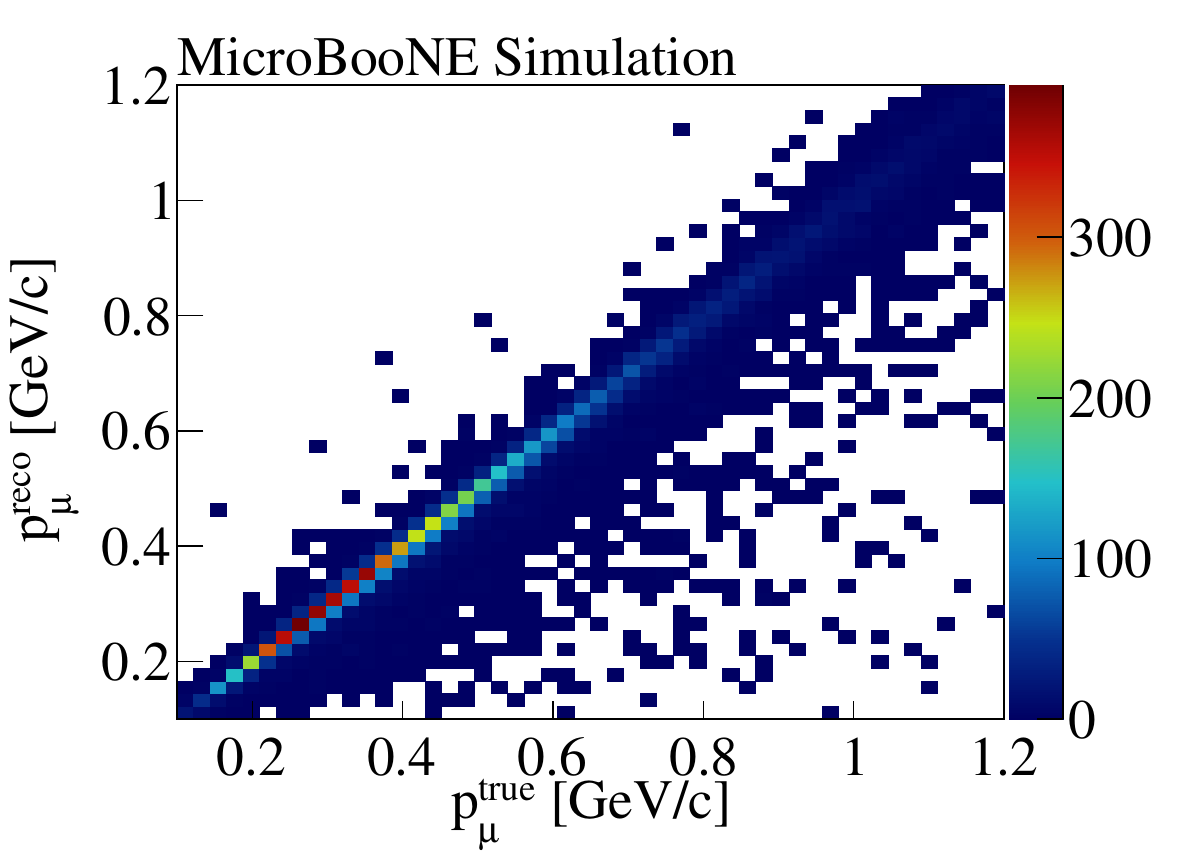}
	\includegraphics[width=0.49\linewidth]{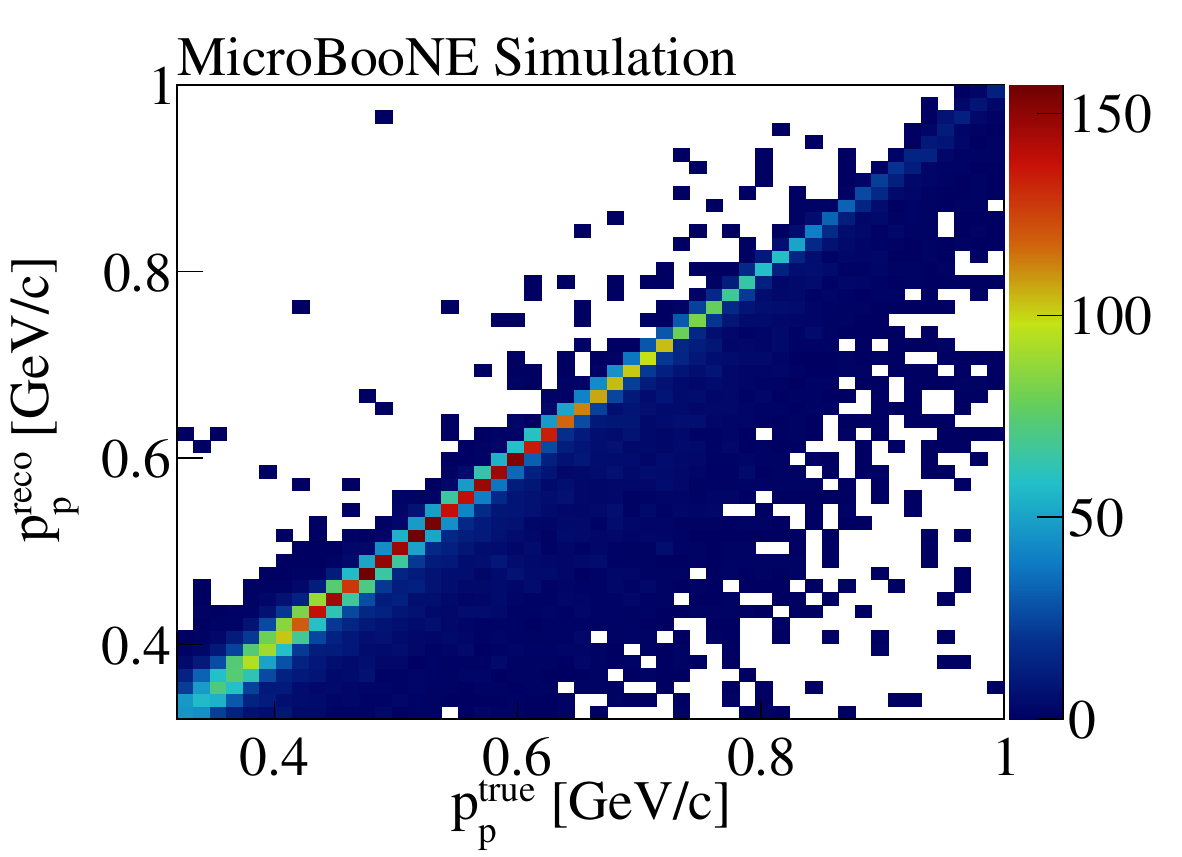}    
    \caption{
    (Top) The two-dimensional correlation between the reconstructed and true cos$\theta$ (left) for the muon and (right) for the proton.
    (Bottom) The two-dimensional correlation between the reconstructed and true momenta (left) for the muon and (right) for the proton.    
    }
    \label{resofigmuonproton}
\end{figure}

Figure~\ref{resofigmuonproton} shows the two-dimensional correlation between the reconstructed and true angle and momenta for the muon and the proton.
The muon ($\theta_{\mu}$) and proton ($\theta_{p}$) angles are defined with respect to the $z$-coordinate of the detector, which also corresponds to the direction of the incoming neutrinos.
They both demonstrate that minimal biases are observed in the angular and momentum reconstruction of the two particles.

%%%%%%%%%%%%%%%%%%%%%%%%%%%%%%%%%%%%%%%%%%%%%%%%%%%%

\clearpage
\section{Covariance Matrices}\label{cov}

The figures below present the covariance matrices (Cov) for the cross section results presented in this work expressed in units of 10$^{-76}$.
They are also included in the DataRelease.root file.
More details on how to manipulate the covariance matrices in order to calculate a $\chi^{2}$ GoF metric can be found in the README file.

\begin{figure}[H]
\centering 
\includegraphics[width=0.48\linewidth]{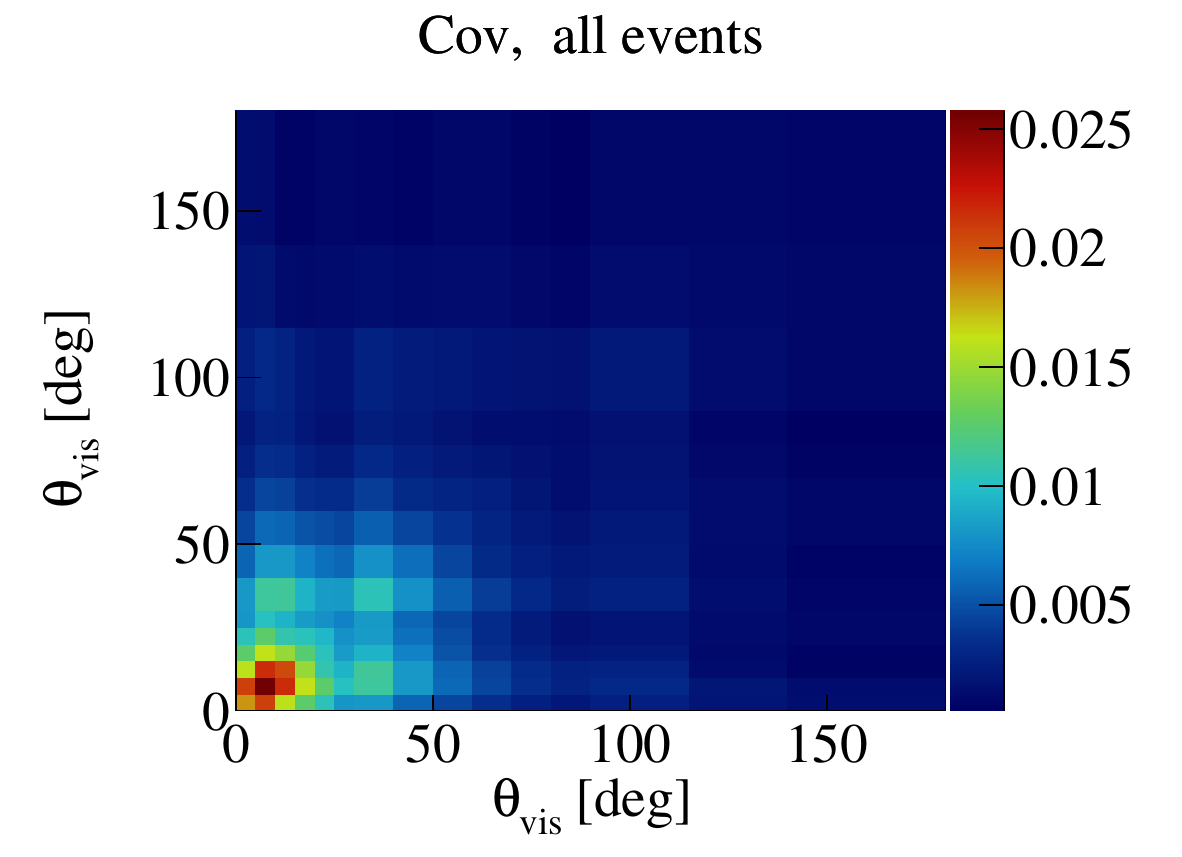}
\includegraphics[width=0.48\linewidth]{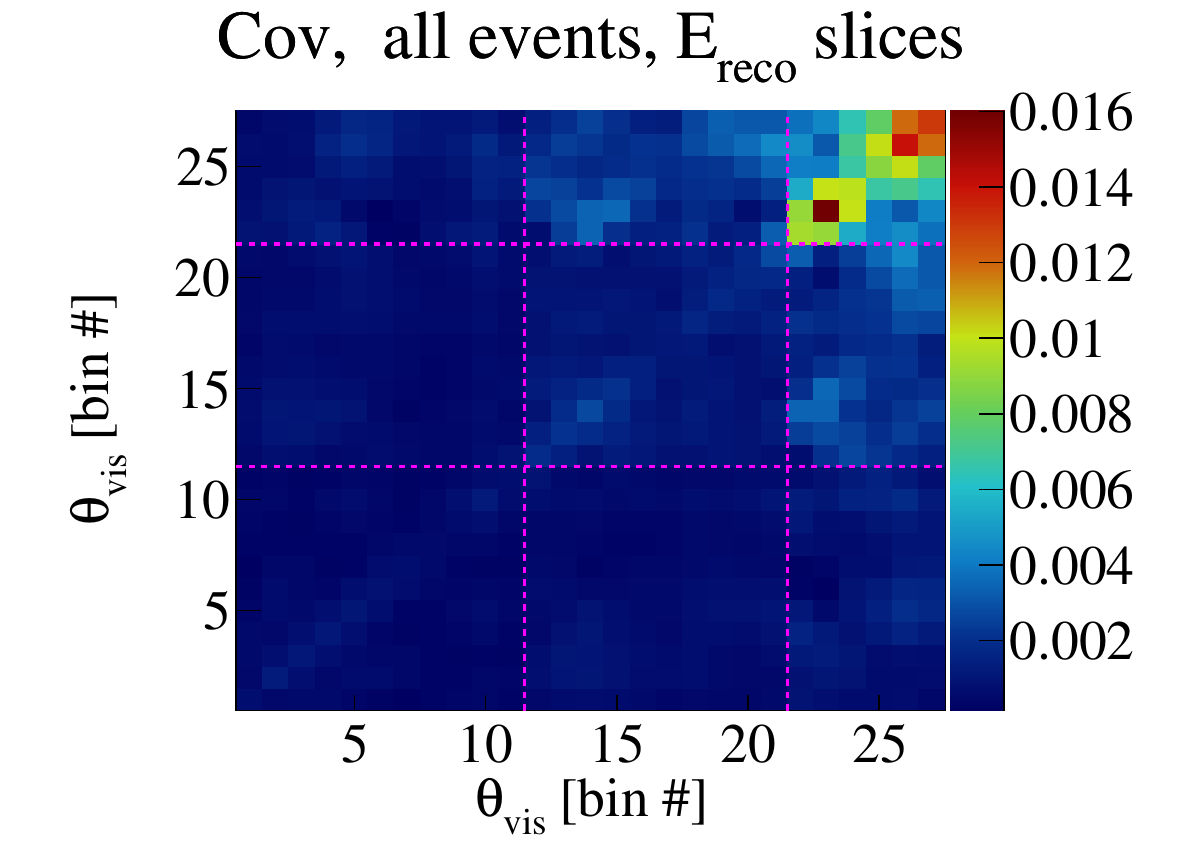}
\includegraphics[width=0.48\linewidth]{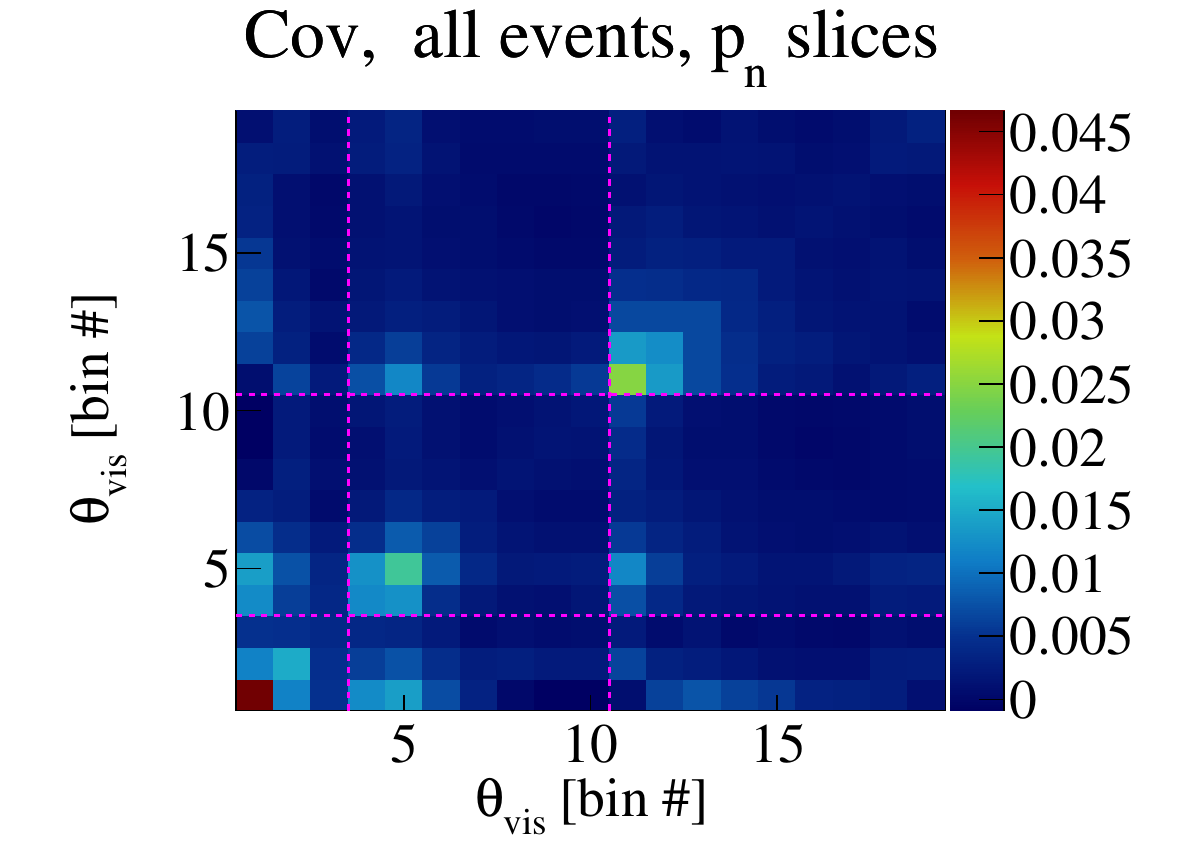}
\includegraphics[width=0.48\linewidth]{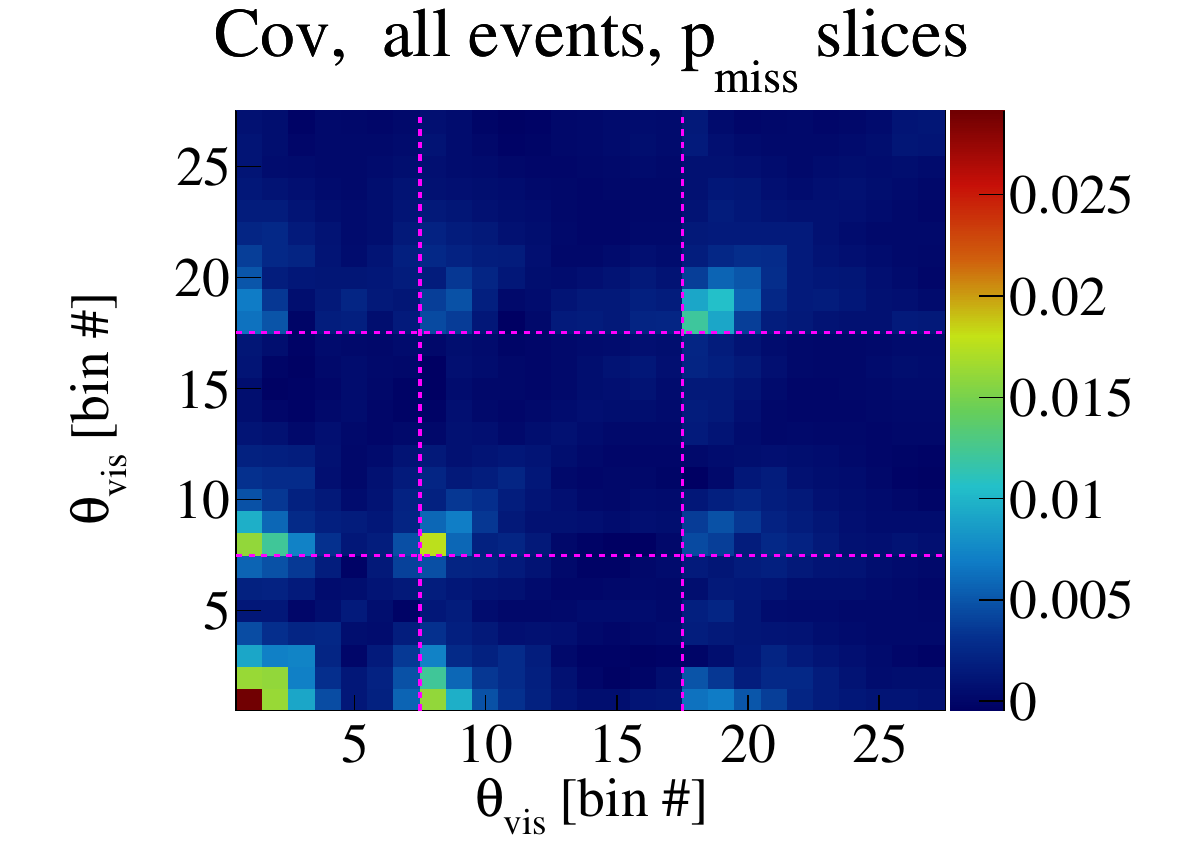}
\caption{
Covariance matrix for (top left) $\theta_{\mathrm{vis}}$, (top right) $\theta_{\mathrm{vis}}$ in reconstructable energy slices, (bottom left) $\theta_{\mathrm{vis}}$ in total struck nucleon momentum slices, and (bottom right) $\theta_{\mathrm{vis}}$ in missing momentum slices.
The double-differential covariances are expressed as a function of the universal bin number defined in the bin scheme file and are separated by the magenta dashed lines.
}
\label{CovDeltaPn}
\end{figure}

%%%%%%%%%%%%%%%%%%%%%%%%%%%%%%%%%%%%%%%%%%%%%%%%%%%%

\clearpage
\section{Regularization Matrices}\label{smear}

The figures below present the regularization matrices (A$_{C}$) for the cross section results presented in this work.
They are also included in the DataRelease.root file.

\begin{figure}[H]
\centering 
\includegraphics[width=0.48\linewidth]{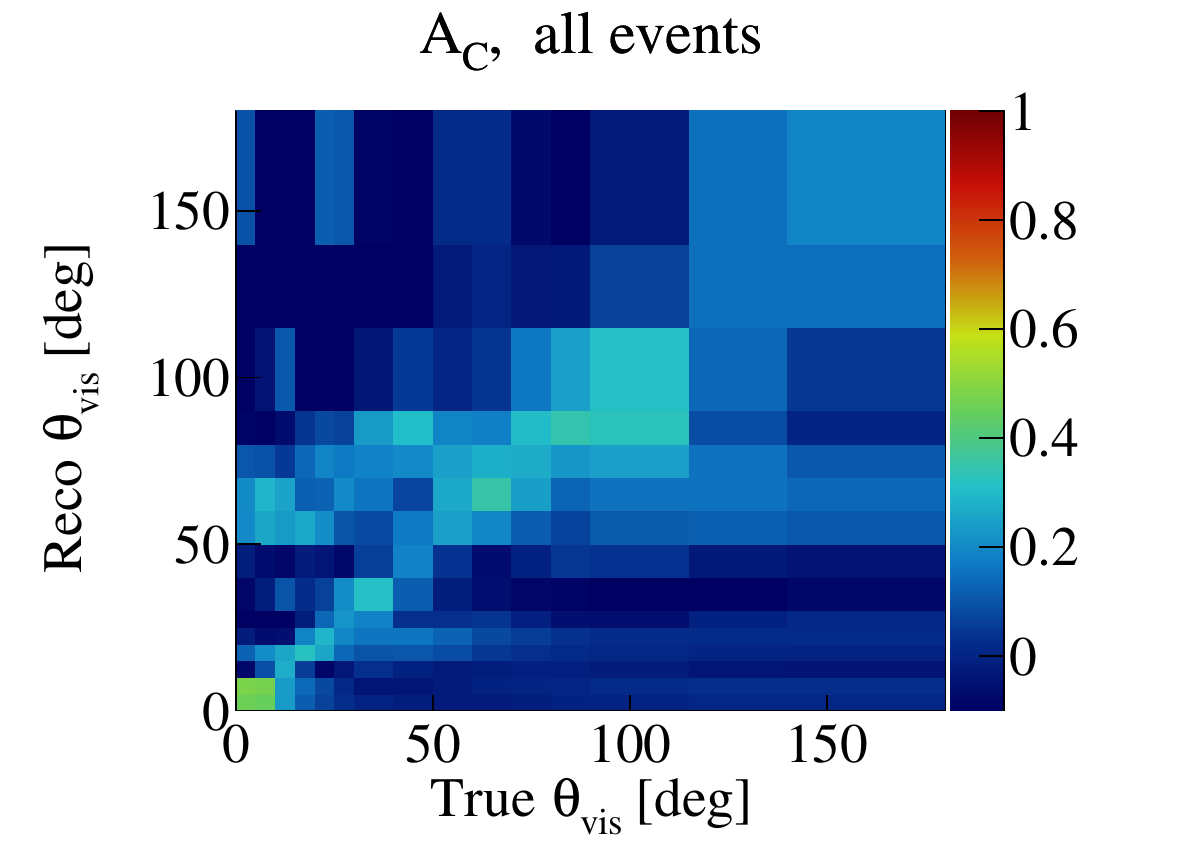}
\includegraphics[width=0.48\linewidth]{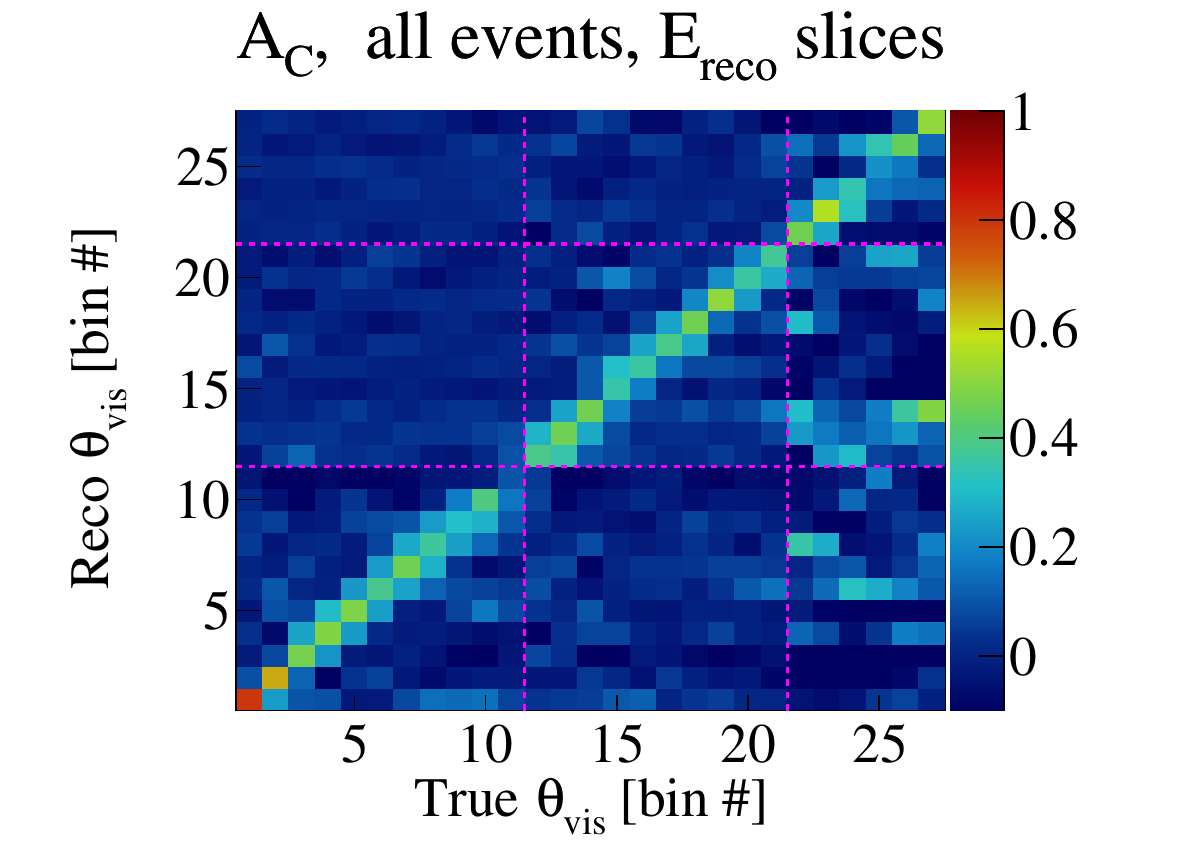}
\includegraphics[width=0.48\linewidth]{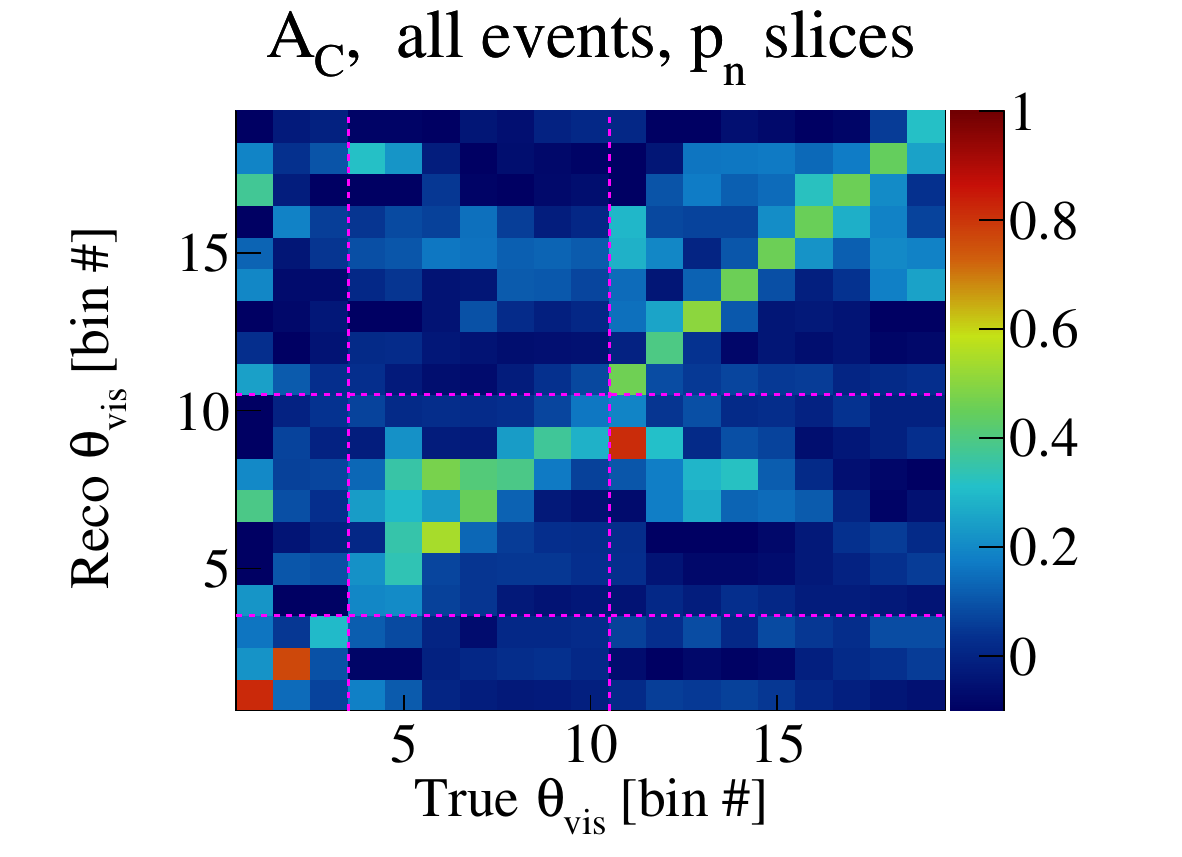}
\includegraphics[width=0.48\linewidth]{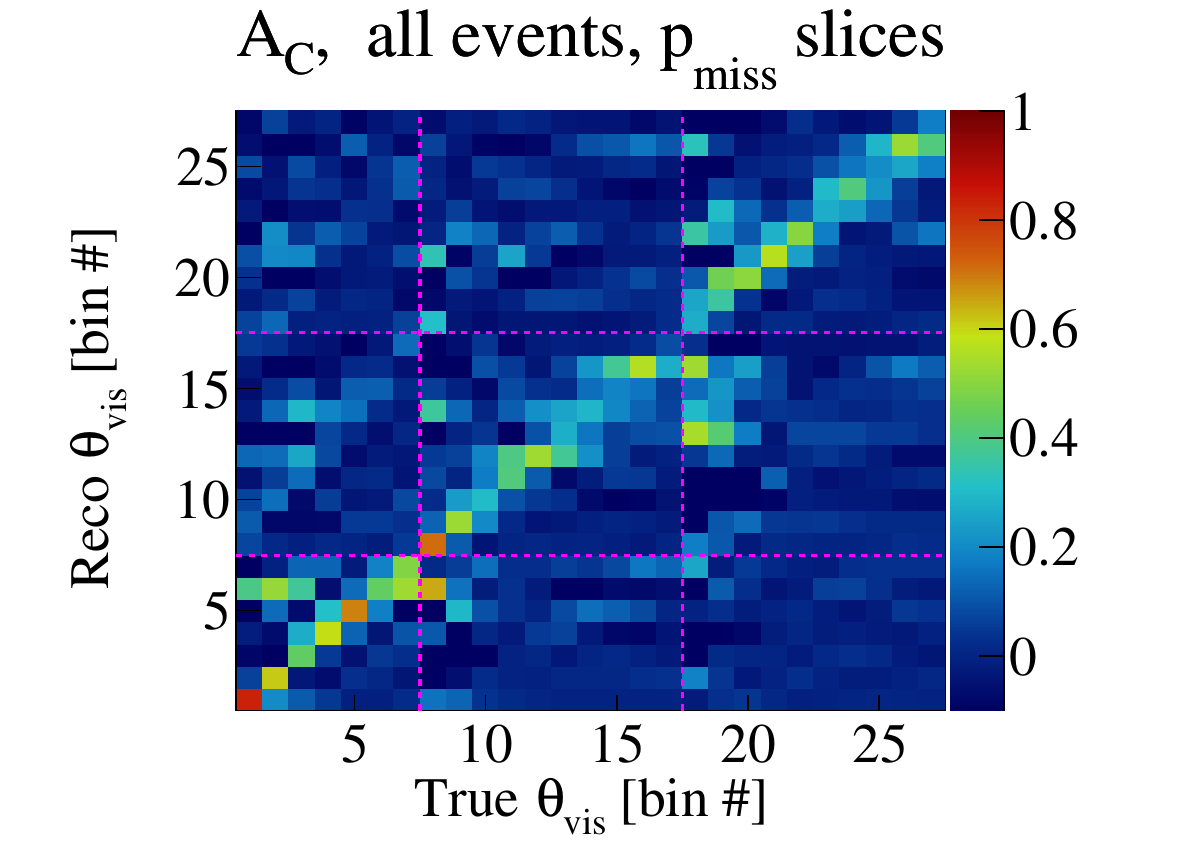}
\caption{
Additional smearing matrix for (top left) $\theta_{\mathrm{vis}}$, (top right) $\theta_{\mathrm{vis}}$ in reconstructable energy slices, (bottom left) $\theta_{\mathrm{vis}}$ in total struck nucleon momentum slices, and (bottom right) $\theta_{\mathrm{vis}}$ in missing momentum slices.
The double-differential covariances are expressed as a function of the universal bin number defined in the bin scheme file and are separated by the magenta dashed lines.
}
\label{AcDeltaPn}
\end{figure}

%%%%%%%%%%%%%%%%%%%%%%%%%%%%%%%%%%%%%%%%%%%%%%%%%%%%

\clearpage
\section{Neutron breakdown}\label{neutron}

\begin{figure}[H]
\centering
    \includegraphics[width=0.45\textwidth]{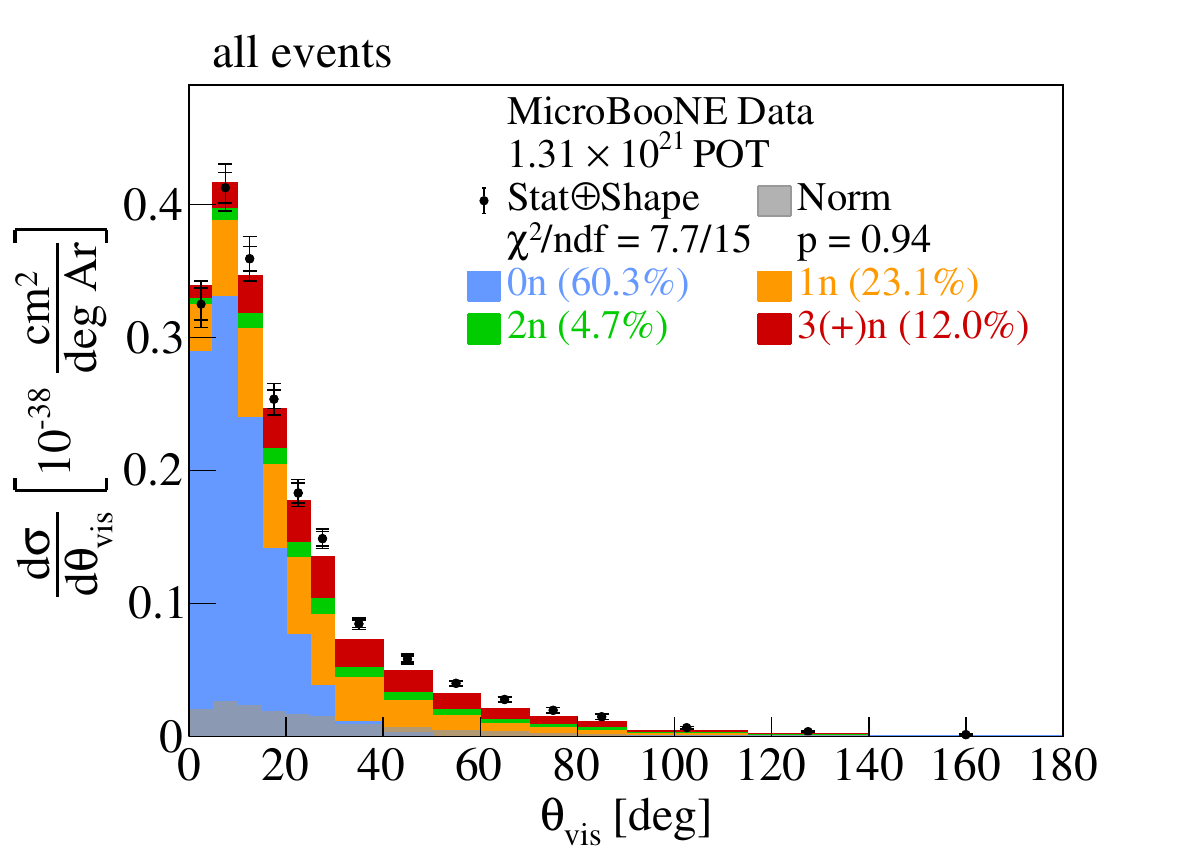}	
    \includegraphics[width=0.45\textwidth]{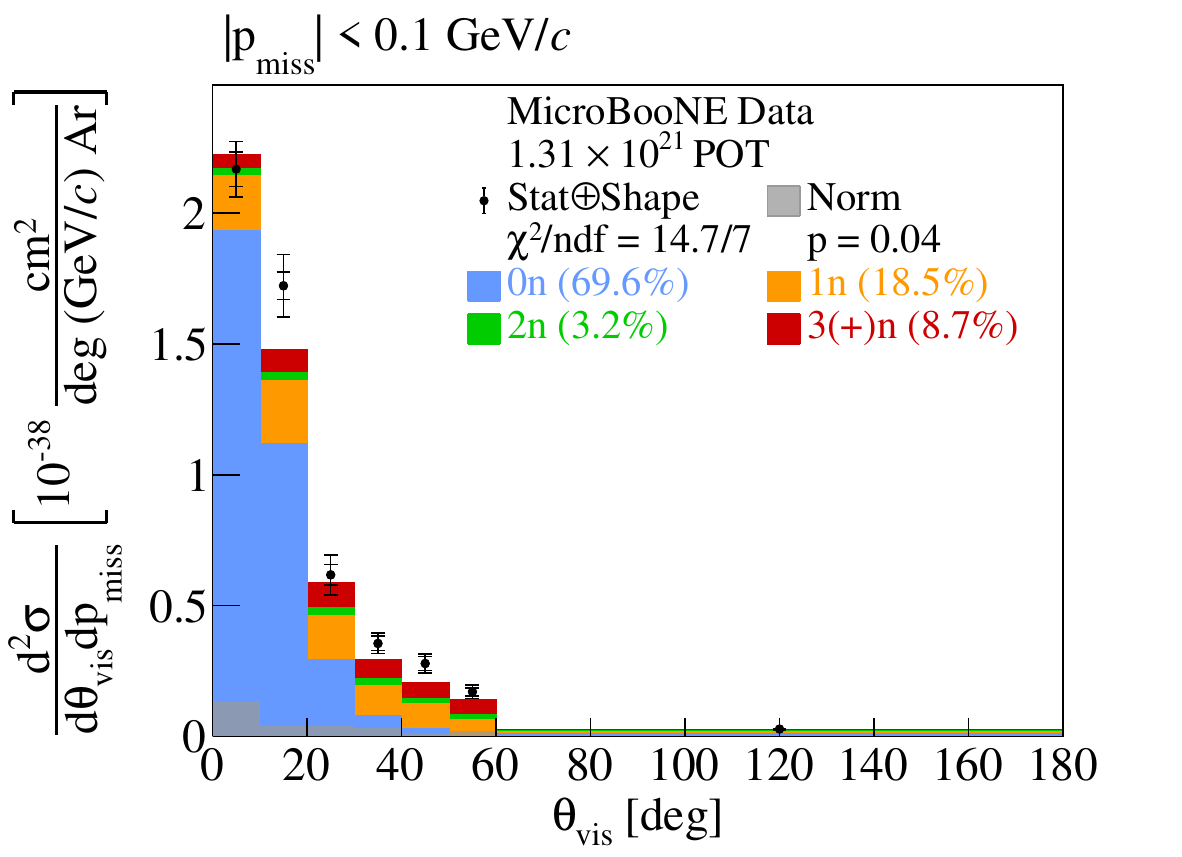}
    \includegraphics[width=0.45\textwidth]{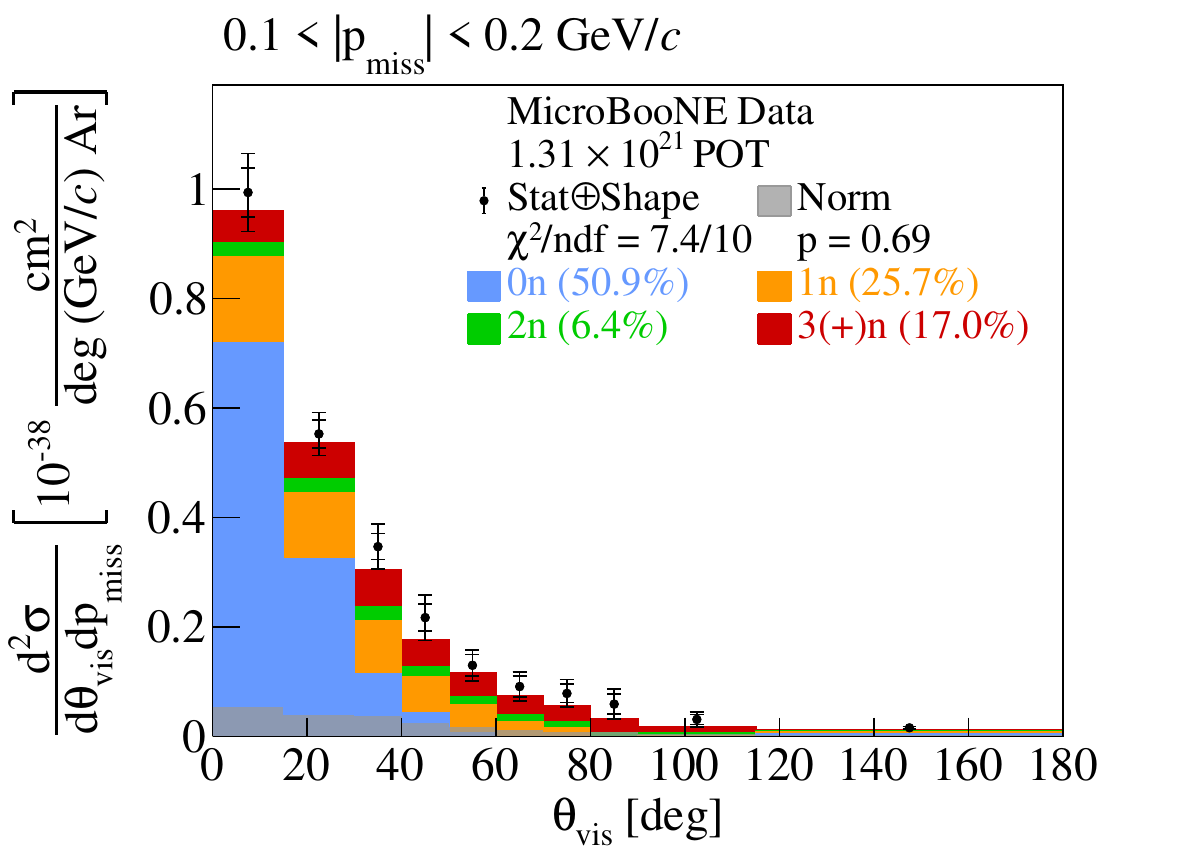}    
    \caption{
    The flux-integrated double-differential cross sections as a function of $\theta_{\mathrm{vis}}$ using (top left) all the selected events, (top right) events with $|p_{\mathrm{miss}}| <$ 0.1\,GeV/$c$, and (bottom) events with 0.1 $< |p_{\mathrm{miss}}| <$ 0.2\,GeV/$c$.  
Colored stacked histograms show the results of $\texttt{G18T}$ theoretical cross section calculations for events with 0 neutrons (light blue), 1 neutron (orange), 2 neutrons (green), and at least 3 neutrons (red).
Inner and outer error bars show the statistical and the statistical$\oplus$shape uncertainty at the 1$\sigma$, or 68\%, confidence level. 
The gray band shows the normalization systematic uncertainty.
The numbers in parentheses show the $\chi^{2}$/ndf calculation for each of the predictions.
    }
    \label{xsecThetaVisInSlices}
\end{figure}

The neutron breakdown of the $\theta_{\mathrm{vis}}$ cross section is presented in Fig.~\ref{xsecThetaVisInSlices} using (top left) all the selected events, (top right) events with low missing momentum, and (bottom) events with medium missing momentum.

%%%%%%%%%%%%%%%%%%%%%%%%%%%%%%%%%%%%%%%%%%%%%%%%%%%%

\clearpage
\section{Muon and proton angular precision}\label{muonprot}

We characterize the performance of the LArTPC reconstruction by comparing the true ($t$) and reconstructed ($r$) angle of the muon and proton events with respect to the $z$-direction of the detector, which also corresponds to the direction of the incoming neutrinos.
These are constructed using the three-dimensional true and reconstructed values for the muon and proton momentum vectors.
This comparison is performed using the selected Monte Carlo (MC) events that satisfy the CC1p0$\pi$ signal definition.
The difference between the true and reconstructed azimuthal angles for the muon ($\theta_{\mathrm{\mu,rt}} = \theta_{\mu,\mathrm{true}} - \theta_{\mu,\mathrm{reco}}$) and the proton ($\theta_{\mathrm{p,rt}} = \theta_{p,\mathrm{true}} - \theta_{p,\mathrm{reco}}$) is found to be better than 5$^{\mathrm{o}}$ for the majority of the events.
This demonstrates the excellent LArTPC reconstruction capabilities.
The peak location ($p^{\prime}$), median ($m$), mean value ($\mu$), and standard deviation ($\tilde{\sigma}$) describing the distributions are also shown.

\begin{figure}[H]
\centering
    \includegraphics[width=0.45\textwidth]{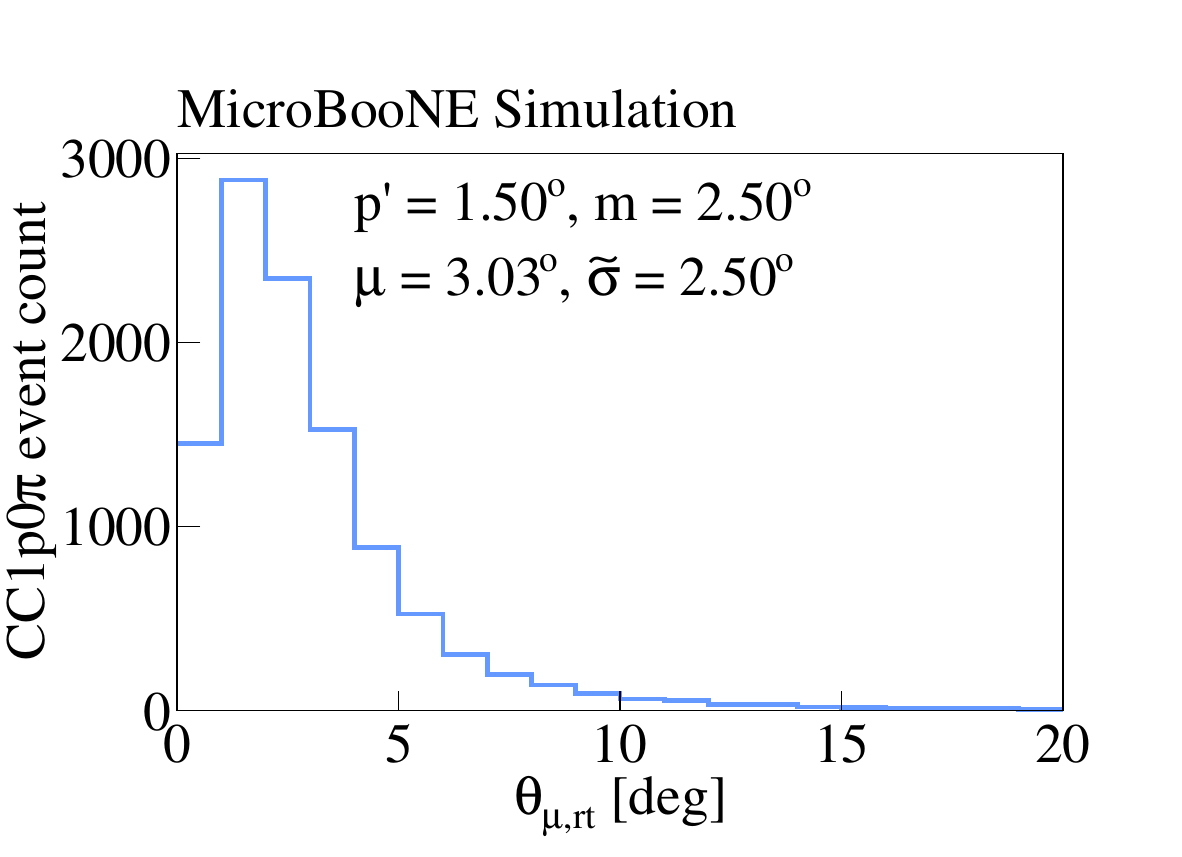}	
    \includegraphics[width=0.45\textwidth]{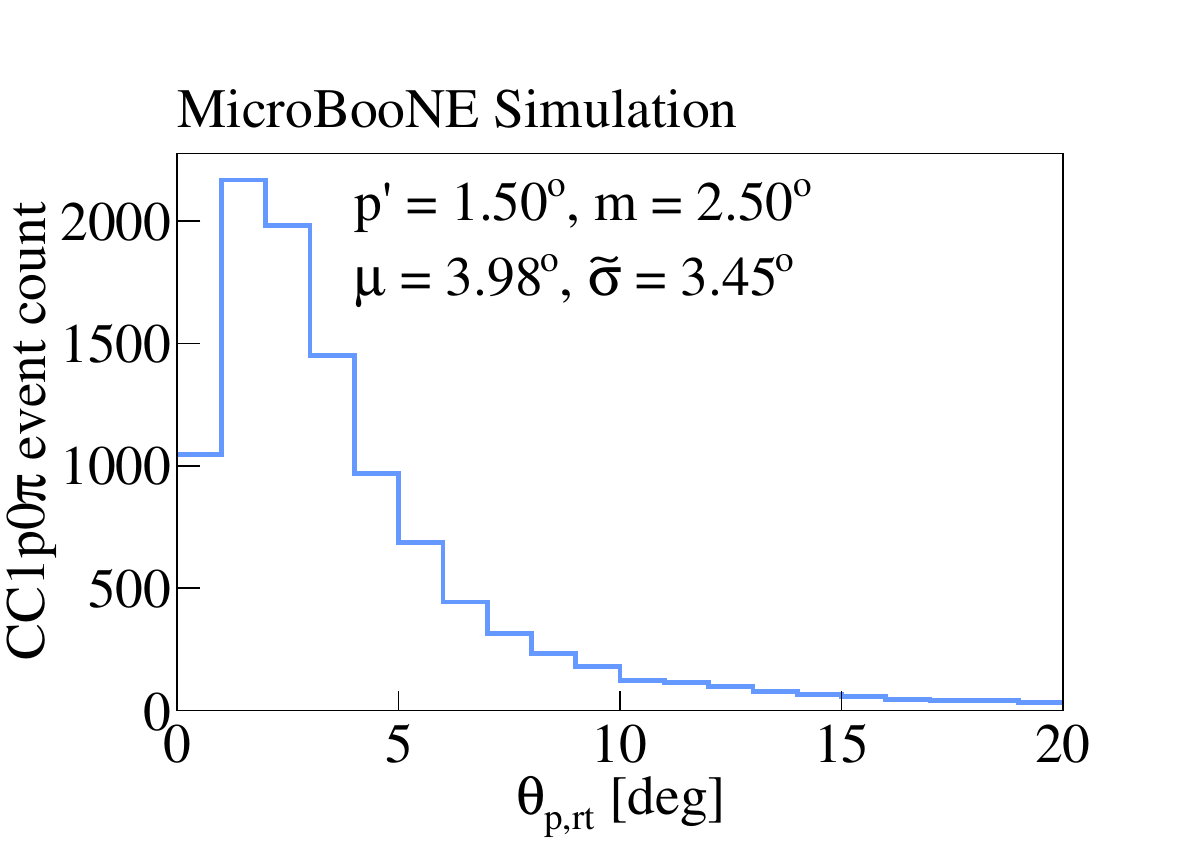} 
    \caption{
    Angular $\theta_{\mathrm{rt}}$ distribution using the selected signal CC1p0$\pi$ simulated events for (left) muons and (right) protons. The peak location ($p^{\prime}$), median ($m$), mean value ($\mu$), and standard deviation ($\tilde{\sigma}$) describing the distribution are also shown.
    }
    \label{muonprotonangle}
\end{figure}	

%%%%%%%%%%%%%%%%%%%%%%%%%%%%%%%%%%%%%%%%%%%%%%%%%%%%

%\clearpage
%\bibliographystyle{unsrt}
\bibliography{main}

%%%%%%%%%%%%%%%%%%%%%%%%%%%%%%%%%%%%%%%%%%%%%%%%%%%%